\documentclass[twocolumn,twocolappendix]{aastex63}
\def\simgeq{{\raise.0ex\hbox{$\mathchar"013E$}\mkern-14mu\lower1.2ex\hbox{$\mathchar"0218$}}} 

\def\Mdyn25{$M_{\rm dyn, R_{25}}$}

\usepackage{amsmath}
\usepackage{enumerate}
\usepackage{color}
\usepackage{longtable}
\usepackage{epsfig}
\usepackage{hyperref}
\usepackage{amsmath}
\usepackage{amssymb}
\usepackage{natbib}
\usepackage{graphicx}

\newcommand\incp{\ifmmode {i_\mathrm{phot}}\else \mbox{$i_\mathrm{phot}$}\fi}
\newcommand\inck{\ifmmode {i_\mathrm{kin}}\else \mbox{$i_\mathrm{kin}$}\fi}
\newcommand\pap{\ifmmode {\phi_\mathrm{phot}}\else \mbox{$\phi_\mathrm{phot}$}\fi}
\newcommand\pak{\ifmmode {\phi_\mathrm{kin}}\else \mbox{$\phi_\mathrm{kin}$}\fi}

\begin{document}

%
\title{PHANGS CO kinematics: disk orientations and rotation curves at 150\,pc resolution}

\author{Philipp Lang}
\affiliation{Max-Planck-Institut f\"ur Astronomie, K\"{o}nigstuhl 17 D-69117 Heidelberg, Germany}

\author{Sharon E. Meidt}
\affiliation{Sterrenkundig Observatorium, Universiteit Gent, Krijgslaan 281 S9, B-9000 Gent, Belgium}

\author{Erik Rosolowsky}
\affiliation{Department of Physics, University of Alberta, Edmonton, AB, Canada} 

\author{Joseph Nofech}
\affiliation{Department of Physics, University of Alberta, Edmonton, AB, Canada} 

\author{Eva Schinnerer}
\affiliation{Max-Planck-Institut f\"ur Astronomie, K\"{o}nigstuhl 17 D-69117 Heidelberg, Germany}

\author{Adam K. Leroy}
\affiliation{Department of Astronomy, The Ohio State University, 140 West 18th Avenue, Columbus, Ohio 43210, USA}

\author{Eric Emsellem}
\affiliation{European Southern Observatory, Karl-Schwarzschild Stra{\ss}e 2, D-85748 Garching bei M{\"u}nchen, Germany}
\affiliation{Univ Lyon, Univ Lyon1, ENS de Lyon, CNRS, Centre de Recherche Astrophysique de Lyon UMR5574, F-69230 Saint-Genis-Laval France}

\author{Ismael Pessa}
\affiliation{Max-Planck-Institut f\"ur Astronomie, K\"{o}nigstuhl 17 D-69117 Heidelberg, Germany}

\author{Simon C. O. Glover}
\affiliation{Instit\"ut  f\"{u}r Theoretische Astrophysik, Zentrum f\"{u}r Astronomie der Universit\"{a}t Heidelberg, Albert-Ueberle-Strasse 2, 69120 Heidelberg, Germany}

\author{Brent Groves}
\affiliation{Research School of Astronomy and Astrophysics, Australian National University, Canberra, ACT 2611, Australia}
\affiliation{International Centre for Radio Astronomy Research, The University of Western Australia, Crawley, WA 6009 Australia}

\author{Annie Hughes}
\affiliation{Universit\'{e} de Toulouse, UPS-OMP, 31028 Toulouse, France}
\affiliation{CNRS, IRAP, Av. du Colonel Roche BP 44346, 31028 Toulouse cedex 4, France}

\author{J.~M.~Diederik Kruijssen}
\affiliation{Astronomisches Rechen-Institut, Zentrum f\"{u}r Astronomie der Universit\"{a}t Heidelberg, M\"{o}nchhofstra\ss e 12-14, 69120 Heidelberg, Germany}

\author{Miguel Querejeta}
\affiliation{Observatorio Astron{\'o}mico Nacional (IGN), C/Alfonso XII 3, Madrid E-28014, Spain}

\author{Andreas Schruba}
\affiliation{Max-Planck-Institut f{\"u}r Extraterrestrische Physik, Giessenbachstra{\ss}e 1, D-85748 Garching bei M{\"u}nchen, Germany}


\author{Frank Bigiel}
\affiliation{Argelander-Institut f{\"u}r Astronomie, Universit\"{a}t Bonn, Auf dem H\"ugel 71, 53121 Bonn, Germany}

\author{Guillermo A. Blanc}
\affiliation{The Observatories of the Carnegie Institution for Science, 813 Santa Barbara St., Pasadena, CA, 91101}
\affiliation{Departamento de Astronom\'{i}a, Universidad de Chile, Camino del Observatorio 1515, Las Condes, Santiago, Chile}

\author{M\'{e}lanie Chevance}
\affiliation{Astronomisches Rechen-Institut, Zentrum f\"{u}r Astronomie der Universit\"{a}t Heidelberg, M\"{o}nchhofstra\ss e 12-14, 69120 Heidelberg, Germany}

\author{Dario Colombo}
\affiliation{Max-Planck-Institut f{\"u}r Radioastronomie, Auf dem H\"ugel 9, D-53121 Bonn, Germany}

\author{Christopher Faesi}
\affiliation{Max-Planck-Institut f\"ur Astronomie, K\"{o}nigstuhl 17 D-69117 Heidelberg, Germany}

\author{Jonathan D. Henshaw}
\affiliation{Max-Planck-Institut f\"ur Astronomie, K\"{o}nigstuhl 17 D-69117 Heidelberg, Germany}

\author{Cinthya N. Herrera}
\affiliation{Institut de Radioastronomie Millim\'etrique, 300 rue de la Piscine, 38406 Saint-Martin-d'H\`eres Cedex}

\author{Daizhong Liu}
\affiliation{Max-Planck-Institut f\"ur Astronomie, K\"{o}nigstuhl 17 D-69117 Heidelberg, Germany}

\author{J\'{e}r\^{o}me Pety}
\affiliation{Institut de Radioastronomie Millim\'etrique, 300 rue de la Piscine, 38406 Saint-Martin-d'H\`eres Cedex}
\affiliation{Sorbonne Universit\'e, Observatoire de Paris, Universit\'e PSL, CNRS, LERMA, 75014, Paris, France} 

\author{Johannes Puschnig}
\affiliation{Argelander-Institut f{\"u}r Astronomie, Universit\"{a}t Bonn, Auf dem H\"ugel 71, 53121 Bonn, Germany}

\author{Toshiki Saito}
\affiliation{Max-Planck-Institut f\"ur Astronomie, K\"{o}nigstuhl 17 D-69117 Heidelberg, Germany}

\author{Jiayi Sun}
\affiliation{Department of Astronomy, The Ohio State University, 140 West 18th Avenue, Columbus, Ohio 43210, USA}

\author{Antonio Usero}
\affiliation{Observatorio Astron{\'o}mico Nacional (IGN), C/Alfonso XII 3, Madrid E-28014, Spain}

\begin{abstract}
We present kinematic orientations and high resolution (150\,pc) rotation curves for 67 main sequence star-forming galaxies surveyed in CO (2-1) emission by PHANGS-ALMA. Our measurements are based on the application of a new fitting method tailored to CO velocity fields. Our approach identifies an optimal global orientation as a way to reduce the impact of non-axisymmetric (bar and spiral) features and the uneven spatial sampling characteristic of CO emission in the inner regions of nearby galaxies. The method performs especially well when applied to the large number of independent lines-of-sight contained in the PHANGS CO velocity fields mapped at $1''$~resolution. The high resolution rotation curves fitted to these data are sensitive probes of mass distribution in the inner regions of these galaxies. We use the inner slope as well as the amplitude of our fitted rotation curves to demonstrate that CO is a reliable global dynamical mass tracer. From the consistency between photometric orientations from the literature and kinematic orientations determined with our method, we infer that the shapes of stellar disks in the mass range of log($\rm M_{\star}(M_{\odot})$)=9.0-10.9 probed by our sample are very close to circular and have uniform thickness. 
\end{abstract}


\section{Introduction}
\label{Sect1.sec}
Modern extragalactic CO surveys measure the properties of molecular gas at cloud ($50{-}100$\,pc) scales with high completeness across large areas \citep[e.g.,][]{Schinnerer2013,freeman,Leroy2017,Sun2018}.  This makes it possible to place cloud-scale molecular gas properties in the context of the global kinematic response to the galactic potential, offering a view of the factors that influence the organization of the molecular gas and the star formation that occurs within it.  One of the strengths of this approach is the ability to use one data set to characterize the density distribution, the local gas motions, and the galaxy dynamical mass for any single galaxy over a range of spatial scales.  

There are several factors that make the cold molecular phase of the interstellar medium (ISM) a good tracer of the global dynamics of a galaxy. For nearby galaxies, information about the molecular gas kinematics can typically be obtained out to the edge of the optical disk \citep[e.g.,][report extents in the range $\sim0.3$ to 1.3 $R_{{\rm 25}}$]{Schruba2011}.  
They reveal a projection of motion that can be used to infer the orientation of the galaxy with respect to the line-of-sight, as well as the speed at which material orbits around the galaxy center.  The projected velocities in this case are an almost direct probe of the galaxy's underlying mass distribution given the dissipative nature of molecular gas, which results in characteristically low velocity dispersions ($1{-}20$ km~s$^{-1}$) in the disks of typical nearby galaxies \citep[e.g.,][]{Combes1997,Helfer2003,Colombo2014,Sun2018,Boizelle2019}.  This feature has prompted the use of high resolution kinematic maps of the inner portions of CO disks observed with the Atacama Large Millimeter/submillimeter Array (ALMA) to reconstruct the masses of supermassive black holes in the centers of nearby galaxies \citep[e.g.,][]{Davis2017,Onishi2017,Davis2018}.  

The dissipative quality of the gas also leads to a distinctly strong response to underlying non-axisymmetric components in the gravitational potential \citep[e.g.,][]{Roberts1987,Kim2006,Dobbs2009}.  Bar and spiral arm features yield recognizable deviations from circular motions in velocity field maps that provide powerful constraints on the structure of the ISM and its organization by the local galactic environment. 

Yet, as a global dynamics tracer, molecular gas kinematics are confronted with issues that become more obvious at high spatial resolutions:  Because the molecular phase traces the material at densities where gas begins to decouple from the background potential, thus forming weakly self-gravitating clouds \citep[e.g.,][]{Meidt2018}, molecular gas emission is primarily confined to spiral arms, bars, or collections of individual clouds.
The gas in these regions reveals a strong contribution from non-circular streaming motions, more evident at higher spatial resolution \citep[e.g.,][]{Meidt2013,Colombo2014}.  Without additional modelling, measurements of disk orientation based on velocity maps thus exhibit recognizable deviations from their true values. In these cases, rotational velocities also reflect the added contribution from non-circular motions, leading to systematic offsets in the inferred rotation curve \citep[e.g.,][]{Fathi2005,vandeVen2010,Chemin2015,Chemin2016}.  

In this paper, we present global disk orientations and rotation curves from $\sim 1''$ PHANGS-ALMA\footnote{\url{www.phangs.org}} CO data extending on average out to $\sim 0.7\,R_{{\rm 25}}$ for 67 local star-forming galaxies.  This is the largest such compilation of CO rotation curves to date \citep[c.f.,\,][]{Sofue2001} and complements the rich kinematics constraints offered by a growing number of stellar and multi-phase gas kinematics surveys in the local universe  \citep[][]{deBlok2008,Epinat2008,Chung2009,lelli2016,Kalinova2017,Levy2018,Korsaga2018}.

We use an approach that is optimized for kinematic tracers with low spatial filling factor that probe the central regions of galaxies at high spatial resolution.  This makes it ideally suited for the wide-field CO data now regularly produced by ALMA and NOEMA. Our technique incorporates many of the same elements implemented in the wealth of existing algorithms that fit 2D and 3D kinematic data (\citealt{vanAlbada}; \citealt{Begeman1987}; \citealt{Schoenmakers1999}; \citealt{simon03}; \citealt{krajnovic}; \citealt{jozsa2007}; \citealt{Spekkens2007}; \citealt{bouche}; \citealt{diT};  \citealt{kamphuis}; \citealt{peters}; \citealt{Oh2018}).  To overcome the complications of fitting CO velocity fields, we introduce a new element that involves fitting orientations to all lines-of-sight at once.  This maximizes constraints from regions that provide an unbiased probe of circular motion.  

This paper is structured as follows: We discuss the observational data used as well as our sample selection in Section~\ref{Sect2.sec}.  This is followed by a description of our fitting technique in Section~\ref{Sect3.sec}.  Our results are presented in Section~\ref{Sect4.sec}, followed by a discussion of our findings in Section~\ref{Sect5.sec}.  Finally, we present our conclusions in Section~\ref{Sect6.sec}. 
 
\section{Observations and sample}
\label{Sect2.sec}
\subsection{ALMA CO (2-1) observations and sample selection}
\label{Sample_properties.sec}
\begin{figure}[t]
\centering
 \includegraphics[width=0.49\textwidth]{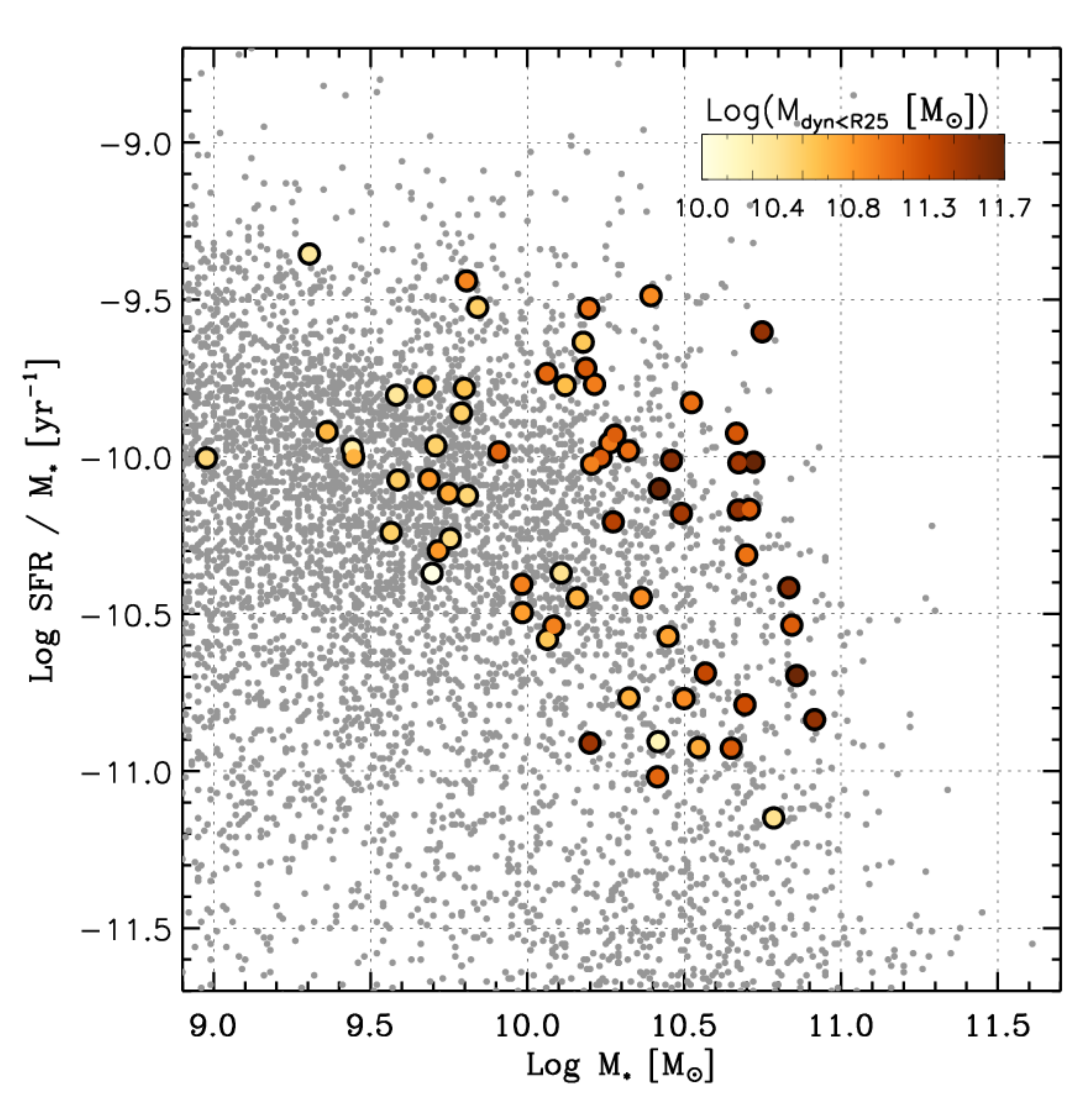}\\
\caption{The subset of the PHANGS-ALMA sample used for this paper (colored symbols) in the $M_\star{-}{\rm sSFR}$ plane.  The color-coding indicates the total dynamical mass within $R_{25}$ (\Mdyn25, see Section~\ref{smooth_fits.sec}).  The underlying population of star-forming galaxies taken from \cite{Leroy2019} is shown as gray symbols (see text for details).  Our kinematic sample spans a large range of stellar masses, SFRs, and ratios of stellar-to-total dynamical masses, while well sampling the underlying population of local main sequence galaxies.}
 \label{MS.fig}
\end{figure}

This work exploits observations of the PHANGS-ALMA sample of nearby galaxies with CO (2-1) emission mapped with ALMA at high ($\sim1''$) angular resolution.  Details of the sample selection, ALMA observations, and data processing are described in detail by A.~K.\ Leroy et al.\ (in prep).  In short, PHANGS-ALMA is designed to target 74 massive ($\log{(M_\star/{\rm M_{\sun}})} > 9.75$), actively star-forming ($\log({\rm \,sSFR}/{\rm Gyr^{-1}}) > -11$), moderately inclined ($i<75^\circ$ ) galaxies that are accessible to ALMA (see Figure~\ref{MS.fig}).  Furthermore, PHANGS targets are chosen to be nearby (distance $< 17$ Mpc) such that the survey's nominal  resolution of $1''$ accesses linear scales of about $100$~pc.  The PHANGS-ALMA sample comprises observations from several ALMA programs: 2013.1.00650.S, P.I.\ E.~Schinnerer; 2013.1.00803.S, P.I.\ D.~Espada; 2013.1.01161.S, P.I.\ K.~Sakamoto; 2015.1.00925, P.I.\ G.~Blanc; 2015.1.00956, P.I.\ A.~K.\ Leroy; 2017.1.00392.S, P.I.\ G.~Blanc; 2017.1.00886.L, P.I.\ E.~Schinnerer; 2018.1.01651.S,  P.I.\ A.~K.\ Leroy.

Each galaxy is observed by one or more rectangular mosaics.  We combine observations of the mosaiced region from all three ALMA arrays to produce fully sampled images, including data from the 12-m array (designated `12m'), the Morita Atacama Compact Array (designated `7m') and the 12-m telescopes configured to observe as single-dish telescopes, providing estimates of the total power (designated `TP').  The 12m data are obtained from array configurations C43-1 or C43-2.  We calibrate the 12m and 7m data using the observatory-provided pipeline, and we calibrate and image the TP data using the scripts developed by \citet{herrera20}.  We imaged the 12m+7m data with CASA v5.4 using an emission-mask based cleaning scheme as described in A.~K.\ Leroy et al.\ (in prep).  The interferometer data are merged with the single-dish data using the {\sc feather} task within CASA, producing a well-sampled map that is then corrected for the primary beam power pattern. We convolve the maps to have circular synthesized beams. The $1\sigma$ noise level in the maps is approximately $0.3$~K in a $2.5$ $\mathrm{km~s^{-1}}$ channel.  There is significant (30\%) variation in the beam size and survey depth across the different galaxies, arising from the sky positions of the galaxy, the atmospheric conditions at ALMA, and the configurations used in observation. The imaging process yields a spectral-line data cube where the spectral axis is radial velocity as determined by using the Doppler effect with the radio convention.  The velocities are referenced to the kinematic local standard of rest (LSRK) frame.

From the spectral-line data cubes, we generate two-dimensional integrated intensity maps as well as estimates for the mean line-of-sight velocity that we will use for rotation curve fitting.  We begin by making an empirical estimate of the noise level at each point in the cube, first measuring the noise as the median absolute deviation of the pixels for each spectrum, iteratively rejecting positive outliers associated with signal.  We then smooth the spatial noise estimate over $\sim 3$ beam FWHM.  At this point we calculate the relative variation of the noise level in each spectral channel that, in combination with the noise map, yields an estimate of the noise level at every point in the data cube.  Next, we identify signal at every point in the map, using a two stage masking process.  We first identify all regions with $T_{\rm B}>3.5\sigma$ in three consecutive channels.  These regions are then dilated in three dimensional space to all connected regions that show $T_{\rm B}>2\sigma$ in two consecutive channels.  We refer to this masking process as the `strict' mask.  This masking process is then repeated on a data cube that is convolved to $500$~pc linear resolution. This latter process identifies low surface brightness emission.  We build a `broad' mask by identifying pixels in the cube that are associated with signal in either the low or high resolution mask.

Using these masks, we calculate the integrated intensity maps ($I_\mathrm{CO}$ or `moment-0' maps) for each position as:
\begin{equation}
    I_\mathrm{CO}(x,y) = \sum_k T_\mathrm{B}(x,y,v_k) \mathcal{M}(x, y, v_k)\, \delta v,
\end{equation}
where $\mathcal{M}$ is the mask (value of 1 for signal, 0 otherwise) and $\delta v$ is the channel width.  We use a moment-based estimator for the line-of-sight velocity:
\begin{equation}
    V_\mathrm{obs}(x,y) = \frac{1}{I_\mathrm{CO}}\sum_k v_k T_\mathrm{B}(x,y,v_k) \mathcal{M}(x, y, v_k)\, \delta v.
\end{equation}
We create strict and broad maps for the integrated intensity and observed velocity. The strict maps show lower flux recovery but high quality $V_\mathrm{obs}$ maps.  The broad maps have better flux recovery but occasionally show spurious features in their velocity maps associated with noise.  We build final $V_\mathrm{obs}$ maps by using the broad maps but we remove positions where (1) the $I_\mathrm{CO, broad}$ is less than $2\times$ its associated uncertainty or (2) the velocity field at high resolution differs from the velocity field measured at 500\,pc resolution by more than $30~\mathrm{km~s^{-1}}$.

Out of the sample of 74 PHANGS-ALMA targets, we select 72 galaxies for which high-resolution moment maps (i.e., either combined 12m+7m+TP or 12m+7m) are available.  We reject one source with a high-resolution moment map that contains no detected pixel above the applied S/N cut (IC~5332), and one source (NGC~3239) that is only detected within the very central region (at radii $\le 2''$).  Three sources (NGC~4424, NGC~4694, and NGC~4731) show $V_\mathrm{obs}$ maps that suggest no ordered rotational motion within the detected area, and are subsequently rejected.  We thus have 67 PHANGS galaxies with high resolution CO maps for analysis.  The position of our sample in the $M_\star{-}\mathrm{sSFR}$ plane is shown in Figure~\ref{MS.fig}, and the basic properties are presented in Table~\ref{priors.tbl}.  Note that, due to updates on distances and stellar mass-to-light ratio ($M/L$), our sample expands to lower stellar masses than implied by the initial stellar mass cut of PHANGS-ALMA quoted above. 

We adopt global galaxy parameters for the galaxies as outlined in A.~K.\ Leroy et al.\ (in prep).  We use the stellar mass of a galaxy ($M_\star$), the global star formation rate (SFR), and distance measurements from the $z{=}0$ Multiwavelength Galaxy Survey (z0MGS) compilation of data \citep{Leroy2019}.  HyperLEDA\footnote{\url{http://leda.univ-lyon1.fr/}} \citep{Paturel2003,LEDA} values are used for the radius of the 25 mag~arcsec$^{-2}$ isophote ($R_{25}$).  Measurements of effective half-light radii ($R_{\rm e}$) are calculated from the cumulative flux distribution determined on the NIR \textit{WISE1} maps \citep{Leroy2019}.

\subsection{Sample demographics}\label{sec:demographics}
\label{demographics.sec}

This section summarizes the properties of the observed ALMA velocity fields used for this study which are important to the design of our fitting method.  We have examined the impact of these properties on the results of our method by performing extensive tests on mock galaxy velocity fields.  These tests have suggested a number of quality assessment criteria that we present and discuss in detail later in Section~\ref{Overview.sec}.  

\subsubsection{Radial extent of the CO emission}\label{sec:RCO}

CO emission is an ideal tracer of the kinematics of the inner disks of galaxies but may provide a limited probe of outer disk dynamics, depending on the (finite) extent of the CO emission.  In practice, the outer probed edge of our CO velocity fields is determined by the drop in CO signal with galactocentric radius and the finite extent of the PHANGS maps.  We quantify the impact of finite spatial extent by measuring the ratio between the maximum galactocentric radius out to which we detect the CO(2-1) line emission, $R_{{\rm CO,max}}$, and the outer radius of the optical disk, $R_{25}$.  To derive $R_{{\rm CO,max}}$, we bin the observed velocity field in elliptical annuli that have a width of three pixels (corresponding to about the observed FWHM).  The annuli are centered on the photometric center (see Table~\ref{priors.tbl}) and oriented according to our best-fit kinematic orientations (see Section~\ref{Orienations.sec}).  We define $R_{{\rm CO,max}}$ as the radius at which the total number of detected pixels within a given annulus falls below 10.  Varying this threshold by a factor of 2 changes $R_{{\rm CO,max}}$ by only 4\,\% on average. 

\begin{figure*}[thb]
\centering
 \includegraphics[width=0.98\textwidth]{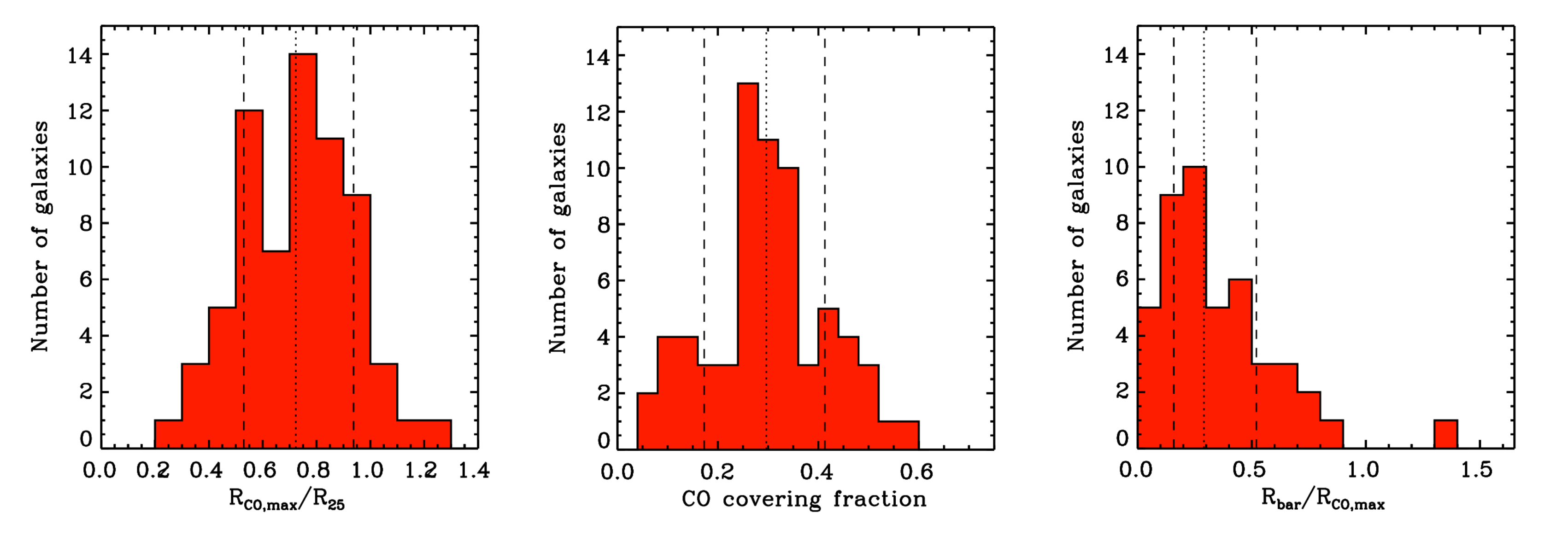}\\
\caption{Left: Distribution of the ratio between the extent of the ALMA CO velocity fields ($R_{{\rm CO,max}}$) and the disk size ($R_{25}$).  Middle: Distribution of spatial coverage filling factor in the CO velocity fields at $\sim 1''$ resolution. Right: Distribution of the ratio between bar length $R_{\rm{bar}}$ and $R_{{\rm CO,max}}$.  Shown is only the subset of barred galaxies (45 out of 67 systems).  In all panels, the median of the distribution is shown as dotted line, and the 16th and 84th percentiles are shown as dashed lines, respectively. }
 \label{demographics.fig}
\end{figure*}

The distribution of $R_{{\rm CO,max}}/R_{25}$ for our sample, shown in the left panel of Figure~\ref{demographics.fig}, has a range of  $0.26{-}1.2$, and median of $0.72$. Our data thus probe out to $\sim$3.6 CO scale lengths \citep[following][]{Schruba2011}, thereby characterizing $\sim$90\% of the total CO emission \citep{Schruba2011,Puschnig2020}.  Moreover, our maps typically probe beyond radii where the rotation curve flattens out (which happens on average within $\rm 0.1\,R_{25}$, as discussed in Section~\ref{RCs.sec}).  For a subset of galaxies, however, the CO emission traces only the part of the galaxy where rotational velocities rise steeply.  The outer, asymptotic rotation curve behavior is thus unconstrained in these cases and fits to the inner solid-body motion alone yield poorly constrained disk orientations and inner rotation curves.  

\subsubsection{Impact of the clumpiness of the CO spatial distribution on velocity field recovery}

The low area filling factor of CO emission can have consequences for the observational recovery of velocity information in PHANGS-ALMA moment maps. Depending on the surface brightness sensitivity limit, using a molecular emission line tracer can favor sightlines that sample narrow, high density contrast features organized by bars and spiral arms.  This leads to non-axisymmetric or incomplete azimuthal coverage especially when, e.g., lines-of-sight in the lower density inter-arm regions (where motion is expected to be more nearly circular) are under-represented.  Clumpiness can further heighten the sparseness of the velocity coverage. 

\begin{figure}[b]
\centering
 \includegraphics[width=0.40\textwidth]{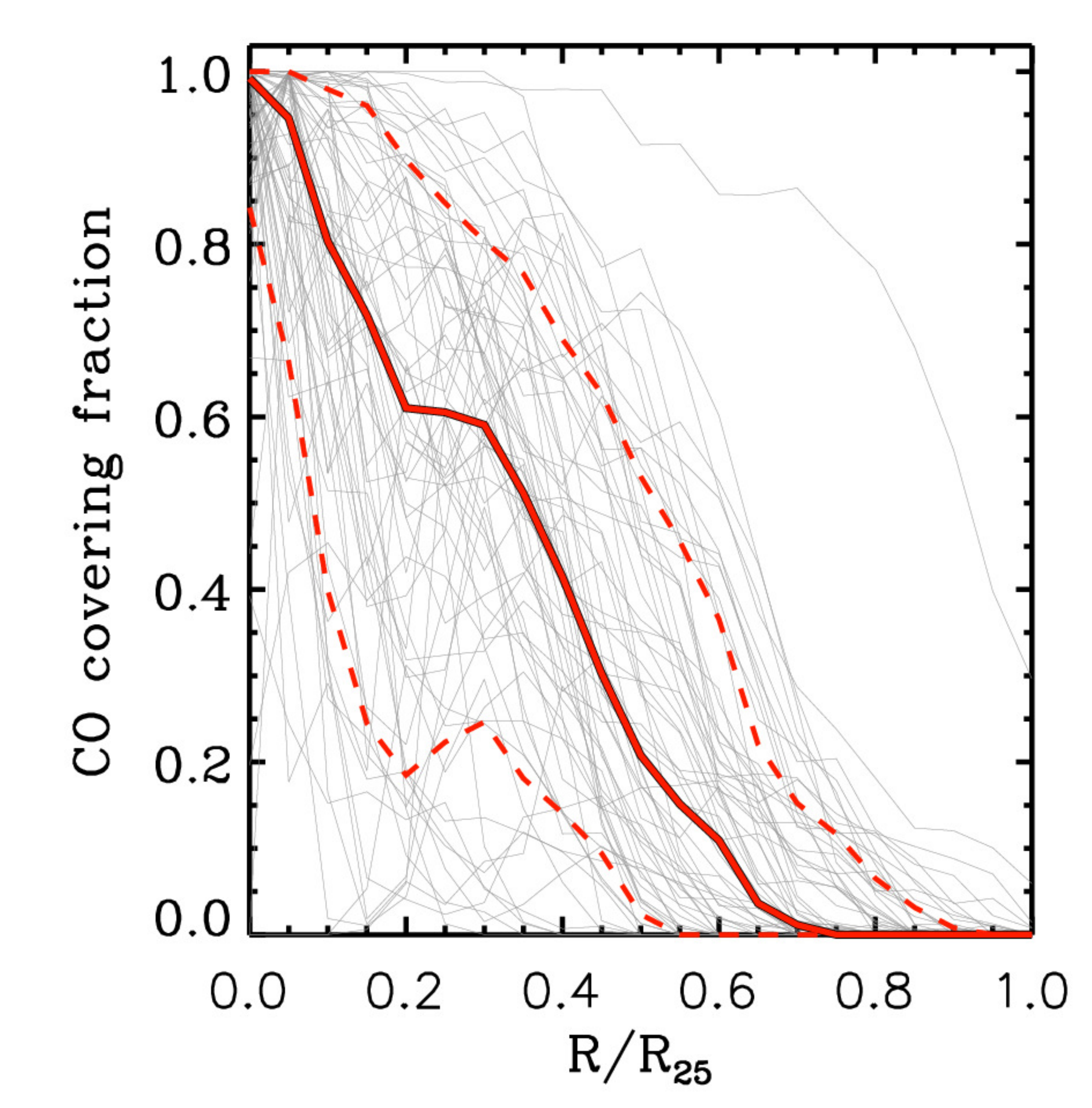}\\
\caption{Spatial CO covering fraction per radial bin versus $R/R_{25}$ for our sample.  The median and scatter based on the 16th and 84th percentiles are shown as solid and dashed red lines, respectively. Individual galaxies are shown as grey lines.}
 \label{filling.fig}
\end{figure}

The middle panel of Figure~\ref{demographics.fig} illustrates the velocity field coverage (see Section~\ref{Sample_properties.sec}) throughout the sample, defined as the fraction of detected pixels inside $R_{{\rm CO,max}}$.  The median CO covering fraction in the sample is about 30\,\%.  This is strongly influenced by typically poorer coverage towards outermost radii probed in each map.  This is also portrayed in Figure~\ref{filling.fig}, which tracks the drop in CO brightness with galactocentric radius. The median covering fraction drops smoothly with $R/R_{\rm 25}$ and typically becomes zero at around $0.6{-}0.8\ R_{\rm 25}$. Figure~\ref{filling.fig} also highlights a strong variation of the CO covering fraction among our sample at a given radius.

\subsection{The structure and orientation of the underlying stellar disk}
For PHANGS targets, extensive imaging data available for the sample provides a complementary view of the inner gravitational potential.  These data provide key information about the disk structure and orientation that we use to inform our fitting and the interpretation of our results.  

\subsubsection{Photometric priors on disk orientation and centering for the sample}\label{sec:priors}

As will be described in more detail in the next section, our method relies on prior information about the disk orientation (i.e. position angle\footnote{The position angle follows the standard convention of being measured east-of-north on the celestial sphere.} $\phi$ and inclination $i$), center position of the disk, and the systemic velocity $V_{{\rm sys}}$.  In our fitting process explained below, we re-determine $\phi$, $i$ and $V_{{\rm sys}}$ for each galaxy while keeping the center position fixed. 

Where possible, we adopt photometric priors from the Spitzer/${\rm S^4G}$ survey \citep{Sheth2010,Munoz2013,Querejeta2015}, covering 54 sources from our total sample.  We made this catalog our primary choice due to combination of good angular resolution and excellent imaging depth. 
We adopt central coordinates, position angles and inclinations from the \cite{Salo2015} catalog based on isophotal analysis on Spitzer/3.6\,$\mu$m images.  Values for $\phi$ and $i$ measured through photometry are hereby referred to as \pap\ and \incp, respectively.  These are determined from profiles of ellipticity and position angle of the outer galaxy isophotes, where the influence of bars and/or spiral structures is minimized.  To convert the axial ratio $q$ implied by the quoted ellipticity $\eta$ (with $q = 1 - \eta$) into \incp, we use the relation

\begin{equation}
\cos{(i_{{\rm phot}})} = \sqrt{\frac{q^2 - q_{{\rm min}}^2}{1 - q_{{\rm min}}^2}}.\label{photinc.eq}
\end{equation}
Here, $q_{{\rm min}}$ is the intrinsic edge-on thickness of the disk, which we fix to a value of $0.25$ \citep[see\,][]{vdW2014}. Adopting a thin disk with $q_\mathrm{min}=0.1$ leads to smaller values of $i_{{\rm phot}}$ by at most $2.5$ degrees for the range of inclinations covered by our sample.  By comparing the 3.6\,$\mu$m ${\rm S^4G}$ orientations with the HyperLEDA orientations (defined in optical bands) for this subset of the sample, we obtain a rough (10 deg) estimate of typical uncertainties on \pap\ and \incp.  

For galaxies not covered by ${\rm S^4G}$ (13~sources), we adopt the central positions from the 2MASS/LGA catalog \citep{Jarrett2003} based on $J$/$H$/$K$-band imaging if available, and from HyperLEDA otherwise.  Galaxy orientations for non-${\rm S^4G}$ sources are adopted from HyperLEDA.  For sources with no \pap\ measurement available within HyperLEDA, we use the respective value from 2MASS/LGA.

As a consistency check on the positional accuracy of our photometric centers, we computed the offsets between the central positions of both ${\rm S^4G}$ and 2MASS catalogs for our covered sources.  We find good agreement between these central positions, with a $1\sigma$ scatter of about $1''$.  We also confirm by visual inspection that there is consistency between the adopted photometric center and the kinematic center of our CO velocity fields. Prior systemic velocities for our sources are taken from the collection of {\sc HI} measurements from the HyperLEDA data base. 

Table~\ref{priors.tbl} lists the basic properties of our sample including the photometric priors.  

\subsubsection{Central bar and bulge properties}
One of the main goals of this work is to correlate the inner stellar structures hosted by star-forming galaxies with the kinematics of their molecular gas disks. Thus, we collect estimates of stellar bar length ($R_{{\rm bar}}$), and bulge-to-total (B/T) ratio mainly based on Spitzer/IRAC 3.6\,$\mu$m images.  $R_{{\rm bar}}$ measurements for the PHANGS-ALMA sample are compiled by M.~Querejeta et al.\ (in prep.), who combine previous catalogs from \cite{Herrera2015} with additional 3.6\,$\mu$m imaging for a subset of PHANGS-ALMA targets not covered by ${\rm S^4G}$, as well as existing literature measurements.  B/T ratios are available from the two-component decompositions performed on ${\rm S^4G}$ images by \cite{Salo2015}.

\subsubsection{The prominence of stellar bars}

Next, we inspect the frequency and properties of stellar bars in our sample, as the strongest and most prominent of these can have a major impact on the appearance of inner stellar and gaseous velocity fields due to their tumbling non-axisymmetric potential \citep[e.g.,][]{Wong2004,Fathi2005,boone,hunt,Sormani2015}.  

The majority (67\,\%) of our sample are barred systems, with the more massive galaxies hosting a higher fraction of bars.  
In the right panel of Figure~\ref{demographics.fig}, we show the distribution of $R_{{\rm bar}}/R_{{\rm CO,max}}$ for the strongly barred subset of our sample (i.e., bar ellipticity $\epsilon > 0.4$; \citealt{jogee2004}).  Most of the CO fields extend beyond the bar, offering good constraints of the CO kinematics from within the disk-dominated portion of the map.  However, we note that galaxies for which the bar is strong and covers a large portion of the observed CO map pose strong limitations to our method for constraining the disk orientation, and are therefore treated with care in our subsequent analysis.

\section{Methods}
\label{Sect3.sec}
In this section, we present an overview of the method used to determine the kinematic properties of the PHANGS galaxies.  We first outline the considerations that led to the development of the method and then we present the details of the implementation.

\subsection{Considerations for Modeling CO velocity fields}
\label{sec:motivation}

Our implementation is tailored to three features of our data.  First, our observations are confined to the central part of the galactic disk, where a flat circular disk is a good approximation to the galactic geometry.  Second, the CO emission is clumpy, requiring a method that is robust to missing data.  Finally, the molecular gas emission highlights spiral arms and bars where non-circular motions can be significant.  We address these considerations by using a Bayesian framework, sampling a probability density function describing the data using a Markov Chain Monte Carlo (MCMC) sampler.  

\startlongtable
\begin{deluxetable*}{DDDDDDDDDDD}
\tablecaption{Basic properties of the kinematic PHANGS sample}
\tablehead{
\multicolumn2c{ID} & \multicolumn2c{RA$^1$} & \multicolumn2c{DEC$^1$} & \multicolumn2c{Distance} & \multicolumn2c{$V_{\rm {sys}}$$^2$} & \multicolumn2c{$\phi_{{\rm phot}}$$^3$} & \multicolumn2c{$i_{{\rm phot}}$$^4$} & \multicolumn2c{$f_{{\rm CO}}$$^5$}& \multicolumn2c{$\frac{R_{{\rm bar}}}{R_{{\rm CO,max}}}$} & \multicolumn2c{$\frac{R_{{\rm CO,max}}}{R_{{\rm 25}}}$}\\
\multicolumn2c{} & \multicolumn2c{[deg]} & \multicolumn2c{[deg]} & \multicolumn2c{[Mpc]} & \multicolumn2c{[$\mathrm{km~s^{-1}}$]} & \multicolumn2c{[deg]} & \multicolumn2c{[deg]} & \multicolumn2c{[arcsec]} & \multicolumn2c{} & \multicolumn2c{}
}
\decimals
\startdata
${\rm IC~1954  }$     &      52.87971  &    -51.90486   &  15.2  &    1062.9 &  61.0   &   57.7  & 0.46 &  0.09 & 0.84  \\
${\rm IC~5273  }$     &     344.86118  &    -37.70284   &  14.7  &    1293.9 &  229.0  &   44.3  & 0.27 &  0.22 & 0.97  \\
${\rm NGC~0628 }$     &      24.17385  &     15.78364   &   9.8  &     658.5 &  0.1    &   28.9  & 0.29 &  0.00 & 0.55  \\
${\rm NGC~0685 }$     &      26.92845  &    -52.76198   &  16.0  &    1359.7 &  99.2   &   38.6  & 0.21 &  0.25 & 0.87  \\
${\rm NGC~1087 }$     &      41.60492  &     -0.49872   &  14.4  &    1522.7 &  1.4    &   51.4  & 0.36 &  0.20 & 0.94  \\
${\rm NGC~1097 }$     &      41.57896  &    -30.27467   &  14.2  &    1269.4 &  126.5  &   48.6  & 0.25 &  0.52 & 0.57  \\
${\rm NGC~1300 }$     &      49.92081  &    -19.41111   &  26.1  &    1578.2 &  286.1  &   31.8  & 0.10 &  0.68 & 0.67  \\
${\rm NGC~1317 }$     &      50.68454  &    -37.10379   &  19.0  &    1948.5 &  200.3  &   35.0  & 0.48 &  0.19 & 0.35  \\
${\rm NGC~1365 }$     &      53.40152  &    -36.14040   &  18.1  &    1638.3 &  210.7  &   55.4  & 0.14 &  0.52 & 0.49  \\
${\rm NGC~1385 }$     &      54.36901  &    -24.50116   &  22.7  &    1497.3 &  172.3  &   49.7  & 0.34 &  0.00 & 0.72  \\
${\rm NGC~1433 }$     &      55.50619  &    -47.22194   &  16.8  &    1075.3 &  204.3  &   28.6  & 0.09 &  0.05 & 0.75  \\
${\rm NGC~1511 }$     &      59.90246  &    -67.63392   &  15.6  &    1334.9 &  304.0  &   71.1  & 0.31 &  0.20 & 0.95  \\
${\rm NGC~1512 }$     &      60.97557  &    -43.34872   &  16.8  &     897.7 &  254.5  &   42.5  & 0.12 &  0.62 & 0.45  \\
${\rm NGC~1546 }$     &      63.65122  &    -56.06090   &  18.0  &    1252.9 &  147.4  &   67.4  & 0.46 &  0.00 & 0.72  \\
${\rm NGC~1559 }$     &      64.40238  &    -62.78341   &  19.8  &    1295.3 &  245.1  &   54.0  & 0.31 &  0.09 & 1.00  \\
${\rm NGC~1566 }$     &      65.00159  &    -54.93801   &  18.0  &    1501.8 &  200.9  &   30.4  & 0.26 &  0.27 & 0.61  \\
${\rm NGC~1672 }$     &      71.42704  &    -59.24726   &  11.9  &    1338.8 &  148.5  &   23.7  & 0.19 &  0.48 & 0.83  \\
${\rm NGC~1792 }$     &      76.30969  &    -37.98056   &  12.8  &    1210.6 &  316.1  &   60.7  & 0.34 &  0.00 & 0.81  \\
${\rm NGC~1809 }$     &      75.52066  &    -69.56794   &  15.0  &    1300.6 &  144.8  &   80.5  & 0.10 &  0.00 & 0.57  \\
${\rm NGC~2090 }$     &      86.75787  &    -34.25060   &  11.8  &     922.1 &  196.5  &   71.7  & 0.44 &  0.00 & 0.53  \\
${\rm NGC~2283 }$     &     101.46997  &    -18.21080   &  10.4  &     840.8 &  0.1    &   44.2  & 0.18 &  0.14 & 1.05  \\
${\rm NGC~2566 }$     &     124.69003  &    -25.49952   &  23.7  &    1637.9 &  235.6  &   48.5  & 0.15 &  0.46 & 1.02  \\
${\rm NGC~2775 }$     &     137.58396  &      7.03807   &  17.0  &    1354.1 &  163.5  &   34.2  & 0.30 &  0.00 & 0.58  \\
${\rm NGC~2835 }$     &     139.47044  &    -22.35468   &  10.1  &     886.5 &  0.3    &   56.4  & 0.07 &  0.23 & 0.48  \\
${\rm NGC~2903 }$     &     143.04212  &     21.50084   &   8.5  &     555.3 &  199.6  &   63.5  & 0.28 &  0.25 & 0.75  \\
${\rm NGC~2997 }$     &     146.41164  &    -31.19109   &  11.3  &    1086.9 &  96.6   &   54.3  & 0.42 &  0.06 & 0.62  \\
${\rm NGC~3059 }$     &     147.53400  &    -73.92219   &  19.8  &    1258.0 &  -5.0   &   21.2  & 0.30 &  0.22 & 0.77  \\
${\rm NGC~3137 }$     &     152.28116  &    -29.06430   &  14.9  &    1106.3 &  3.9    &   47.2  & 0.30 &  0.00 & 0.53  \\
${\rm NGC~3351 }$     &     160.99064  &     11.70367   &  10.0  &     778.0 &  188.4  &   45.1  & 0.25 &  0.41 & 0.58  \\
${\rm NGC~3507 }$     &     165.85573  &     18.13552   &  20.9  &     975.6 &  80.1   &   27.5  & 0.27 &  0.34 & 0.85  \\
${\rm NGC~3511 }$     &     165.84921  &    -23.08671   &   9.9  &    1105.4 &  258.4  &   72.1  & 0.30 &  0.11 & 0.94  \\
${\rm NGC~3521 }$     &     166.45240  &     -0.03595   &  11.2  &     800.9 &  342.3  &   66.2  & 0.44 &  0.00 & 0.89  \\
${\rm NGC~3596 }$     &     168.77581  &     14.78707   &  10.1  &    1192.4 &  108.2  &   16.9  & 0.27 &  0.00 & 0.66  \\
${\rm NGC~3621 }$     &     169.56792  &    -32.81260   &   6.6  &     730.1 &  341.2  &   67.5  & 0.34 &  0.00 & 0.83  \\
${\rm NGC~3626 }$     &     170.01588  &     18.35684   &  20.0  &    1479.5 &  163.2  &   46.6  & 0.40 &  0.88 & 0.26  \\
${\rm NGC~3627 }$     &     170.06252  &     12.99150   &  10.6  &     720.9 &  166.2  &   55.7  & 0.40 &  0.37 & 0.58  \\
${\rm NGC~4207 }$     &     183.87681  &      9.58493   &  16.8  &     608.6 &  121.3  &   70.5  & 0.27 &  0.00 & 0.89  \\
${\rm NGC~4254 }$     &     184.70680  &     14.41641   &  16.8  &    2407.9 &  76.6   &   36.2  & 0.33 &  0.00 & 1.13  \\
${\rm NGC~4293 }$     &     185.30346  &     18.38257   &  16.0  &     901.6 &  64.0   &   65.0  & 0.16 &  1.32 & 0.32  \\
${\rm NGC~4298 }$     &     185.38651  &     14.60611   &  16.8  &    1128.4 &  311.1  &   56.4  & 0.55 &  0.00 & 1.21  \\
${\rm NGC~4303 }$     &     185.47888  &      4.47374   &  17.6  &    1566.9 &  323.6  &   24.2  & 0.56 &  0.36 & 0.49  \\
${\rm NGC~4321 }$     &     185.72886  &     15.82230   &  15.2  &    1573.9 &  158.2  &   33.2  & 0.30 &  0.04 & 1.03  \\
${\rm NGC~4457 }$     &     187.24593  &      3.57062   &  15.6  &     884.9 &  66.4   &   17.4  & 0.48 &  0.78 & 0.50  \\
${\rm \,\,NGC~4496a}$ &     187.91358  &      3.93962   &  14.9  &    1730.4 &  62.4   &   29.1  & 0.12 &  0.29 & 0.80  \\
${\rm NGC~4535 }$     &     188.58459  &      8.19797   &  15.8  &    1962.1 &  202.8  &   22.4  & 0.33 &  0.28 & 0.54  \\
${\rm NGC~4536 }$     &     188.61278  &      2.18824   &  15.2  &    1807.3 &  305.5  &   67.7  & 0.25 &  0.19 & 0.84  \\
${\rm NGC~4540 }$     &     188.71193  &     15.55172   &  16.8  &    1284.1 &  56.0   &   11.8  & 0.34 &  0.41 & 0.68  \\
${\rm NGC~4548 }$     &     188.86024  &     14.49633   &  16.2  &     481.5 &  149.2  &   38.3  & 0.25 &  0.52 & 0.69  \\
${\rm NGC~4569 }$     &     189.20759  &     13.16287   &  16.8  &    -219.5 &  24.8   &   69.7  & 0.33 &  0.61 & 0.58  \\
${\rm NGC~4571 }$     &     189.23492  &     14.21733   &  14.9  &     335.7 &  264.1  &   14.1  & 0.28 &  0.00 & 0.74  \\
${\rm NGC~4579 }$     &     189.43138  &     11.81822   &  16.8  &    1516.8 &  92.1   &   41.3  & 0.49 &  0.37 & 0.72  \\
${\rm NGC~4654 }$     &     190.98575  &     13.12672   &  16.8  &    1043.4 &  120.7  &   58.6  & 0.37 &  0.22 & 0.91  \\
${\rm NGC~4689 }$     &     191.93990  &     13.76272   &  16.8  &    1608.4 &  163.7  &   35.3  & 0.41 &  0.00 & 0.70  \\
${\rm NGC~4781 }$     &     193.59916  &    -10.53712   &  15.3  &    1262.3 &  290.8  &   61.5  & 0.33 &  0.18 & 0.92  \\
${\rm NGC~4826 }$     &     194.18184  &     21.68308   &   4.4  &     407.6 &  295.9  &   58.2  & 0.22 &  0.00 & 0.38  \\
${\rm NGC~4941 }$     &     196.05461  &     -5.55154   &  14.0  &    1111.2 &  203.5  &   39.4  & 0.46 &  0.00 & 0.83  \\
${\rm NGC~4951 }$     &     196.28213  &     -6.49382   &  12.0  &    1177.6 &  91.5   &   76.1  & 0.24 &  0.00 & 0.83  \\
${\rm NGC~5042 }$     &     198.87920  &    -23.98388   &  12.6  &    1389.1 &  195.9  &   57.4  & 0.28 &  0.30 & 0.53  \\
${\rm NGC~5068 }$     &     199.72808  &    -21.03874   &   5.2  &     671.2 &  300.1  &   27.0  & 0.05 &  0.17 & 0.94  \\
${\rm NGC~5134 }$     &     201.32726  &    -21.13419   &  18.5  &    1755.7 &  300.0  &   4.9   & 0.24 &  0.79 & 0.70  \\
${\rm NGC~5248 }$     &     204.38336  &      8.88519   &  12.7  &    1152.3 &  103.9  &   37.8  & 0.25 &  0.00 & 0.98  \\
${\rm NGC~5530 }$     &     214.61380  &    -43.38826   &  11.8  &    1193.1 &  306.5  &   66.5  & 0.30 &  0.00 & 0.70  \\
${\rm NGC~5643 }$     &     218.16991  &    -44.17461   &  11.8  &    1190.0 &  320.6  &   29.9  & 0.41 &  0.45 & 0.76  \\
${\rm NGC~6300 }$     &     259.24780  &    -62.82055   &  13.1  &    1108.7 &  118.6  &   53.2  & 0.32 &  0.16 & 0.79  \\
${\rm NGC~6744 }$     &     287.44208  &    -63.85754   &  9.5   &     851.4 &  15.4   &   53.5  & 0.14 &  0.36 & 0.53  \\
${\rm NGC~7456 }$     &     345.54306  &    -39.56941   &   7.9  &    1200.7 &  22.9   &   75.2  & 0.20 &  0.00 & 0.69  \\
${\rm NGC~7496 }$     &     347.44703  &    -43.42785   &  18.7  &    1650.3 &  213.2  &   9.0   & 0.37 &  0.44 & 0.80  \\
 \enddata
\tablecomments{1: Adopted photometric galaxy center; 2: prior of systemic velocity; 3: prior of photometric disk position angle; 4: prior of photometric disk inclination; 5: spatial CO covering fraction.}
\label{priors.tbl}
\end{deluxetable*}

Most existing techniques to determine the kinematic properties of galaxies build on the basic tilted-ring model of the line-of-sight velocity field first introduced by \cite{Rogstad1974} \citep[see also ][]{Bosma1987,Begeman1987}.  This yields measurements of $\phi$, $i$ and rotational velocity in a series of radial bins.  This approach works well for smooth, extended {\sc Hi} emission \citep{deBlok2008,Trachternach2008,Schmidt2016} and can also be formulated using a Bayesian approach \citep{Oh2018}.  For applications to kinematic tracers characterized by sparseness or non-circular motions, the tilted-ring approach has tended to require careful supervision  \citep[e.g.,][]{Fathi2005,jozsa2007,Spekkens2007,vandeVen2010,Colombo2014}.  
As a kinematic tracer, CO emission differs from atomic gas.  Even though the CO emission is sparser than the atomic gas, it has the advantage of tracing a thinner disk and arises from a more centrally concentrated area, so that it avoids the warps and radial flows that are seen in outer atomic gas disks \citep[see, e.g.,][]{Schmidt2016}.  We can thus approximate the CO emission as arising from a thin, circular disk, which simplifies our approach to determining disk orientation parameters ($\phi$ and~$i$).  

On the other hand, CO emission is frequently organized into features associated with complex non-circular motions, such as spiral arms and bars characteristic of the inner disks of nearby galaxies.  The sharpness of these features in CO requires more careful treatment than the smoother, puffier atomic gas would across the same region \citep[e.g.,\,][]{Colombo2014}.  For kinematic tracers like CO that emphasize non-circular motions, fitting orientation parameters in independent radial bins following the tilted ring model can be biased by the sparse coverage of the mapped area and an over-representation of lines-of-sight dominated by non-circular motion.  

To control for these issues, in this paper we prefer to find globally optimal orientation parameters rather than fitting within individual radial bins.  This should be appropriate for the inner disk environment traced by molecular gas, where genuine warps and twists characteristic of outer disks are uncommon.  Thus, in fitting, the orientation parameters $\phi$ and $i$ are held constant with radius in our method.  The rotational velocities for a given orientation, however, are fitted in a series of radial bins.  

This approach works because globally, most lines-of-sight will probe regions dominated by circular motion, even though non-circular motions may bias any given radial bin.  Furthermore, the influence of biased bins tends to be less prominent when averaged across the entire disk, given the change in the magnitude and projected orientation of non-circular motions across typical bar and spiral components.  

\subsection{Fitting model CO velocity fields with MCMC} 

Our fitting method assumes that the CO emission arises from a thin, circular disk where the geometry of that disk is described in polar coordinates by a radius ($R$) and angle measured from the kinematic major axis ($\theta$).  At a given radius, we assume that the rotation speed as a function of angle $\theta$ is well described by a Fourier series \citep[following ][]{Franx1994}.  

Thus, the line-of-sight velocity observed in the plane of the sky is then:
\begin{equation}
V_{{\rm los}} = V_{{\rm sys}} + \displaystyle\sum_{j=1}^{N} [c_{j} \cos{(j\theta)} + s_{j} \sin{(j\theta)}]\sin{i},
\label{HD.eq}
\end{equation}
where the coefficients $c_{j}$ and $s_{j}$ vary as a function of galactocentric radius, $V_\mathrm{sys}$ is the systemic velocity and the harmonic expansion is truncated at some integer $N$.  This type of expansion has been incorporated into a number of algorithms to date (i.e., \textit{Reswri}, \citealt{Schoenmakers1999}; \textit{Ringfit}, \citealt{simon03}; \textit{Kinemetry}, \citealt{krajnovic};  \textit{DiskFit}, \citealt{Spekkens2007} and \citealt{sellsan2010}) in order to account for the complex array of (non-circular) motions present in the stellar and gaseous disks of galaxies.

The results in this paper are derived under the assumption of purely circular rotation, with $s_{1}=0$ and with $s_{j}=0$ and $c_{j}=0$ for $j>1$.  
In this case, the model reduces to :
\begin{equation}
  V_{{\rm los}} = V_{{\rm sys}} + V_{{\rm rot}}(R)\,\cos{\theta}\sin{i}.
  \label{Model_o1.eq}
\end{equation}
As we will show in the next section, this basic model performs well using all lines-of-sight at once to constrain the disk orientation, as in our approach.  
For the sake of generality, below we refer to the generic order-N harmonic expansion incorporated into our algorithm.  

We assume that the relationship between the coordinates in the disk and those observed on the sky are described by single values of $i$ and $\phi$.  For a given center position of the disk measured in sky coordinates ($X_\mathrm{cen}, Y_\mathrm{cen}$), we can relate the sky position to the disk position using that:
\begin{equation}
\cos{\theta} = \frac{-(X - X_{{\rm cen}}) \sin{\phi} + (Y - Y_{{\rm cen}}) \cos{\phi} } {R}
\end{equation}
and
\begin{equation}
\sin{\theta} = \frac{-(X - X_{{\rm cen}}) \cos{\phi} - (Y - Y_{{\rm cen}}) \sin{\phi} } {R\cos{i}}.
\label{trans.eq}
\end{equation}

Given this model, we view the problem from a Bayesian perspective, seeking the parameters that produce the model rotation surface, $V_\mathrm{los}$, that provides the most credible representations of the observed data, $V_\mathrm{obs}$.  Given our model, $V_\mathrm{los}$ is determined by the parameter set $\{i, \phi, V_{{\rm sys}}, X_\mathrm{cen}, Y_{\mathrm{cen}}, c_{j}, s_{j}\}$.  We could proceed by sampling a likelihood function given the observed data using this full set of parameters to determine the posterior distribution of these parameters.  However, this leads to a large number of parameters since $c_{j}$ and $s_{j}$ vary as a function of galactocentric radius, increasing the parameter space by $2N\times k$ parameters, where $k$ is number of parameters required to represent the variation of these harmonic coefficients with radius.  We reduce the dimensionality of the problem by optimizing $c_{j}$ and $s_{j}$ given $\{i, \phi, X_\mathrm{cen}, Y_{\mathrm{cen}}\}$ through a least-squares regression of these harmonic coefficients with radius (Section~\ref{sec:harmdec}).

In summary, our procedure adopts the following two elements:
\begin{itemize}
\item Bayesian optimization to calculate the posterior distributions for parameters $i$, $\phi$, $X_\mathrm{cen}$, $Y_{\mathrm{cen}}$, $V_{\rm sys}$ and $\sigma_0$;

\item Linear least-square fit for all additional parameters (including all the harmonic amplitudes) at a given set of ($i$, $\phi$, $X_\mathrm{cen}$, $Y_{\mathrm{cen}},V_{\rm sys}$), embedded within the above optimization.
\end{itemize}
This scheme is efficient as it isolates the linear part of the fitting process, reducing the dimensionality of the solution space probed by the MCMC process.

\subsubsection{The Posterior Distribution}
We adopt a log-likelihood for the posterior distribution of the parameters given our data:
\begin{eqnarray}
l(i, \phi,V_{{\rm sys}}, X_\mathrm{cen}, Y_{\mathrm{cen}}, \sigma_0^2) & \propto & 
-\sum_{X,Y}\left[\frac{\left( V_{{\rm obs}} - V_{{\rm los}} \right)^2}{2w(\sigma_{{\rm obs}}^2+\sigma_0^2)}\right. \nonumber \\
&& - \left.\frac{1}{2} \ln (w(\sigma_{\rm obs}^2+\sigma_0^2)) \right]\nonumber  \\
&& +P(i, \phi,  V_{{\rm sys}}, X_\mathrm{cen}, Y_{\mathrm{cen}}, \sigma_0^2).\nonumber \\
\label{Likelyhood.eq}
\end{eqnarray}
Here, $P$ represents the prior distributions we assume for the parameters.  The quantity $\sigma_\mathrm{obs}$ (discussed more below) accounts for the observational uncertainties and the factor $1/w$ represents the weight given to each pixel.  We have also introduced an excess variance term $\sigma_0^2$ that accounts for motions in excess of the measurement uncertainties which are not produced by a rotational model.  To estimate the posterior distributions, we sample this log-likelihood using Markov-Chain Monte Carlo methods (Section~\ref{sec:mcmc}).

The error $\sigma_{\mathrm{obs}}$ in Equation~\ref{Likelyhood.eq} above represents the measurement uncertainty on each $V_\mathrm{los}$ in the PHANGS CO first moment map.  We compute the median of all pixels in the uncertainty map to define a global $\sigma_\mathrm{obs}$ for each galaxy.  By assigning a uniform error we effectively let all pixels in the observed velocity contribute equally (modulo the weight $1/w$) to the fitting process without altering the median S/N of our data. 

A more explicit weighting scheme is introduced with the factor $1/w$.  We have adopted $1/w=\cos(\theta_\mathit{PA})$ weights for all results presented in this paper, where $\theta_\mathit{PA}$ is the angle measured with respect to the kinematic major axis.  This weighting is used to down-weight pixels near the minor axis that only weakly constrain circular motion.  A number of variations on this weighting scheme that are used in the literature, including a zero-weight wedge around the minor axis and radial weighting, are also incorporated (but not used).  We have found that our results are not particularly sensitive to the choice of weighting scheme; orientations measured with uniform weighting do not change significantly ($\lesssim 3\sigma$) compared to our nominal results.

\subsubsection{Harmonic Decomposition}
\label{sec:harmdec}

To avoid the computational expense of using MCMC to determine the rotation curve, we  perform a least-squares fit to determine the harmonic coefficients $c_j, s_j$ in Equation~\ref{HD.eq}.  For a given $\phi$ and $i$, $V_{{\rm sys}}$, $X_\mathrm{cen}$, $Y_{\mathrm{cen}}$ (and $\sigma_0$), we determine these harmonic amplitudes by performing a least-squares regression to the $k=1\cdots n$ observed values of the velocity within a series of radial bins:
\begin{eqnarray}
\underset{c_j, s_j}{\mathrm{argmin}} \sum_{k=1}^n \frac{1}{{\sigma_{\mathrm{obs},k}}^2} && \Big(V_{\mathrm{obs},k}-V_{{\mathrm{sys}}} \nonumber \\ 
&& - \sum_{j=1}^{N} [c_j \cos{(j\theta_k)} \nonumber \\
&& + s_j \sin{(j\theta_k)}]\sin{i}\Big)^2 \label{eq:harmlsq}
\end{eqnarray}
Here, the coefficients $c_j,s_j$ represent the amplitudes of the $j$th mode, $n$ is the number of pixels in a given bin and the polar angle of each datum within the disk ($\theta_k$) is set by the orientation parameters.  Since the optimization is linear in the coefficients, this calculation is efficient and reduces the dimensionality of the parameter space that the MCMC sample needs to explore.  

The algorithm performs this radial decomposition in concentric, annular bins generated from the orientation parameters, $\phi$ and $i$.  The radial bin width along the kinematic major axis is chosen to correspond to the observed FWHM of the circularized ALMA beam.  Even at the highest spatial resolutions of our data set, maps contain sufficient numbers of lines-of-sight per radial bin to yield statistically well-constrained least-squares solutions. 

The velocity field model extends out to the edge of the observed velocity field so that, in practice, the least-squares fit (Equation~\ref{eq:harmlsq}) occurs across all detected pixels in the PHANGS first moment map.  For each new set of orientation parameters sampled in the full parameter space probed with MCMC, a unique set of radial bins is defined.

Depending on the azimuthal coverage of the outermost bins (i.e., depending on the organization of the emission near the edge of the map), adopting this approach can make the outer bins overly sensitive to minor axis information, which is only weakly constraining of circular motion.  We find that this affects the fitted orientations in only two cases (NGC~4303 and NGC~4826).  We therefore truncate the respective velocity fields at outer radii ($60''$ for NGC~4303 and $75''$ for NGC~4826) before we perform our fitting for these galaxies.  The impact of incomplete coverage in the outer radial bins has a significant impact on the outer rotation curve, which we address in a later step of the analysis. 

\subsubsection{MCMC Implementation}
\label{sec:mcmc}

Our regression process is performed using MCMC as implemented specifically in the  {\sc python} {\sc emcee} package \citep{Foreman2013}.  In this work, we restrict our harmonic decomposition to the lowest-order cosine coefficient, $c_1$.  We have found that this basic model fit is well-suited to constrain the global $\phi$ and $i$ given the demographics of the CO emission in the PHANGS-ALMA sample (see Section~\ref{sec:demographics}) in particular when all lines-of-sight are included in the fit. 

We sample a four-dimensional parameter space defined by $\phi$, $i$, $V_{{\rm sys}}$ and excess variance $\sigma^2_{0}$, leaving the center position $(X_{{\rm cen}},Y_{{\rm cen}})$ fixed for this analysis.  The $\sigma^2_{0}$ parameter captures systematic differences between the observed velocity field and the assumed model and effectively accommodates for deviations in the adopted measurement uncertainties on each mapped $V_{{\rm los}}$.  We assume an inverse-Gamma distribution for the excess variance with a characteristic scale $2.5~\mathrm{km^2~s^{-2}}$ so that $\sigma_0^2 \sim \mathrm{InvGamma}(N, 2.5)$, where $N$ is the number of data.

Our approach allows for intrinsic differences between the kinematic and photometric orientation by assuming flat, uninformative priors for $\phi$, $i$, and $V_\mathrm{sys}$.  Furthermore, we initialize the walkers for $\phi$ and $i$ in the {\sc emcee} sampler in a uniform distribution with a conservative width of 50 degrees centered at the photometric values \pap\ and \incp\ presented in Table~\ref{priors.tbl}.  The walkers sampling $V_{{\rm sys}}$ are initialized around the value adopted from HyperLEDA also listed there.  This initialization ensures that the parameter space spanned by $\phi$ and $i$ is sampled to find a global maximum for the log-probability function. Based on extensive testing we have found that our fits converge stably to the highest probability density regions within the parameter space after an initial 350 `burn-in' steps.  Burn-in steps are not used for generating the posterior parameter distribution functions.  In total, the {\sc emcee} sampler sampler is run for 1000 steps with 50 walkers.

\subsection{Outputs}

\subsubsection{Orientation Parameters and their uncertainties}
\label{jackknife.sec} 

After the MCMC run has converged, we compute the marginalized posterior distribution functions (PDFs) for our four free parameters from the MCMC chains after the burn-in phase.  To determine the best-fit parameters and its respective uncertainties, we compute the median as well as the 16th and 84th percentile of the cumulative PDF.  The median values are taken as the best-fit $\phi$ and $i$, and are hereafter referred to as \pak\ and \inck\, to distinguish those from the photometry-based \pap\ and \incp.  The uncertainties implied by the width of the posterior distributions are typically quite small, $\lesssim 0.2$ degrees, representing the formal statistical error in the limit of our applied model.  

The statistical uncertainty does not capture systematic uncertainties that arise through the combination of the PHANGS-ALMA survey coverage and the presence of dynamical structures in the disk (i.e., bar and/or spiral structure).  When the motions in the gas associated with these features are left unfitted, they mimic spatial variations in disk orientation that manifest as well-known variations in tilted-ring determinations of $\phi$ and $i$ with radius in a galaxy, as we also show below.  Therefore, we characterize the uncertainties in our analysis using a jackknifing technique, which sub-samples the data that is incorporated into the analysis for a given galaxy.

We first divide each observed velocity field into four radial bins using ellipses that are oriented based on the photometric values.  Each radial bin corresponds to an annulus of the same width (i.e., $R_{{\rm CO,max}}/4$).  We then create three jackknife realizations of each velocity field by eliminating two of the four radial bins for a given realization.  Next, we re-run our kinematic fits on all three jackknife realization for a given galaxy. We then compute the mean absolute difference of the best-fit parameters from the jackknifes and our best-fit fiducial \pak\ and \inck.  Finally, we convert the values of mean absolute difference to a representative Gaussian $1\sigma$ error by applying a multiplication with a factor of $1.25$.

The jackknife uncertainties overall yield median uncertainties of 1.3 and 5.6 degrees on \pak\ and \inck, respectively.  This corresponds to an increase in uncertainty by factors of about 18 and 20 for $\phi$ and $i$, respectively, with respect to our formal fitting errors.  Thus, our jackknife uncertainties provide a good representation of the significant systematic variations in \pak\ and \inck\ not captured by our formal MCMC errors.

In addition, we determine the uncertainties of our best-fit $V_{{\rm sys}}$ using the jackknife realizations as done for \pak\ and \inck\ above.  We find an average error in $V_{{\rm sys}}$ of about 1\,$\mathrm{km~s^{-1}}$, which represents an increase by a factor of about $10$ compared to the formal fit uncertainty.  

\subsubsection{Final Rotation Curves}
\label{final_rcs.sec}
Once the best-fit orientation is determined for a given galaxy, we perform an independent least-squares fit of the velocity field model to the PHANGS-ALMA $V_\mathrm{obs}$ map to determine the best-fit rotation curve $V_{{\rm rot}}$(R) following the procedure described in Section~\ref{sec:harmdec}, but with $\phi$, $i$, and $V_{{\rm sys}}$ fixed to their best-fit global values.  To derive uncertainties, we create 100 realizations of each rotation curve in an iterative process, each time perturbing $\phi$, $i$, and $V_{{\rm sys}}$ by their respective jackknife error.  The center position $(X_{{\rm cen}},Y_{{\rm cen}})$ is also perturbed by $1''$ in each iteration.  The error on rotation velocity at a given radial bin is then computed based on the 16th and 84th percentile of the 100 realizations.  The rotation curves presented and discussed here are based on fits of our first-order model with $c_1$ determined from harmonic decomposition and all other terms in the expansion set to zero.

For a subset of 15 galaxies (identified and described in Section~\ref{Orienations.sec}), \pak\ and \inck\ are significantly affected by imperfect spatial sampling and/or the presence of either strong or spatially extended bars and spirals.  In these cases, we adopt the photometric disk inclination \incp\ to derive the rotation curve.  

While the MCMC optimization uses radial bins with width set by the ALMA synthesized beam, our final rotation curves are generated with a fixed 150~pc wide bins in radius, which generates rotation curves that probe a consistent physical scale in all galaxies.  This scale is chosen so that it is resolved even for the most distant galaxies (and the bin width corresponds to $\sim 1.5$ times the median resolution of our data).  For many galaxies, this choice has the added advantage of reducing radial variations in the rotation curves that stem from limited spatial sampling of the velocity field.    
The spatial sampling becomes increasingly inhomogeneous towards large galactocentric radii, which in turn induces correlated structures in our rotation curves.  Thus, we eliminate bins where the fraction of detections within a bin falls below a threshold number of pixels.  Furthermore, we eliminate bins where the average angular distance of detections from the kinematic major axis exceeds an angle $\delta$. Based on our inspection of the quality of rotational velocity measurements along with the coverage per ring, we adopt $\delta=55$ degrees and a 10\% threshold for our sample.

\begin {figure*}[htb]
\centering
\includegraphics[width=0.95\textwidth]{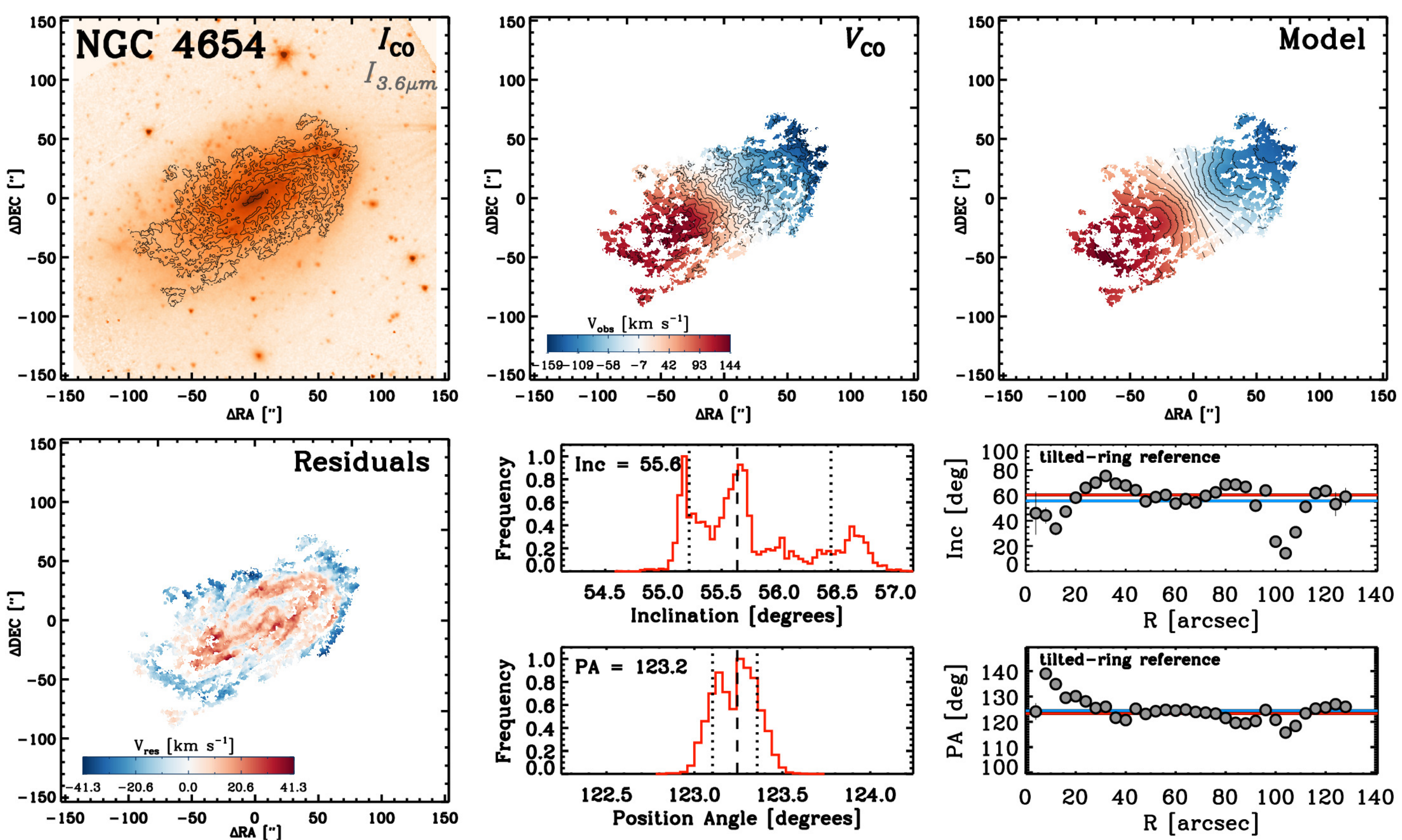}
\caption{One example case for our modelling results.  Top: Spitzer/3.6\,$\mu$m image with the PHANGS CO contours overlaid, moment-1 velocity field, best-fit velocity model.  Bottom: Best-fit residual field with fixed scaling of $\pm 50 {\rm km s^{-1} \sin{(i)}}$, $\phi$ and $i$ posterior distributions.  The final best-fit inclination \inck\ and position angle \pak\ determined from the posterior distributions are indicated as dashed vertical lines, with the 16th and 84th percentiles shown as vertical dotted lines.  Finally, we show the $\phi$ and $i$ profiles as a function of radius R as inferred from tilted-ring fits to the velocity fields.  Our global best-fit as well as the median value derived from all tilted rings are indicated by the red and blue horizontal lines, respectively.  The best-fit model and residual fields shown are derived from the final orientations \pak\ and \inck; note that the latter is replaced by the photometric \incp\ in case of unreliable fits.}
 \label{Showcase.fig}
\end {figure*}

\section{Results}
\label{Sect4.sec}
In this section we present the results of our modelling technique applied to the high resolution PHANGS-ALMA CO velocity fields of 67 galaxies.  Our assessment of the quality of these results is motivated by extensive testing of the method on mock galaxy velocity fields.  These tests confirm the overall accuracy of the method, with the exceptions described in detail in Section~\ref{Overview.sec}.  As a general assessment of the performance of our technique, we find good consistency with a number of kinematic orientations in the literature derived using similar techniques (see Appendix~\ref{orientationcomp.sec}).

\subsection{Example of results for one galaxy, NGC~4654}

As an example of our results, we highlight the application of our technique to one galaxy, NGC~4654.  This galaxy exemplifies the performance of our method on velocity fields with outer clumpy spatial sampling and modest non-circular motions.

Figure~\ref{Showcase.fig} shows the observed zeroth and first moment map, our best-fit model, and the respective residual field for NGC~4654.  The zeroth moment map compared to the Spitzer/3.6\,$\mu$m image is shown to illustrate the underlying stellar structure of the galaxy relative to the regions sampled by our CO kinematic maps.  Further, marginalized PDF posterior distributions for $\phi$ and $i$ are presented with an indication of the adopted best-fit value and the formal MCMC fit error.  For reference, we plot radial profiles of $\phi$ and $i$ as determined from a tilted-ring analysis with the {\sc ROTCUR} routine in the Groningen Image Processing System ({\sc GIPSY}; \citealt{vdhulst1992}).

The posterior distributions (middle, bottom) for the global \pak\ and \inck\ for NGC~4654 are well-sampled and reflect good convergence of the fit.  The posterior distributions are typically single-peaked and close to Gaussian, but multiple peaks can also occur.  In the latter case, the median of the distribution may occasionally fall between the peaks.  However, our jackknife uncertainty encompasses the width of the full multi-peaked distribution.  These well-behaved posteriors suggest that, despite the presence of non-circular motions, even our most basic model with only circular motion provides a sufficiently good match to the data for obtaining robust global disk orientations. 
We also note that the posterior distributions of both systemic velocity and excess variance for our fits (not shown) resemble well-sampled Gaussians. 

We attribute this behavior to our use of an excess variance term while fitting, which accommodates for differences between the data and fitted model.  These differences are apparent in the residual velocity field (bottom left) where the systematic deviations from our model are on the order of $10{-}15$ km~s$^{-1}$.  The excess variance term in this case converges to a similar level.  

For NGC~4654, as for the majority of galaxies, non-zero residuals reflect the presence of non-circular motions in the gas not included in our model of pure circular motion.  In the specific case of NGC~4654, residuals at radii dominated by the stellar bar ($a_B=29''$) in detail show the monopolar red-blue pattern characteristic of bars (red along one side, blue along the other; \citealt{Canzian1993}).  At larger radii, the velocity field exhibits the 3-armed pattern expected for an $m=2$ pattern outside corotation \citep{Canzian1993}. 

It is worth noting that the structure in the residual velocity map signifies that the observed velocity field is more complex than our chosen model of basic circular motion (rather than that the best fit model of this kind has not been identified).  Non-zero residuals also imply that the global orientation and rotation curve determined with our strategy are not strongly biased by non-circular motions. Alternative strategies that minimize residuals and compensate for non-circular motions by allowing \pak~and \inck~to unrealistically vary with radius (see the bottom right two panels of Figure~\ref{Showcase.fig}, for example) can yield, on average, an orientation that differs considerably from the global value fitted with our approach.  In the case of NGC~4654, the tilted-ring $\phi$ and $i$ averaged over all radii are consistent with our global solutions, although testing on other galaxies shows that this is sensitive to the area fitted and averaged.

\subsection{Overview of fit performance and assessment of reliability}
\label{Overview.sec}

To assess the reliability of our best-fit parameters, we carefully inspect the performance of each fit with regard to three primary criteria:

1) \textit{Physically realistic rotational velocities} -- We compare the outermost rotational velocities $V_{{\rm rot}}$ implied by the de-projected rotation curve associated with the fitted \inck to the velocity implied by the Tully-Fisher relation measured by \cite{McGaugh2015} using our integrated stellar mass estimates (Leroy et al in prep.). Unreliable fits for the galaxies in our sample (with stellar masses $9.0 < \log\,M/M_\odot < 10.9$) are easily recognizable for their unphysically low ($V_{{\rm rot}} < 50$ $\mathrm{km~s^{-1}}$) or high ($V_{{\rm rot}} > 500$ $\mathrm{km~s^{-1}}$) average de-projected outer rotational velocities.  Furthermore, the \inck\ measured for galaxies with an outer maximum $V_{\rm rot}$ that differs by more than $100$ $\mathrm{km~s^{-1}}$ from the Tully-Fisher value are deemed unreliable. 

2) \textit{Reasonable consistency with the galaxy's photometric appearance} -- Galaxies with a salient contrast between the fitted \inck\ and our \incp\ prior (with $\lvert i_{{\rm kin}} - i_{\rm phot} \rvert \gtrsim 30$ degrees) are flagged as unreliable.  

3) \textit{Not strongly affected by large stellar bars} -- Galaxies with large strong bars (as quantified by $\epsilon>0.4$ and $R_{{\rm bar}} / R_{{\rm CO,max}}\gtrsim 0.45$) are also flagged as unreliable as we find that in those cases our \inck\ measurements are likely compromised by bar-induced features in the observed velocity field. See Section~\ref{bars.sec} for more details. 

According to these criteria, we obtain reliable results for the majority (78\,\%) of our kinematic sample.  We adopt the best-fit values of \pak\ and \inck\ measured with our technique as the final orientation for  52 galaxies in total, and use them for all of our subsequent analysis.  For all unreliable fit results, we replace our best-fit \inck\ with our prior \incp\ when measuring the rotation curve and during all subsequent analysis.  We assign an uncertainty of 2 and 6 degrees on $\phi$ and $i$, respectively, for these galaxies based on our jackknife analysis (see Section\ \ref{Orienations.sec}). 

We also perform consistency checks between the rotation velocities implied by \incp\ and the expectation from the Tully-Fisher relation, and find that for two galaxies in our sample (NGC~1672 and NGC~5134), \incp\ implies rotational velocities in stark contrast to what is expected from the Tully-Fisher relation.  In these cases, we have traced the discrepancy to an issue with the 3.6\,$\mu$m disk axis ratio used to infer \incp, which seems to be affected by the presence of the outer (pseudo-)ring identified at 3.6\,$\mu$m in each of these two galaxies \citep[][]{Herrera2015}.  Thus, we replace our \incp\ values with the inclinations that are implied by the \cite{McGaugh2015} Tully-Fisher relation, using the total stellar mass and observed projected rotation curve of each of these two galaxies.

For the remaining galaxies, the characteristics of our fit performance vary significantly.  The fits that are deemed unreliable according to the above criteria are impacted by the spatial filling of the CO velocity field (see Section~\ref{impact_sampling.sec}; 4~galaxies in total) and/or the presence of strong non-axisymmetric bar and/or spiral structure (Section~\ref{bars.sec}; 11~galaxies in total). Galaxies with problematic fits are flagged in Table~\ref{Results.tbl}. Five additional showcases are presented in Appendix~\ref{Examples.sec} and discussed in detail in the upcoming sections.

\subsubsection{Impact of imperfect spatial sampling}
\label{impact_sampling.sec}
The spatial sampling of our data varies significantly across our sample (see Section~\ref{demographics.sec}) and impacts the coverage of the mapped velocity field.  This can adversely affect the quality of our MCMC fit, depending on how well the major axis information is sampled in the map.  When lines-of-sight close to the minor axis are significantly over-presented in the PHANGS-ALMA CO velocity field, rotational velocities and orientations are weakly constrained.  CO velocity fields with limited radial extents, in which velocities along the major axis never reach the turnover or flat part of the rotation curve, are also compromised.

As an example of how velocity field coverage affects the quality of our MCMC fits, we show the results for NGC~5042 and NGC~4293 in Figure~\ref{Showcase_sampling.fig}.  NGC~5042 (top panels) has a covering fraction of 27\,\%, but our fit still yields reliable results given that the coverage is sparse but fairly uniform.  
The CO emission in NGC~4293 (bottom panels in Figure~\ref{Showcase_sampling.fig}), in contrast, is only detected within the central $\sim 2$\,kpc and, as a result, the rotation curve can only be traced within its inner rising part.  The velocity field thus appears like a `solid-body' rotator and the degeneracy between $i$ and $V_{{\rm rot}}$ in this case prevents reliable measurement of \inck.

\subsubsection{Impact of stellar bars and spiral arms}
\label{bars.sec}
For galaxies in the sample with especially strong and prominent stellar bars and/or spirals, our fitting technique is unable to reliably measure the disk inclination angle.  This is expected, as strong non-circular streaming motions in bars and spiral arms as well as, e.g., gas inflow along bar dust lanes, lead to systematic deviations from our basic model of purely circular motion. 

When covering a large fraction of the mapped area (as quantified by $R_{{\rm bar}} / R_{{\rm CO,max}}\gtrsim 0.45$), strong bars (with $\epsilon > 0.4$) are especially problematic and yield \inck\ estimates that are unreliable based on our criteria discussed above. We therefore use $R_{{\rm bar}} / R_{{\rm CO,max}}\gtrsim 0.45$ as a guideline for rejecting kinematic orientations. We emphasize, however, that rejection is, in the end, decided based on detailed inspection of the solution, keeping in mind, e.g., the orientation of the bar with respect to the major axis, the morphology of the gas within the bar, the overally CO filling factor and the magnitude of streaming motions suggested by the size of velocity residuals.  Good kinematic solutions can still be obtained, e.g. for long bars that are oriented along the galaxy major axis, which thus present few strongly biased lines-of-sight, or when bar orbits are not uniformly populated by gas (given the gas morphology), so that the bar zone contributes only a small fraction of the total number of lines-of-sight in the velocity field.  

The filling factor of the CO outside the bar can also determine the impact of streaming motions from the bar zone on the fitted orientation.  Solutions for some systems below our guideline threshold $R_{{\rm bar}} / R_{{\rm CO,max}}\sim 0.45$ may thus be compromised.  However, in our sample, we have found that even when a small bar leads to obvious deviations from circular motion in the observed velocity field, typical CO filling factors are high enough that there are sufficient unbiased lines-of-sight to obtain an unbiased global fit.

Overall, we find that our technique is able to perform well on the majority of barred systems.  Of the 67\,\% of the total sample that is barred, only 11 host the most prominent troublesome bars (see Figure \ref{demographics.fig}).  Solutions for these are categorized as bar-dominated and flagged as unreliable.   Good solutions are obtained for the remaining barred galaxies.  As an example, Figure~\ref{Showcase_bars.fig} shows the case of NGC~4535 (top panels).  Within the region of the stellar bar ($R_{{\rm bar}} = 38''$), strong non-circular motions lead tilted ring measurements (bottom right two panels) to deviate strongly from the global best-fit $\phi$ and $i$, whereas the outer measured orientation (in bins where the velocity field better traces circular motion) matches the global MCMC fit.  In these cases, our \pak\ and \inck\ are robust and are only mildly affected by the presence of the bar and/or spiral arms (see Section~\ref{Orienations.sec}).

The results for NGC~5643 (bottom panels in Figure~\ref{Showcase_bars.fig}) are a good illustration of a failed case, where a prominent stellar bar with strong non-circular motions yields a problematic fit.  The failure appears related to the extent of the bar (which reaches 46\,\% of the total CO radial extent) and the high CO covering fraction throughout the bar area.   %

As an example of the global impact of spiral arms, Figure~\ref{Showcase_4254.fig} highlights results for the unbarred galaxy NGC~4254.  The CO velocity field covers far out into the disk, reaching radii where the rotation curve flattens out.  The non-circular motions associated with the extended spiral arms lead to small apparent deviations in $\phi$ and $i$ from their global average values.  However, as the overall spatial filling factor of the CO velocity field is quite high, the global impact of these narrow features is low, and we are able to obtain a robust MCMC fit.

\subsection{Galaxy orientations}
\label{Orienations.sec}

Table~\ref{Results.tbl} lists our final adopted orientation parameters and their associated uncertainties measured with our jackknife analysis, together with our best-fit systemic velocities.  For the 15 fits that are deemed unreliable, the \inck\ measurement is replaced by the respective photometric inclination \incp, indicated in Table~\ref{Results.tbl}.  We assign a typical uncertainty in those cases (and in cases where we adopt an inclination based on the Tully-Fisher relation) of 6~degrees for~$i$.  These values are adopted based on the median jackknife error for our kinematic sample (see Section~\ref{jackknife.sec}). 

\startlongtable
\begin{deluxetable}{DDDDDD}
\tablecaption{Best-fit results and final orientation parameters for our kinematic sample. Galaxies with fits that are deemed as unreliable are marked with an asterisk ($*$).}
\tablehead{
\multicolumn2c{ID} & \multicolumn2c{$\phi$} & \multicolumn2c{$\sigma_{\phi}$} & \multicolumn2c{$i$} & \multicolumn2c{$\sigma_i$} & \multicolumn2c{$V_{{\rm sys}}$}\\ \multicolumn2c{} & \multicolumn2c{[deg]} & \multicolumn2c{[deg]} & \multicolumn2c{[deg]} & \multicolumn2c{[deg]} & \multicolumn2c{[$\mathrm{km~s^{-1}}$]}
}
\decimals
\startdata
${\rm IC~1954\;\:}$    &    63.4  &       0.2  &     57.1  &    0.7  &  1039.9         \\     
${\rm IC~5273\;\:}$    &   234.1  &       2.0  &     52.0  &    2.1  &  1285.6         \\     
${\rm NGC~0628\;\:}$     &    20.7  &       1.0  &      8.9  &   12.2  &   650.9         \\   
${\rm NGC~0685\;\:}$   &   100.9  &       2.8  &     23.0  &   43.4  &  1347.0         \\     
${\rm NGC~1087\;\:}$   &    -0.9  &       1.2  &     42.9  &    3.9  &  1501.6         \\     
${\rm NGC~1097^*}$     &   122.4  &       3.6  &     48.6$^1$  &    6.0  &  1257.5         \\ 
${\rm NGC~1300^*}$     &   278.1  &       0.9  &     31.8$^1$  &    6.0  &  1545.4         \\     
${\rm NGC~1317 \;\:}$  &   221.5  &       2.9  &     23.2  &    7.7  &  1930.0         \\     
${\rm NGC~1365^*}$     &   201.1  &       7.5  &     55.4$^1$  &    6.0  &  1613.3         \\ 
${\rm NGC~1385 \;\:}$  &   181.3  &       4.8  &     44.0  &    7.6  &  1476.7         \\     
${\rm NGC~1433^*}$     &   199.7  &       0.3  &     28.6$^1$  &    6.0  &  1057.4         \\ 
${\rm NGC~1511\;\: }$  &   297.0  &       2.1  &     72.7  &    1.2  &  1329.8         \\     
${\rm NGC~1512^*}$     &   261.9  &       4.3  &     42.5$^1$  &    6.0  &   871.4         \\ 
${\rm NGC~1546\;\: }$  &   147.8  &       0.4  &     70.3  &    0.5  &  1243.7         \\     
${\rm NGC~1559\;\: }$  &   244.5  &       3.0  &     65.3  &    8.4  &  1275.0         \\     
${\rm NGC~1566\;\: }$  &   214.7  &       4.1  &     29.6  &   10.7  &  1482.5         \\     
${\rm NGC~1672^*}$     &   134.3  &       0.5  &     42.6$^2$  &    6.0  &  1319.2         \\ 
${\rm NGC~1792\;\: }$  &   318.7  &       0.8  &     65.1  &    1.1  &  1176.9         \\     
${\rm NGC~1809\;\: }$  &   138.2  &       8.9  &     57.6  &   23.6   & 1288.8          \\    
${\rm NGC~2090\;\: }$  &   192.6  &       0.6  &     64.5  &    0.2  &   897.7         \\     
${\rm NGC~2283\;\: }$  &    -4.1  &       1.0  &     43.7  &    3.6  &   822.1         \\     
${\rm NGC~2566^*}$     &   312.3  &      75.9  &     48.5$^1$  &    6.0  &  1609.6         \\ 
${\rm NGC~2775\;\: }$  &   156.5  &       0.1  &     41.2  &    0.5  &  1339.2         \\     
${\rm NGC~2835\;\: }$  &     1.0  &       1.0  &     41.3  &    5.3  &   866.9         \\     
${\rm NGC~2903\;\: }$  &   203.7  &       2.0  &     66.8  &    3.1  &   547.1         \\     
${\rm NGC~2997\;\: }$  &   108.1  &       0.7  &     33.0  &    9.0  &  1076.0         \\     
${\rm NGC~3059\;\: }$  &   -14.8  &       2.9  &     29.4  &   11.0  &  1236.3         \\     
${\rm NGC~3137\;\: }$  &    -0.3  &       0.5  &     70.3  &    1.2  &  1087.1         \\     
${\rm NGC~3351^*}$     &   192.7  &       0.4  &     45.1$^1$  &    6.0  &   774.7         \\ 
${\rm NGC~3507\;\: }$  &    55.8  &       1.3  &     21.7  &   11.3  &   969.7         \\     
${\rm NGC~3511\;\: }$  &   256.8  &       0.8  &     75.1  &    2.2  &  1096.7         \\     
${\rm NGC~3521\;\: }$  &   343.0  &       0.6  &     68.8  &    0.3  &   798.1         \\     
${\rm NGC~3596\;\: }$  &    78.4  &       0.9  &     25.1  &   11.0  &  1187.9         \\     
${\rm NGC~3621\;\: }$  &   343.7  &       0.3  &     65.8  &    1.8  &   723.3         \\     
${\rm NGC~3626^*}$     &   165.4  &       0.7  &     46.6$^1$  &    6.0  &  1470.7         \\ 
${\rm NGC~3627\;\: }$  &   173.1  &       3.6  &     57.3  &    1.0  &   717.9         \\     
${\rm NGC~4207\;\: }$  &   121.9  &       2.0  &     64.5  &    6.0  &   606.2         \\     
${\rm NGC~4254\;\: }$  &    68.1  &       0.5  &     34.4  &    0.9  &  2388.5         \\
${\rm NGC~4293^*}$     &    46.6  &       1.1  &     65.0$^1$  &    6.0  &   926.2         \\ 
${\rm NGC~4298\;\: }$  &   313.9  &       0.7  &     59.2  &    0.7  &  1137.8         \\     
${\rm NGC~4303\;\: }$  &   312.4  &       2.5  &     23.5  &    9.2  &  1560.2         \\     
${\rm NGC~4321\;\: }$  &   156.2  &       1.7  &     38.5  &    2.4  &  1572.4         \\     
${\rm NGC~4457^*}$     &    78.0  &       1.7  &     17.4$^1$  &    6.0  &   886.0         \\ 
${\rm NGC~4496a}$      &    51.1  &       4.1  &     53.8  &    3.5  &  1722.1         \\     
${\rm NGC~4535\;\: }$  &   179.7  &       1.6  &     44.7  &   10.8  &  1953.1         \\     
${\rm NGC~4536\;\: }$  &   305.6  &       2.3  &     66.0  &    2.9  &  1795.1         \\     
${\rm NGC~4540\;\: }$  &    12.8  &       4.4  &     28.7  &   28.7  &  1286.6         \\     
${\rm NGC~4548^*}$     &   136.6  &       0.5  &     38.3$^1$  &    6.0  &   482.7         \\ 
${\rm NGC~4569^*}$     &    18.1  &       4.4  &     70.0$^1$  &    6.0  &  -226.6         \\ 
${\rm NGC~4571\;\: }$  &   217.5  &       0.6  &     32.7  &    2.1  &   343.2         \\     
${\rm NGC~4579\;\: }$  &    91.3  &       1.6  &     40.2  &    5.6  &  1516.0         \\     
${\rm NGC~4654\;\: }$  &   123.2  &       1.0  &     55.6  &    5.9  &  1052.6         \\     
${\rm NGC~4689\;\: }$  &   164.1  &       0.2  &     38.7  &    2.5  &  1614.0         \\     
${\rm NGC~4781\;\: }$  &   290.0  &       1.3  &     59.0  &    3.9  &  1248.5         \\     
${\rm NGC~4826\;\: }$  &   293.6  &       1.2  &     59.1  &    0.9  &   409.9         \\     
${\rm NGC~4941\;\: }$  &   202.2  &       0.6  &     53.4  &    1.1  &  1115.9         \\     
${\rm NGC~4951\;\: }$  &    91.1  &       0.5  &     70.2  &    2.3  &  1176.6         \\     
${\rm NGC~5042\;\: }$  &   190.6  &       0.8  &     49.4  &   18.1  &  1386.2         \\     
${\rm NGC~5068\;\: }$     &   342.4  &       3.2  &     35.7  &   10.9  &   667.1      \\     
${\rm NGC~5134^*}$     &   308.9  &       2.8  &     22.1$^2$  &    6.0  &  1749.1      \\    
${\rm NGC~5248\;\:}$   &   109.2  &       3.5  &     47.4  &   16.3  &  1162.0         \\     
${\rm NGC~5530\;\: }$  &   305.4  &       0.9  &     61.9  &    2.3  &  1183.8         \\     
${\rm NGC~5643^* }$    &   317.0  &       2.4  &     29.9$^1$  &    6.0  &  1191.3      \\    
${\rm NGC~6300\;\: }$  &   105.4  &       2.3  &     49.6  &    5.9  &  1102.2         \\     
${\rm NGC~6744\;\: }$  &    14.0  &       0.2  &     52.7  &    2.2  &   832.4         \\     
${\rm NGC~7456\;\: }$  &    16.0  &       2.9  &     67.3  &    4.3  &  1192.4         \\     
${\rm NGC~7496\;\: }$     &   193.7  &       4.2  &     35.9  &    0.9  &  1636.9     \\
\enddata
\tablecomments{$^1$: Final inclination adopted from photometric prior; $^{2}$: final inclination based on the Tully-Fisher relation from \cite{McGaugh2015}}
\label{Results.tbl}
\end{deluxetable}

\subsubsection{Comparison to photometric orientations}
\label{comp_phot.sec}
\begin{figure*}[tb]
\centering
 \includegraphics[width=0.90\textwidth]{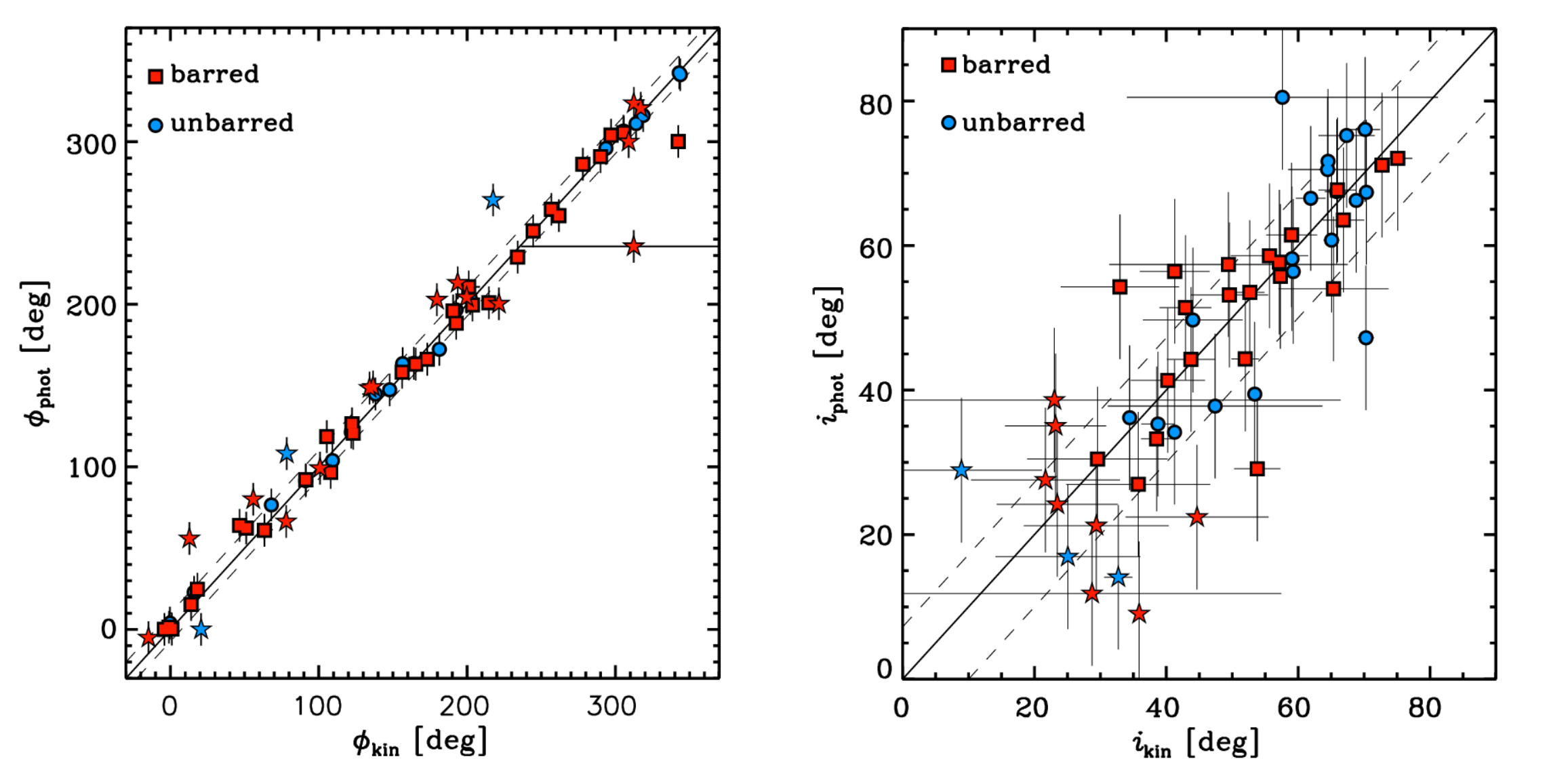}
\caption{Best-fit position angles $\phi$ and inclinations $i$ determined from kinematics versus photometry.  We split our sample in barred (red) and unbarred (blue) systems to highlight the impact of bars on measured galaxy orientation.  In addition, nearly face-on systems with either \inck\ or $\incp < 25$ degrees are marked with a different symbol. The right panel only includes the subset of 52 galaxies with reliable \inck\ measurement (see Section~\ref{Overview.sec}).  The shown error bars represent the uncertainties listed in Table~\ref{Results.tbl}, and are derived from our jackknife analysis. Solid and dashed lines represent a one-to-one correlation and the scatter in the relations, respectively (see Table~\ref{Offsets.tbl} for tabulated values).}
 \label{Kin_phot.fig}
\end{figure*}

Figure~\ref{Kin_phot.fig} compares our kinematic orientations to the photometric orientations throughout the sample.  The left panel shows the measurements of \inck~and \incp~for all galaxies in  our sample.  The right panel only shows measurements for the subset of 52 galaxies with reliable \inck\ measurements.  Both panels separate barred and unbarred subsets.

Overall, we find good agreement between the kinematic and photometric orientations.  The median offsets and scatter based on the 16th and 84th percentiles from a one-to-one relation in both panels of Figure~\ref{Kin_phot.fig} are listed in Table~\ref{Offsets.tbl}. 

\begin{table}[hbt]
\centering
\caption{Median offsets and scatter (taken from the 16th and 84th percentiles) of photometric versus kinematic orientations presented in Figure~\ref{Kin_phot.fig}.}
\begin{tabular}{l*{3}{c}r}
\hline
\hline
Sample\,\,\, & $\langle \phi_{{\rm kin}} - \phi_{{\rm phot}} \rangle$\,\,\, & $\langle i_{{\rm kin}} - i_{{\rm phot}} \rangle$   \\
 $\,$ & [deg]  & [deg]  \\
\hline
All      & $-1.3^{+7.4}_{-9.9}$ & $-0.5^{+10.0}_{-7.2}$ \\ 
Barred   & $-0.9^{+17.6}_{-12.4}$ & $-0.6^{+10.4}_{-7.5}$ \\ 
Unbarred & $-0.7^{+3.5}_{-6.2}$ & $ 1.7^{+7.4}_{-8.5}$  \\

\hline
\hline
\end{tabular}
\label{Offsets.tbl}
\end{table}

The basic agreement between the photometric and kinematic orientations implies an underlying consistency between the orientation probed by the CO-emitting disk and the orientation of the outer stellar disk as traced at 3.6\,$\mu$m, as considered in more detail in Section~\ref{Shapes.sec}. The agreement suggests that the quality of our fits is robust to variations in CO filling factor across the sample, except in extreme cases, as discussed in Section~\ref{impact_sampling.sec}.  

Furthermore, orientations agree equally well for unbarred and barred galaxies (which show similarly small offsets and scatter in Figure~\ref{Kin_phot.fig} and Table~\ref{Offsets.tbl}).  Since the photometric orientations we adopt are largely obtained from the outer disk, this suggests that our kinematics-based \pak\ are only mildly affected by the presence of bars. The same appears to hold true for our measures of \inck, although the quality of the agreement with \incp\ is partly biased by our omission of the systems with the most prominent bars (see Section~\ref{sec:demographics}).

In detail, differences between photometric and kinematic orientations are present, but these can mostly be related to the limitations of our adopted model of uniform galaxy flatness and circularity assumed in Equation~\ref{photinc.eq}, as well as the uncertainties inherent at low inclination.  Three galaxies have $\vert\pak{-}\pap\vert > 3\sigma$, where $\sigma$ is the combined uncertainty on the kinematic and photometric angles\footnote{The differences $\vert\inck{-}\incp\vert$ are never larger than $3\sigma$.}. These are almost exclusively systems with either low \inck\ or low \incp.  For such nearly face-on galaxies that appear nearly circular in projection (axis ratios $\gtrsim 0.75$), the \pap\ is inherently poorly constrained. 
At present, however, our adopted errors either on on \pap\ or on \incp\ represent the average of the differences in inclinations from ellipticities measured in different photometric bands across the sample and therefore do not reflect this systematic trend with inclination. 

The systematic disagreement between photometric and kinematic orientations in face-on systems is also evident in the right panel of Figure~\ref{Kin_phot.fig}, where the scatter about the one-to-one line increases towards the bottom left of the plot.  In a small majority of these nearly face-on cases, \incp\ underestimates \inck\ (many systems scatter below the one-to-one relation). 
This does not appear to be an issue with the fitted \inck, as we find no correlation between the difference $\vert\inck{-}\incp\vert$ and most other galaxy properties, including stellar mass, specific star-formation rate, $R_{25}$, B/T, bar presence, bar length, size of the bulge and disk, fitted rotation curve transition radius $r_{\rm t}$.  It seems much more likely that the \incp~for these systems is biased low, due to a mismatch between the intrinsic flatness of the galaxy and the assumed flatness (with underestimated thickness leading to underestimated \incp), or due to an intrinsic, but modest, departure from circularity in the outer disk shape.  These factors, which are discussed more in Section~\ref{Shapes.sec}, introduce discrepancies in \incp\ that systematically increase in low-inclination systems. 

A number of possible physical mechanisms can induce the variations in galaxy shapes that yield discrepant \inck\ and \incp\ (see  Section~\ref{Shapes.sec}).  Here we note that, of the few galaxies with $\vert\inck{-}\incp\vert > 1\sigma$ and $\inck > 45$ degrees, the majority are unbarred, making it unlikely that strong non-circular motions are responsible for the offset.

\subsubsection{Comparison with lower-resolution measurements}

\begin {figure*}[htb]
\centering
 \includegraphics[width=0.90\textwidth]{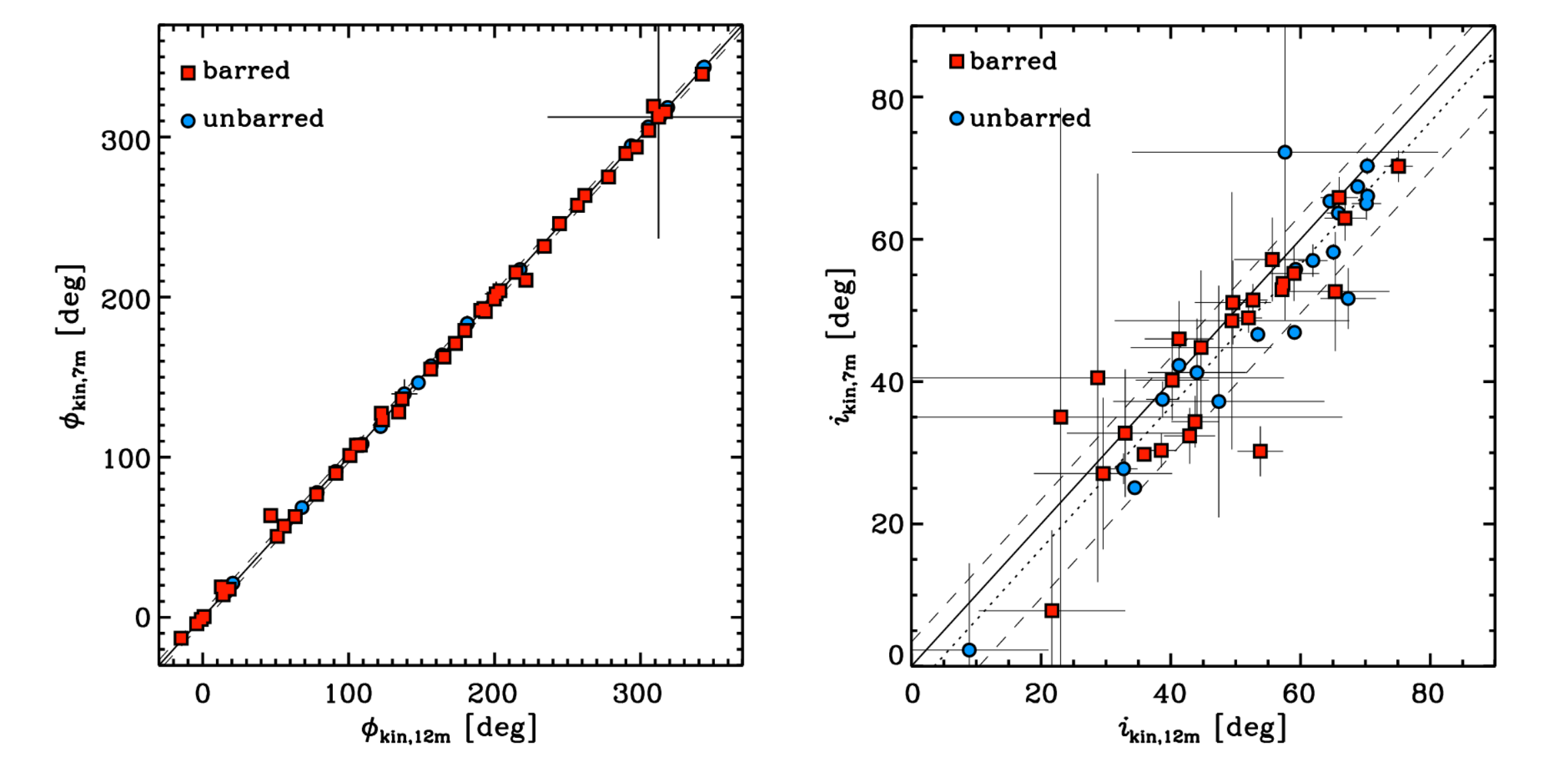}
\caption{Comparison between the best-fit orientations derived from our 12m PHANGS-ALMA data set to the same measurements on lower-resolution 7m ALMA data.  Uncertainties are adopted from our jackknife method based on the 12m data set.  Solid, dotted and dashed lines represent the one-to-one relation, measured offset, and scatter in the relations, respectively.  We split our sample in barred and unbarred systems, indicated by red and blue symbols, respectively. The right panel shows the 44 galaxies with reliable \inck\ measured from both the 12m and 7m data sets.}
 \label{Orientations_7m.fig}
\end {figure*}

Next, we explore the impact of spatial resolution on our orientation measurements. To do so, we repeat our kinematic fits to the PHANGS-ALMA first moment maps derived from the 7m(+TP) observations (see A.~K.\ Leroy et al., in prep., for details) for the same set of galaxies. These maps are characterized by lower spatial resolution ($\sim 650$\,pc on average) and increased surface brightness sensitivity (and thus higher CO covering fraction) compared to our fiducial high-resolution data set.  Our comparison of fitted inclinations is restricted to the subset of the sample (44 galaxies) with reliable \inck\ measured from both the 12m and 7m data sets.   

In principle, measurements of orientations and rotation curves from fits to lower angular resolution velocity fields should be less prone to biases introduced by non-circular motions, which become less prominent with spatial averaging.  An increase in beam smearing on the velocity field at lower angular resolution, though, may be expected to affect our ability to constrain the inclination, possibly leading to measurement biases. 

In practice, we find that the values we derive for $\phi$ and $i$ from the 12m maps are mostly comparable to the values derived from the 7m data as shown in Figure~\ref{Orientations_7m.fig} (and tabulated in Table\ \ref{7m.tbl} in Appendix \ref{7m.sec}). For both $\phi$ and $i$, we find only small median systematic offsets of $0.0^{+1.0}_{-0.9}$ and $3.5^{+5.8}_{-4.3}$ degrees, respectively.  The scatter in both relations is in agreement with the uncertainties we have inferred for our measurements. 

The tight correspondence between galaxy inclination measured at different spatial resolution highlights the design of our approach, which maximizes constraints from lines-of-sight not strongly biased by non-circular motions. The small offset to lower
inclination when using lower angular resolution data
plausibly stems from the impact of beam smearing, which not only reduces the inner rotational velocity gradient but also circularizes the distribution of emission.  We have found that, with Gaussian smoothing, basic models of projected circular motion are best fit with rings that are systematically less inclined than the adopted inclination by an amount that increases with the size of the Gaussian ‘beam’.  At resolutions typical of the 7m data, the fitted inclination is lowered by $1{-}4$ degrees, which is consistent with the typical offset found in the right panel of Figure~\ref{Orientations_7m.fig}.

\begin {figure*}[htb]
\centering
 \includegraphics[width=0.95\textwidth]{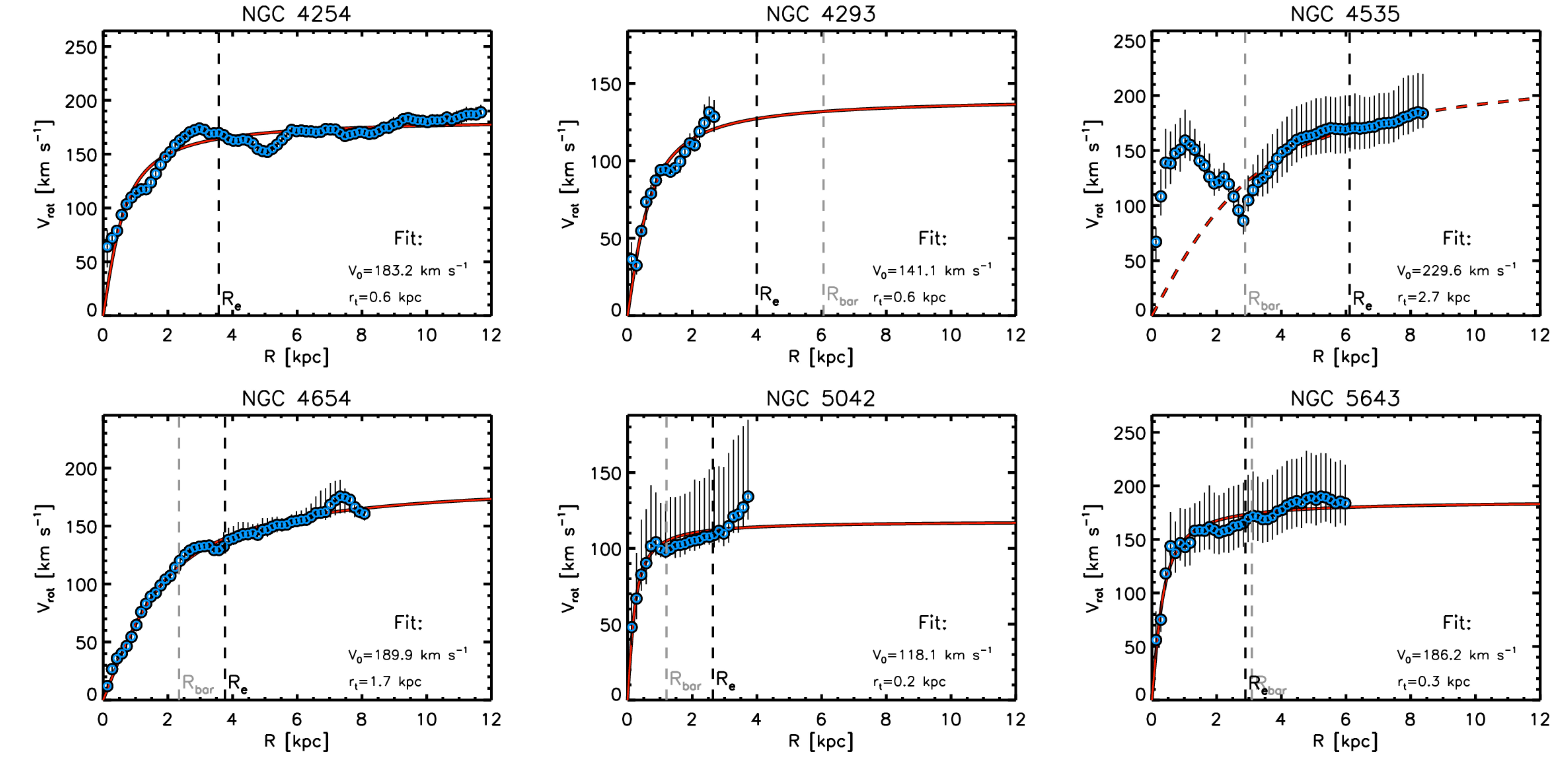}
\caption{Examples of rotation curves for the subset of galaxies shown in Figure\ \ref{Showcase.fig} and the Appendix \ref{Examples.sec}, sampled at a fixed scale of 150\,pc.  The errors in $V_{{\rm rot}}$ mainly stem from uncertainties in our orientation parameters.  Smooth fits are shown as red lines, together with their best-fit parameters and the asymptotic velocity $V_{0}$.  In some panels the red lines are dashed to indicate cases where the central 2\,kpc region have been masked before fitting a smooth model.  Effective radii and bar lengths are  indicated with dashed vertical lines.}
 \label{RC_examples.fig}
\end {figure*}     

\subsection{Rotation curves}
\label{RCs.sec}

\subsubsection{General characteristics}
\label{RCs.sec}
Figure~\ref{RC_examples.fig} shows examples of our rotation curves for the set of 6~galaxies showcased above and in Appendix~\ref{Examples.sec}.  We also present the rotation curves for our full set of 67 galaxies in Appendix~\ref{full_set.sec} (see Figures \ref{RC_all1.fig}, \ref{RC_all2.fig}, \ref{RC_all3.fig}, as well as Table~\ref{Vrot.tbl} for all tabulated rotation curves).  The error on the inclination dominates the uncertainty on the measured rotation curves.  Note that we do not account for bin covariance in our assignment of an uncertainty to each individual radial bin.  In each panel, the result of our smooth model fit, performed as described below in Section~\ref{smooth_fits.sec}, is shown for reference.  The fitted asymptotic rotation velocity ($V_0$) and its uncertainty is also reported.

Several key features (typical of the full sample) are evident:  first, the CO rotation curve can often be reliably determined out to $\sim 0.7\,R_{\rm 25}$ (or roughly $\sim 2.2\,R_{\rm e}$; \citealt{Schruba2011}) and thus captures the characteristic flattening in circular velocity after an initial steep rise.  This transition happens on average within $0.1 R_{25}$, according to the rotation curve fits introduced later in Section~\ref{smooth_fits.sec}. The properties of the inner rotation curve gradient are commented on in Section~\ref{inner_slope.sec}.

Many rotation curves exhibit localized wiggles that occur on $<\,500$\,pc scales (see \citealt{henshaw20}).  The smallest of these ($5{-}10$ $\mathrm{km~s^{-1}}$) likely reflect bin-to-bin differences in azimuthal coverage, given the way the morphology of the gas affects the completeness of line-of-sight sampling. Slightly larger wiggles (on the order of $10{-}15$ $\mathrm{km~s^{-1}}$) may reflect an unaccounted contribution from non-circular motions in spiral arms, which appear on $\sim0.3{-}1$~kpc scales, depending on the pitch angle of the spiral.

\begin{figure*}[htb]
\centering
  \includegraphics[width=0.8\textwidth]{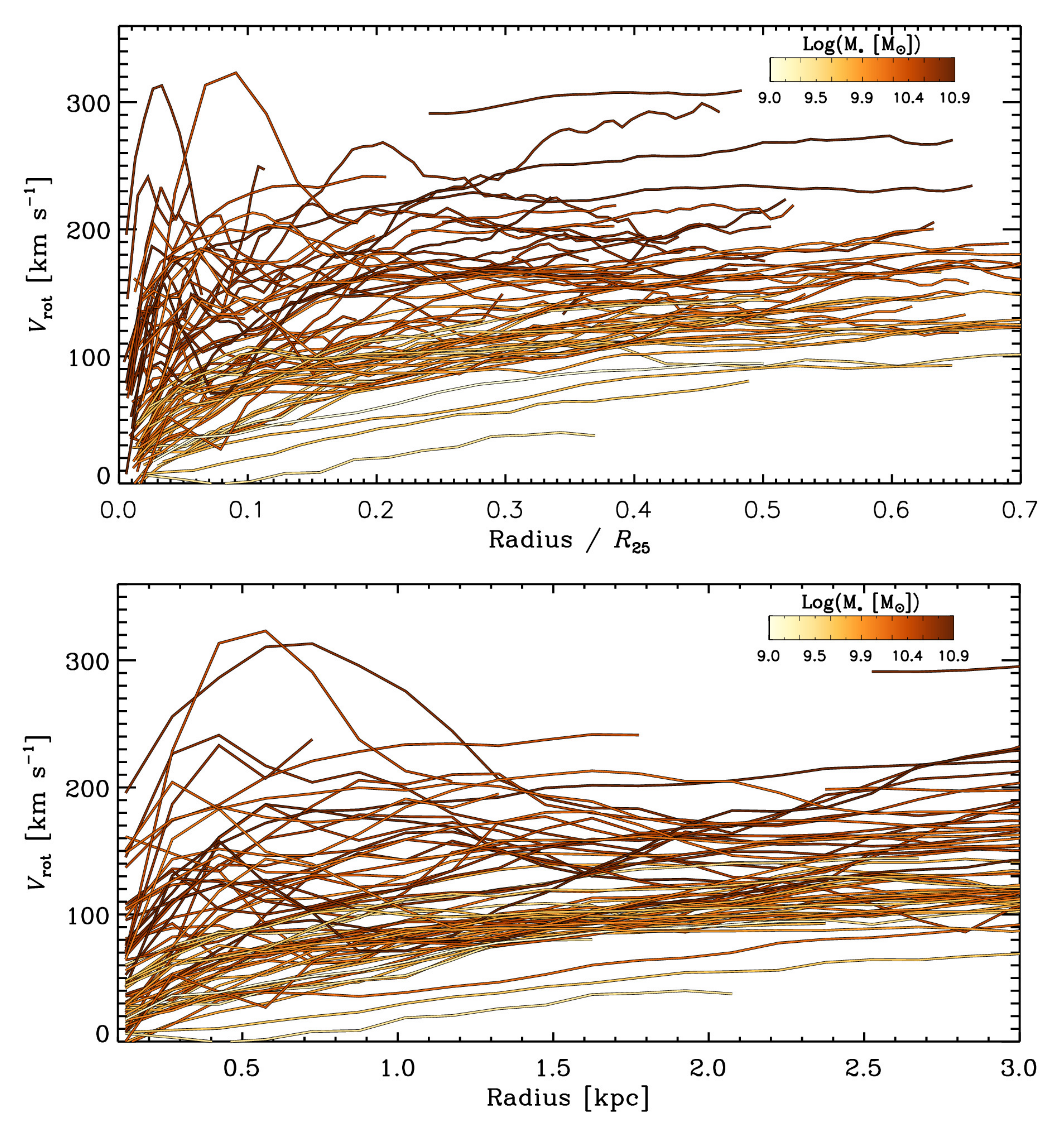}
  \caption{The rotation curves for our sample of 67 galaxies.  Top: Radii are normalized by $R_{\rm 25}$.  Bottom: Zoom-in on the inner 3\,kpc to highlight the variety of inner rotation curves shapes.  The color coding indicates total stellar mass.}
   \label{RCs.fig}
\end{figure*}

In obviously barred galaxies, larger wiggles ($20{-}40$ $\mathrm{km~s^{-1}}$) are evident on scales that are consistent with the length of the bar itself.  NGC~4535 is a good example (see top panels of Figure~\ref{Showcase_bars.fig}), where the bar is oriented at $\sim$35 degrees with respect to the galaxy's kinematic major axis, so that non-circular bar-induced streaming motions project strongly into the line-of-sight.  This contributes to the bump in rotational velocity at radii within the bar.  

The inner peak in NGC~4535's rotation curve also reflects a genuine central mass concentration in the form of the inner bulge that occupies the central ${\sim}1$\,kpc \citep{Yoshino2008}.  As we will examine in more detail in the next section, the PHANGS CO rotation curves are optimal tracers of central mass concentrations in the sample.  

\subsubsection{Rotation curves as tracers of galaxy mass}
Figure~\ref{RCs.fig} plots the rotation curves for our full set of 67 galaxies, with the radius normalized by $R_{\rm 25}$.  
PHANGS-ALMA maps typically extend out to $\sim 0.7\,R_{\rm 25}$ (or $\sim2.2\,R_{\rm e}$; \citealt{Schruba2011}) and the fitted rotation curves regularly trace out to radii where rotational velocities flatten out, thus offering a good view of the enclosed dynamical mass.  We see that the amplitude of the outer flat rotational velocity clearly increases with increasing stellar mass.  Figure~\ref{RCs.fig} also demonstrates that lower mass galaxies tend to have more slowly rising rotation curves than higher mass galaxies and reach lower $V_{{\rm rot}}$ within $0.2\,R_{\rm 25}$.  These trends in rotation curve shape depending on stellar mass have been observed by, e.g., \citet{Kalinova2017}, and are described by global galaxy scaling relations \citep[][]{Dutton2009,Meidt2018}.  We will examine the correlation of rotation curve shape with stellar mass with the aid of smooth analytic fits further in the next section. 

At the innermost radii, rotation curves exhibit a variety of behaviors, particularly in the more massive galaxies in our sample.  Given the greater prominence of bulges in more massive systems \citep[e.g.,][]{kormkenn2004}, which is quantified as an increase in bulge-to-total mass ratio (B/T) with increasing stellar mass \citep[see also][]{Bluck2014}, some of the inner behavior appears to trace genuine variation in the mass distribution.  In other galaxies, the bar and/or bar streaming motions may contribute to the recovered rotational velocities.  Indeed, prominent bars appear preferentially in the high mass PHANGS-ALMA targets.

\subsubsection{Smooth analytic fits}
\label{smooth_fits.sec}
We perform smooth two-parameter analytic fits to the measured rotation curves in the PHANGS-ALMA sample with three main goals.  First, we aim to describe the outer asymptotic behavior of the rotation curve, to serve as a guide for modelling circular velocities at and beyond the edge of the CO field-of-view.  The asymptotic behavior is of particular interest for understanding the accuracy with which CO kinematics can be used as a dynamical mass tracer.  Secondly, smooth model fits minimize fluctuations that most likely arise with non-circular motions, rather than track variations in the  underlying mass distribution.  Thus, smooth analytic fits can be used for fitting the contribution from non-circular motions to the observed velocity field \citep[][]{vandeVen2010,Colombo2014}.  The third reason is for obtaining a desirable analytical derivative of the rotation curve, as an alternative to the often noisy discrete derivative that can be obtained directly from the measured rotation curve.  This is an important quantity when aiming to quantify the influence of shear on, e.g., molecular cloud evolution \citep[][]{Meidt2018,kruijssen19b,chevance20}.

We adopt the empirical parametrization of rotation curves by the function \citep{Courteau1997}: \begin{equation}
V_{{\rm rot}} = V_0 \frac{2}{\pi} \arctan({R/r_{{\rm t}}}),
\label{arctan.eq}
\end{equation}
which describes a smooth rise in rotational velocity up to a maximum $V_0$ reached asymptotically at infinite radius.  The radius $r_{{\rm t}}$ denotes the transition between the rising and flat part of the rotation curve and encodes its inner slope.  We derive uncertainties on $V_0$ and $r_{{\rm t}}$ by iterating our fits over all 100 Monte-Carlo realizations of rotation curves derived in Section~\ref{final_rcs.sec}, and then determining the 16th and 84th percentiles of the resulting distributions of $V_0$ and $r_{{\rm t}}$.  In this way, the uncertainties on the fitted parameters are inherited from the uncertainty on the rotation curve itself, which is dominated by the inclination uncertainty.  The uncertainty on the asymptotic $V_0$ is especially sensitive to the inclination uncertainty (while $r_{{\rm t}}$ is less so) and, as a consequence, the fractional uncertainties on $V_0$ are larger at low inclination.  Our best-fit $V_0$ and $r_{{\rm t}}$ values are presented in Table~\ref{smooth_parameters.tbl}, together with their uncertainties.

Using these model fits we compute the velocity at $R_{25}$, $V_{{\rm rot,25}}$, and calculate the total dynamical mass within the optical disk via $M_{\rm dyn, R_{25}} = {\rm G}^{-1}\,V^{2}_{{\rm rot,25}}R_{25}$.

For a small fraction of the sample, the inner rotation curve strongly deviates from the basic smooth rise parametrized in Equation~\ref{arctan.eq}.  This difference tends to be confined to the central $\sim 2$\,kpc (see Figure~\ref{RCs.fig}) and marks the influence of central dynamical structures (bulges and bars).  Since we are mostly interested in describing the rotation in the disk region, we exclude radii $R<2$\,kpc to prevent this region from dominating our analytic fits. In total, 12 galaxies are masked in this manner, 10 of which are barred. 

Galaxies with rotation curves that do not reach beyond $1R_{\rm e}$ and/or do not flatten out are not fitted.  NGC~2775 and NGC~6744 are also excluded from our smooth fits since rotational velocities could not be measured within the inner galaxy ($\lesssim 2$\,kpc) given the morphology of the CO emission (NGC~2775) and the coverage of the ALMA mosaic (NGC~6744).  In total, a minor set of 6 galaxies are excluded from this exercise. 

\begin{figure*}[tb]
\centering 
  \includegraphics[width=0.85\textwidth]{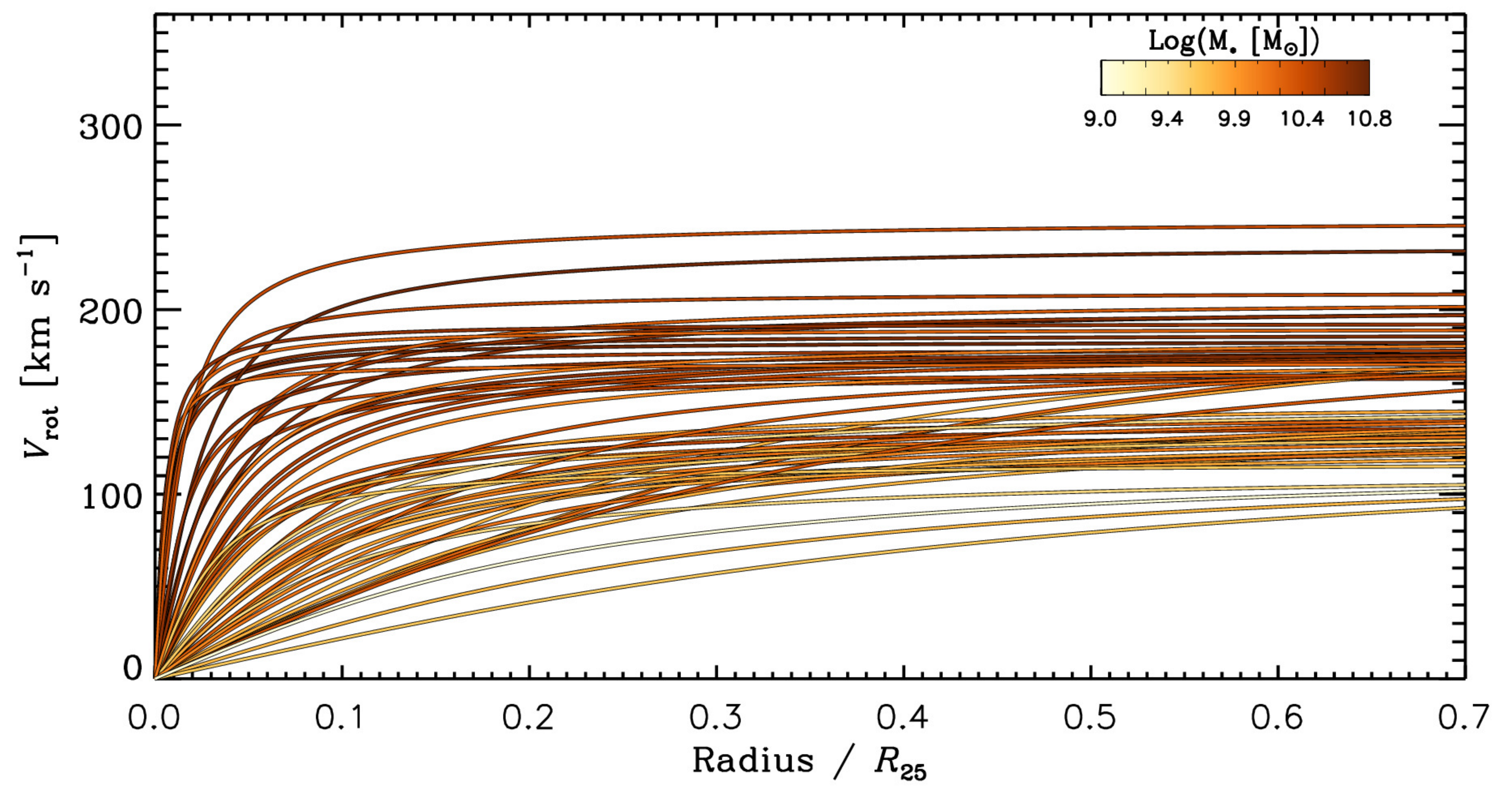}
\caption{Smooth model fits to the rotation curves, with radii normalized to $R_{{\rm 25}}$. The color-coding indicates total stellar mass.  Shown is the subset of 50 galaxies with robust analytic fits, and where the central 2\,kpc have not been masked.}
 \label{Models.fig}
\end{figure*}

Figure~\ref{Models.fig} shows the smooth models for the subset of 50 galaxies that have analytic fits, without masking the central $R<2$\,kpc.  The figure exhibits a similar sorting by stellar mass present in Figure~\ref{RCs.fig}, but now this variation is dominated by the behavior of the rotation curve across the disk-dominated zone.  Thus we see that the mass-dependence in the observed rotation curves is only partly driven by the growing prominence of central dynamical structures with increasing galaxy mass.

\label{RC_shape.sec}

It is worth emphasizing that, for modelling the mass distributions of well-resolved galaxies (and the PHANGS-ALMA sample in particular), the observed rotation curve should be preferred over the smooth model.  In this paper, smooth fits most reliably capture only the outer, asymptotic behavior and are therefore meant to serve as a guide for modelling the behavior of the rotation curve at and beyond the edge of the CO field-of-view.

\begin{table*}[tb]
\centering
\caption{Parameters for smooth analytic fits to our rotation curves (see Equation~\ref{arctan.eq}).}
\begin{tabular}{l*{6}{c}r}
\hline
\hline
ID & $V_0$ & $r_{{\rm t}}$ & ID & $V_0$ & $r_{{\rm t}}$ \\
 $\,$ & $[\mathrm{km~s^{-1}}]$  & [kpc] & $\,$  $\,$ & $[\mathrm{km~s^{-1}}]$  & [kpc] \\
\hline
 ${\rm IC~1954\;\:  }$  &  $137.0^{+  2.9}_{-    1.1}$   &      $    0.69^{+    0.07}_{-    0.04}$ & ${\rm NGC~3626\;\: }$  &  $248.8^{+  33.6}_{-   22.3}$   &      $  0.13^{+  0.16}_{-  0.06}$      \\               
${\rm IC~5273\;\:  }$  &  $150.2^{+  8.5}_{-    5.7}$   &      $    1.30^{+    0.12}_{-    0.11}$ & ${\rm NGC~3627\;\: }$  &  $202.1^{+   6.1}_{-    7.3}$   &      $  0.45^{+  0.10}_{-  0.09}$      \\               
${\rm NGC~0628\;\: }$  &  $144.8^{+108.4}_{-   37.3}$   &      $    0.56^{+    0.05}_{-    0.03}$ & ${\rm NGC~4207\;\: }$  &  $125.0^{+  10.1}_{-    7.4}$   &      $  0.93^{+  0.11}_{-  0.09}$      \\             
${\rm NGC~0685\;\: }$  &  $216.8^{+119.7}_{-   33.2}$   &      $    1.73^{+    0.15}_{-    0.11}$ & ${\rm NGC~4254\;\: }$  &  $183.2^{+   6.5}_{-    4.1}$   &      $  0.59^{+  0.07}_{-  0.04}$      \\                 
${\rm NGC~1087\;\: }$  &  $152.6^{+ 11.5}_{-    7.7}$   &      $    0.35^{+    0.04}_{-    0.02}$ & ${\rm NGC~4293\;\: }$  &  $141.1^{+  15.2}_{-   10.2}$   &      $  0.62^{+  0.17}_{-  0.10}$      \\        
${\rm NGC~1097\;\: }$  &  $328.4^{+ 32.4}_{-   26.0}$   &      $    2.23^{+    0.43}_{-    0.38}$ & ${\rm NGC~4298\;\: }$  &  $138.9^{+   2.7}_{-    1.9}$   &      $  0.81^{+  0.09}_{-  0.06}$      \\         
${\rm NGC~1300\;\: }$  &  $183.3^{+ 45.2}_{-   21.6}$   &      $    0.16^{+    0.14}_{-    0.06}$ & ${\rm NGC~4303\;\: }$  &  $178.2^{+  75.1}_{-   43.0}$   &      $  0.28^{+  0.14}_{-  0.07}$      \\               
${\rm NGC~1317 \;\:}$  &  $165.8^{+ 72.9}_{-   34.2}$   &      $    0.18^{+    0.02}_{-    0.03}$ & ${\rm NGC~4321\;\: }$  &  $181.0^{+   9.0}_{-   10.5}$   &      $  0.45^{+  0.10}_{-  0.05}$      \\               
${\rm NGC~1365\;\: }$  &  $186.7^{+ 26.8}_{-   18.1}$   &      $    0.25^{+    0.20}_{-    0.13}$ & ${\rm NGC~4457\;\: }$  &  --                            &             --                        \\               
${\rm NGC~1385 \;\:}$  &  $136.6^{+ 17.8}_{-    9.2}$   &      $    0.50^{+    0.10}_{-    0.07}$ & ${\rm NGC~4496a    }$  &  $136.9^{+  13.3}_{-    7.0}$   &      $  2.85^{+  0.23}_{-  0.23}$      \\               
${\rm NGC~1433\;\: }$  &  $204.5^{+ 43.3}_{-   27.9}$   &      $    0.59^{+    0.02}_{-    0.02}$ & ${\rm NGC~4535\;\: }$  &  $229.6^{+  44.4}_{-   34.9}$   &      $  2.67^{+  0.20}_{-  0.41}$      \\               
${\rm NGC~1511\;\: }$  &  $228.7^{+ 17.6}_{-   14.6}$   &      $    2.63^{+    0.37}_{-    0.28}$ & ${\rm NGC~4536\;\: }$  &  $170.9^{+   3.9}_{-    2.4}$   &      $  0.06^{+  0.06}_{-  0.03}$      \\               
${\rm NGC~1512\;\: }$  &  $178.9^{+ 20.1}_{-   15.1}$   &      $    0.15^{+    0.04}_{-    0.03}$ & ${\rm NGC~4540\;\: }$  &  $164.3^{+ 108.7}_{-   42.3}$   &      $  1.09^{+  0.32}_{-  0.20}$      \\               
${\rm NGC~1546\;\: }$  &  $180.0^{+  2.5}_{-    1.3}$   &      $    0.36^{+    0.06}_{-    0.03}$ & ${\rm NGC~4548\;\: }$  &  $192.9^{+  23.1}_{-   14.8}$   &      $  0.07^{+  0.03}_{-  0.01}$      \\               
${\rm NGC~1559\;\: }$  &  $237.0^{+ 37.0}_{-   22.0}$   &      $    4.01^{+    1.26}_{-    0.67}$ & ${\rm NGC~4569\;\: }$  &  $273.5^{+  20.6}_{-   23.4}$   &      $  2.89^{+  0.50}_{-  0.46}$      \\               
${\rm NGC~1566\;\: }$  &  $220.6^{+105.9}_{-   54.4}$   &      $    0.00^{+    0.16}_{-    0.00}$ & ${\rm NGC~4571\;\: }$  &  $137.1^{+   7.8}_{-    5.2}$   &      $  0.63^{+  0.04}_{-  0.02}$      \\               
${\rm NGC~1672\;\: }$  &  $160.4^{+ 45.3}_{-   26.5}$   &      $    1.15^{+    0.32}_{-    0.27}$ & ${\rm NGC~4579\;\: }$  &  $314.0^{+  43.3}_{-   28.8}$   &      $  1.50^{+  0.09}_{-  0.10}$      \\               
${\rm NGC~1792\;\: }$  &  $179.4^{+  2.5}_{-    1.3}$   &      $    0.92^{+    0.10}_{-    0.06}$ & ${\rm NGC~4654\;\: }$  &  $189.9^{+   8.5}_{-    6.5}$   &      $  1.67^{+  0.05}_{-  0.03}$      \\               
${\rm NGC~1809\;\: }$  &  --                            &                 --                      & ${\rm NGC~4689\;\: }$  &  $150.6^{+   9.7}_{-    6.2}$   &      $  0.83^{+  0.06}_{-  0.03}$      \\                
${\rm NGC~2090\;\: }$  &  $176.9^{+  3.4}_{-    2.9}$   &      $    0.42^{+    0.06}_{-    0.04}$ & ${\rm NGC~4781\;\: }$  &  $127.5^{+   3.6}_{-    2.8}$   &      $  0.65^{+  0.05}_{-  0.04}$      \\               
${\rm NGC~2283\;\: }$  &  $114.2^{+  6.6}_{-    3.8}$   &      $    0.37^{+    0.02}_{-    0.01}$ & ${\rm NGC~4826\;\: }$  &  $189.9^{+   1.7}_{-    1.3}$   &      $  0.04^{+  0.01}_{-  0.01}$      \\               
${\rm NGC~2566\;\: }$  &  $166.7^{+ 28.4}_{-   43.2}$   &      $    0.21^{+    0.31}_{-    0.21}$ & ${\rm NGC~4941\;\: }$  &  $232.0^{+   2.7}_{-    1.7}$   &      $  1.62^{+  0.04}_{-  0.04}$      \\               
${\rm NGC~2775\;\: }$  &  --                            &                 --                      & ${\rm NGC~4951\;\: }$  &  $137.8^{+   7.9}_{-    4.2}$   &      $  0.38^{+  0.09}_{-  0.07}$      \\              
${\rm NGC~2835\;\: }$  &  --                            &                 --                      & ${\rm NGC~5042\;\: }$  &  $118.1^{+  45.6}_{-    6.8}$   &      $  0.21^{+  0.04}_{-  0.03}$      \\              
${\rm NGC~2903\;\: }$  &  $309.8^{+ 12.7}_{-    6.3}$   &      $    2.49^{+    0.23}_{-    0.14}$ & ${\rm NGC~5068\;\: }$  &  --                             &               --                       \\               
${\rm NGC~2997\;\: }$  &  $210.2^{+ 52.2}_{-   23.0}$   &      $    0.17^{+    0.04}_{-    0.02}$ & ${\rm NGC~5134\;\: }$  &  $210.1^{+  72.7}_{-   38.4}$   &      $  2.17^{+  0.55}_{-  0.42}$      \\               
${\rm NGC~3059\;\: }$  &  $139.3^{+ 54.2}_{-   21.9}$   &      $    0.72^{+    0.13}_{-    0.08}$ & ${\rm NGC~5248\;\: }$  &  $196.5^{+  67.9}_{-   26.4}$   &      $  1.02^{+  0.34}_{-  0.36}$      \\               
${\rm NGC~3137\;\: }$  &  $130.7^{+  2.7}_{-    2.3}$   &      $    0.94^{+    0.08}_{-    0.05}$ & ${\rm NGC~5530\;\: }$  &  $153.7^{+   1.6}_{-    1.4}$   &      $  1.11^{+  0.03}_{-  0.02}$      \\             
${\rm NGC~3351\;\: }$  &  $206.8^{+ 16.9}_{-   13.9}$   &      $    0.30^{+    0.08}_{-    0.03}$ & ${\rm NGC~5643 \;\:}$  &  $186.2^{+  40.8}_{-   23.2}$   &      $  0.32^{+  0.04}_{-  0.03}$      \\               
${\rm NGC~3507\;\: }$  &  $180.9^{+108.6}_{-   61.0}$   &      $    0.44^{+    0.08}_{-    0.02}$ & ${\rm NGC~6300 \;\:}$  &  $199.3^{+  24.8}_{-   13.3}$   &      $  0.66^{+  0.20}_{-  0.12}$      \\               
${\rm NGC~3511\;\: }$  &  $133.2^{+  4.7}_{-    2.2}$   &      $    0.33^{+    0.07}_{-    0.04}$ & ${\rm NGC~6744 \;\:}$  &   --                            &                                 --     \\               
${\rm NGC~3521\;\: }$  &  $236.8^{+  0.7}_{-    1.0}$   &      $    0.32^{+    0.02}_{-    0.02}$ & ${\rm NGC~7456 \;\:}$  &  $121.0^{+   2.3}_{-    2.0}$   &      $  0.84^{+  0.06}_{-  0.04}$      \\               
${\rm NGC~3596\;\: }$  &  $151.9^{+ 99.6}_{-   34.9}$   &      $    0.38^{+    0.06}_{-    0.03}$ & ${\rm NGC~7496 \;\:}$  &  $147.7^{+   4.8}_{-    5.6}$   &      $  1.92^{+  0.12}_{-  0.12}$      \\ 
\hline
\hline
\end{tabular}
\label{smooth_parameters.tbl}
\end{table*}

\section{Discussion/implications}
\label{Sect5.sec}
\subsection{Galaxy shapes}
\label{Shapes.sec}
Differences in the kinematic and photometric orientations of galaxies can have implications for the intrinsic shapes of their underlying stellar disks.  More specifically,  our adopted photometric position angles \pap\ are measured from the projected shape of the outer old stellar disk, whereas the kinematic position angle \pak\ traces the orientation of the inner disk populated by molecular gas.  Small differences could therefore suggest modest twisting of the disk orientation between inner and outer radii.  Differences between \pak\ and \pap\ may also signify a genuine deviation from disk circularity, as quantified by \cite{Franx1994} and \cite{Schoenmakers1997}.  Such deviations from perfect disk circularity will have the most notable impact on position angles measured for galaxies at low inclination (although they may still be present to an equal degree in other systems; see discussion below).  We note that we find no obvious systematic trend when comparing the offset $\pak{-}\pap$ against other global galaxy properties (i.e. stellar mass, specific star formation rate, $\rm R_{25}$, B/T, bar presence, bar length, size of bulge and disk, fitted rotation curve transition radius $\rm r_t$).

Offsets in photometric and kinematic inclinations may
likewise indicate a genuine change in underlying disk structure, in particular to the variation of the inclination
between the inner disk (traced by CO kinematics) and the outer disk (traced at 3.6\,$\mu$m), i.e. warping.  Such warping might arise as part of the same gravitational response that is thought to generate the more readily recognizable outer HI warps of galaxies (\citealt{Sellwood2010}; see also \citealt{Rogstad1974, briggs}).  We find no evidence of residual correlation between $\vert\incp{-}\inck\vert$ and many other global galaxy properties (see list above).

Interestingly, the lack of a strong correlation with stellar mass suggests that large differences $\vert\incp{-}\inck\vert$ are not related to systematic variations in disk flatness, which is another source of discrepancy between \incp\ and \inck.
This may offer an interesting contrast to the increase in light-weighted flatness with increasing mass found by \cite{Yoachim2006} across a broader mass range than studied here, perhaps signifying the growing prominence of a thin disk component with increasing mass.

Any variations from the constant flatness assumed in Equation \ref{photinc.eq} (in the range $0.1{-}0.4$ over the mass range analyzed in this paper; \citealt{vdW2014}) appear more likely a source of scatter between \incp\ and \inck~(i.e. as plotted in Figure \ref{Kin_phot.fig}); any individual galaxy may still reflect departures of the true thickness from the specific value we have adopted.  Scatter in inclinations may also arise as a result of variations in disk circularity.  We emphasize, though, that variations in thickness and/or circularity may be likely to introduce systematic discrepancies, such as highlighted at low inclination in Section \ref{comp_phot.sec}.  This is due to the fact that only modest variations in these shape parameters can cause large discrepancies for face-on systems.  We estimate that, with our adopted flatness $q=0.25$, when $i<25$ degrees the inferred inclination will increase by $20$ degrees with an increase in axis ratio by as little as $0.1$, whereas for $i>45$ degrees, this change to the inclination corresponds to a much larger change ($\gtrsim 0.2$) in the axis ratio.

\subsection{Inner rotation curve shapes at high resolution as tracers of the central mass distributions of galaxies}\label{centralrotcurves.sec}
\label{inner_slope.sec}

\begin {figure}[t]
\centering
 \includegraphics[width=0.45\textwidth]{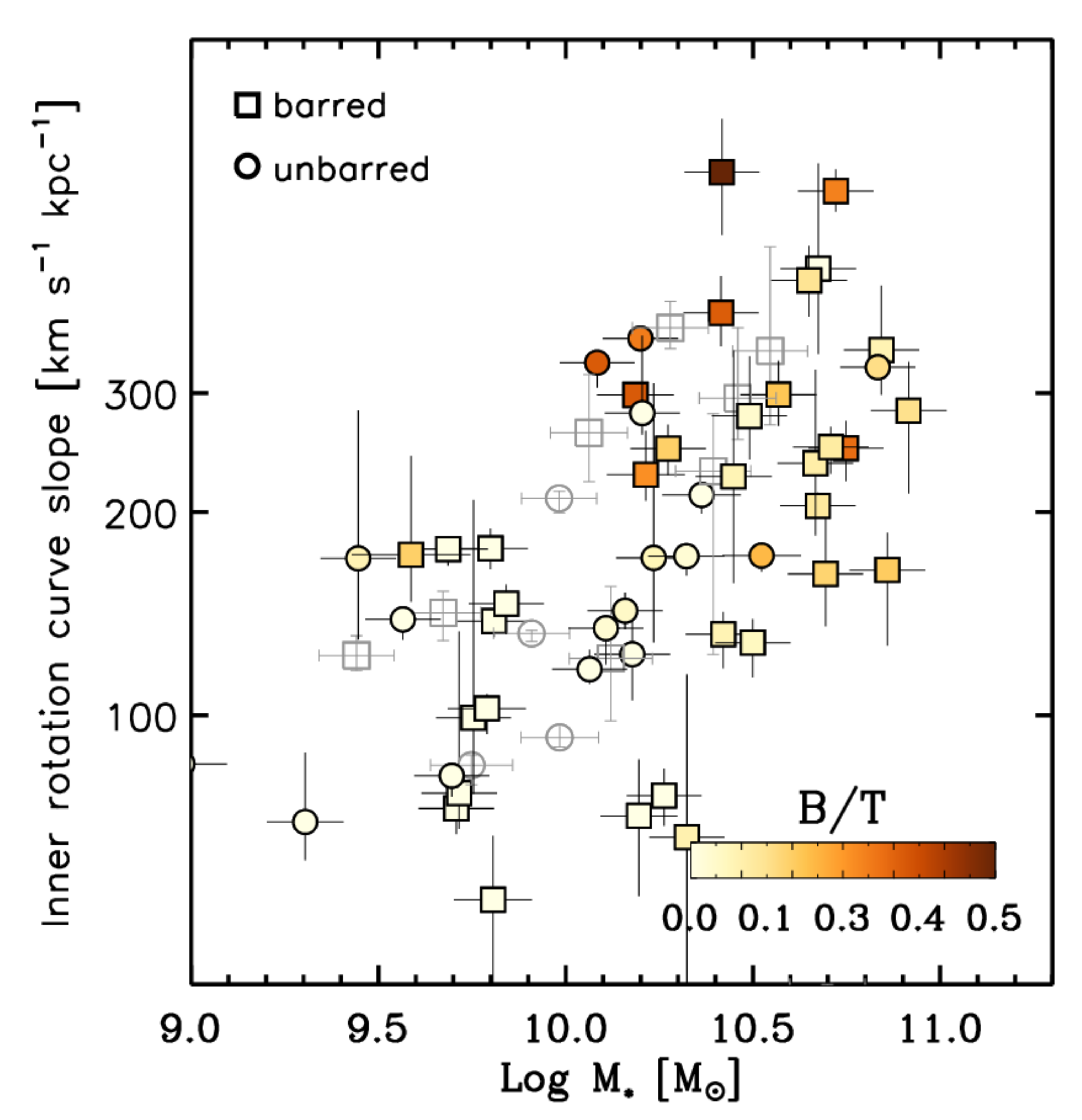}
\caption{Inner ($\le 500$\,pc) rotation curve slope plotted against total stellar mass.  Squares and circles represent barred and unbarred galaxies, respectively. The color-coding encodes the B/T ratio taken from \citet{Salo2015}.  Gray symbols represent systems with no B/T measurement. Galaxies with unconstrained inner rotation curves (NGC~2775, NGC~6744) are omitted. }
 \label{Slopes.fig}
\end {figure}

In Section~\ref{RCs.sec}, we identified a variety of inner rotation curve shapes throughout the PHANGS-ALMA sample.  Here, we consider whether the central rotation curve is shaped by a genuine central mass concentration or by a lingering contribution from non-circular motions in bars.  We do so by quantifying the inner rotation curve gradient of all galaxies and examining whether these gradients are linked with independent measurements of the central mass concentration.  Figure~\ref{Slopes.fig} plots the linear slope measured between $R=0$ and $R=500$\,pc of our rotation curves against galaxy stellar mass.  Shown are all galaxies except for NGC~2775 and NGC~6744, whose rotation curves do not cover the central regions.

Overall, we find a good correlation between inner rotation curve slope and total stellar mass (with a Spearman's rank correlation coefficient of $0.63$).  This is consistent with the well-known dependency of inner galaxy structure (i.e., inner bar and bulge components) and total stellar mass \citep[][]{kormkenn2004,Bluck2014}, and suggests that our high-resolution curves are good probes of the inner mass distributions of galaxies.  To further disentangle the correlation of stellar mass and inner structure, each galaxy is shown color-coded by the stellar mass ratio of the bulge component with respect to the total galaxy, B/T, measured by \citealt{Salo2015} within ${\rm S^4G}$.  (We note that the bulge component measured by \citealt{Salo2015} reflects a genuine central mass concentration and not a bar component, which is fitted independently in their analysis.)  To highlight the potential influence of the non-circular orbital structure of bars on our measured inner rotation curves, we furthermore split the sample into barred and unbarred systems.

Together with the expected increase in B/T with stellar mass \citep[e.g.,][]{Bluck2014}, Figure~\ref{Slopes.fig} also clearly shows an increase in rotation curve slope with B/T at fixed stellar mass: Galaxies with the most pronounced bulges exhibit the steepest inner rotation curves. Correlating B/T with our rotation curve slope estimates yields a Spearman's rank correlation coefficient of 0.60. The correlation of rotation curve slope and stellar mass does not seem to be strongly affected by bar components as barred and unbarred galaxies fall into the same locus of parameter space in Figure~\ref{Slopes.fig}.

\begin {figure}[t]
\centering
 \includegraphics[width=0.45\textwidth]{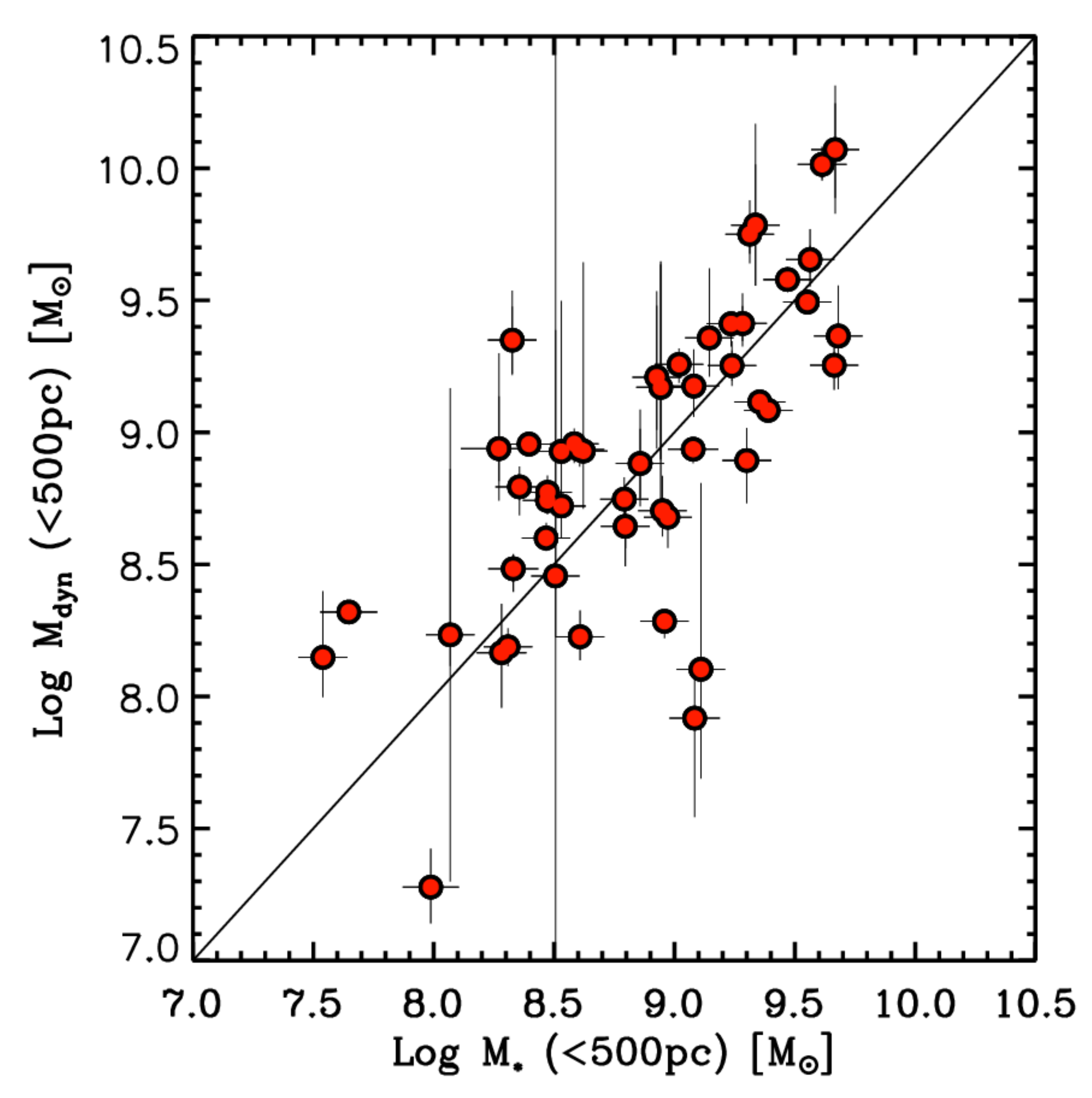}
\caption{Inner dynamical mass versus stellar mass inside radii $\le 500$\,pc. The solid line marks a one-to-one relation. Shown are 51 galaxies that are covered by ${\rm S^4G}$, excluding galaxies with unconstrained inner rotation curves (NGC~2775, NGC~6744).}
 \label{Mdyn.fig}
\end {figure}

To further examine the correspondence between the inner structure in our high-resolution CO rotation curves and the central mass distributions, we show the inferred inner ($\le 500$\,pc) dynamical mass versus the stellar mass within the same radius in Figure~\ref{Mdyn.fig}.  The dynamical mass is calculated according to $M_{\rm dyn} = {\rm G}^{-1}\,V^{2}_{{\rm rot}}\,R$, evaluated at $R=500$\,pc, assuming a spherical geometry.  We expect this mass to be dominated by the stellar component at these radii.  To obtain an independent measure of this enclosed stellar mass, we determine the fraction of the total 3.6 $\mu$m emission arising from within the central 500\,pc using the ${\rm S^4G}$ Sersic models fitted by \cite{Salo2015}, and then multiply this fraction by our fiducial total stellar mass estimates (Section~\ref{Sample_properties.sec}).  Thus, only the ${\rm S^4G}$ subset of our sample are shown in Figure~\ref{Mdyn.fig}.

Overall the two mass estimates track each other fairly well, although there is a significant degree of scatter, most of which can be credited to various sources of systematic uncertainty that are not accounted for in making this coarse comparison. This includes distance uncertainties, for example, and the adopted 3.6\,$\mu$m mass-to-light ratio, which we have assumed is global (thus neglecting local changes in the properties of the stellar population and the amount of contamination by dust emission; e.g., see \citealt[][]{Querejeta2015,Leroy2019}).  A decrease in central M/L as a result of enhanced dust emission in the 3.6\,$\mu$m band would lower $M_\star (<500\,{\rm pc})$ in many of the cases with unrealistic $M_\star > M_{\rm dyn}$, for example.  Instances where $M_{\rm dyn} (<500\,{\rm pc})$ exceeds $M_\star (<500\,{\rm pc})$, on the other hand, likely indicate either that the central enclosed mass distribution is flatter than assumed, or that the mass in gas is non-negligible. 

Even with the coarseness of the approximations adopted in Figure~\ref{Mdyn.fig}, we can recognize that the two mass estimates are reassuringly similar.  We find a median ratio $\langle{\rm log}\,M_{\rm dyn}/{\rm M_{\sun}} - {\rm log}\,M_\star/{\rm M_{\sun}}\rangle = 0.1^{+0.3}_{-0.4}$, where the quoted errors quantify the scatter based on the 16th and 84th percentile. The correlation coefficient of the relation is $0.73$.

\subsection{Rotational velocities as tracers of global mass: the Tully-Fisher relation}
\label{Tully_Fisher.sec}

We now examine the degree to which our high resolution CO kinematics trace global dynamical mass, using the stellar mass Tully-Fisher relation (Tully \& Fisher 1977) as our point of reference.  

\begin {figure}[tb]
\centering
 \includegraphics[width=0.48\textwidth]{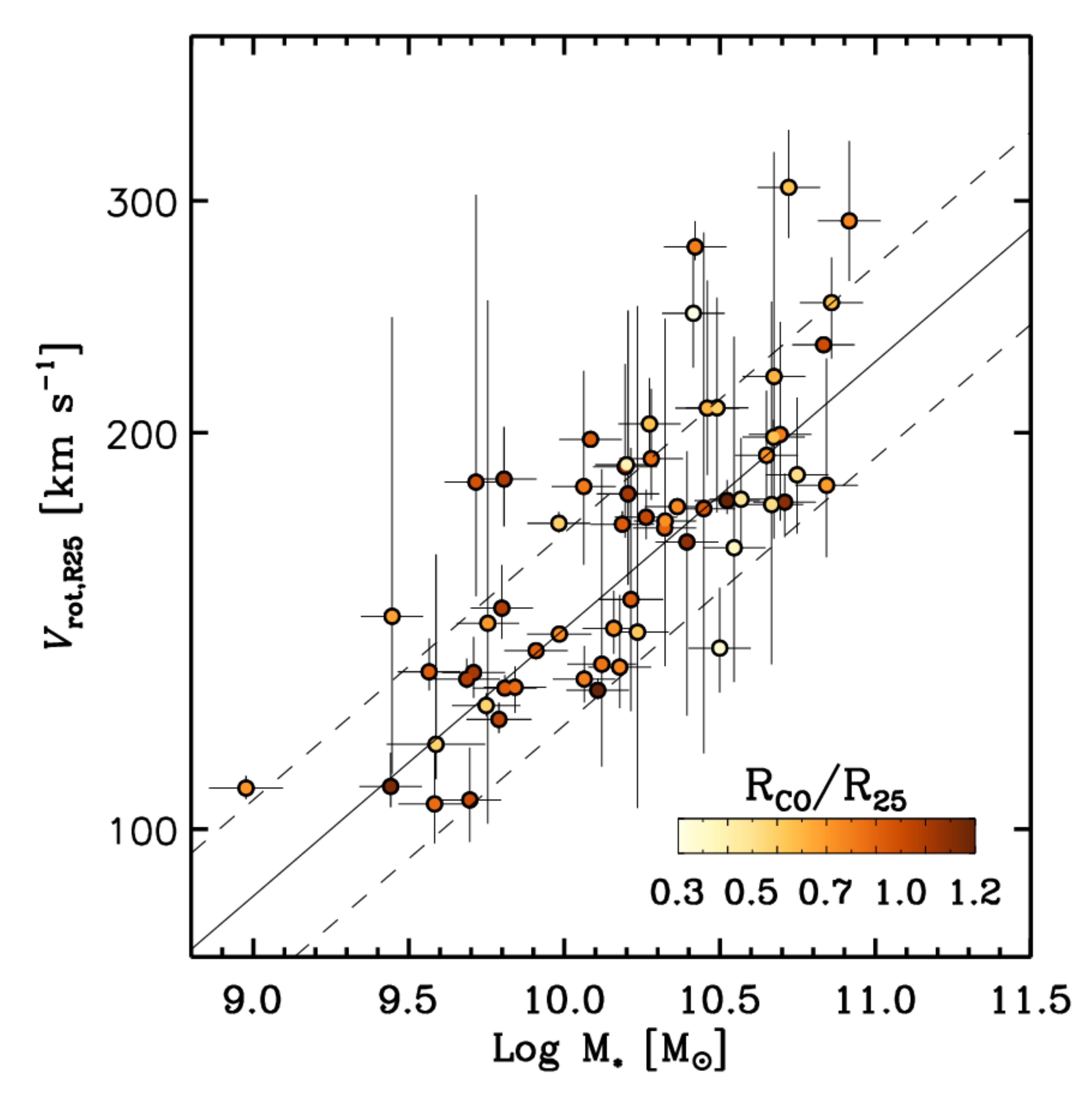}
\caption{Tully-Fisher relation, based on total stellar masses and outer rotation velocities ($V_{{\rm rot,25}}$) of our galaxy sample. Rotation velocities are derived from our smooth analytic fits to rotation curves.  The color-coding encodes the radial coverage of CO relative to the size of the optical disk. The solid and dashed lines show the stellar Tully-Fisher relation and its scatter from \cite{McGaugh2015}, respectively. Only systems with robust kinematic fits are shown (52 galaxies).}
 \label{Tully_Fisher.fig}
\end {figure}

The Tully-Fisher relation implies a tight fundamental relation between rotation velocity and total stellar mass.  Figure~\ref{Tully_Fisher.fig} shows the Tully-Fisher relation for our sample using the outer rotational velocities $V_{{\rm rot,25}}$ estimated by extrapolating our CO rotation curves out to $R_{25}$ using our fitted smooth arctangent models.  The stellar Tully-Fisher relation recently measured in massive spiral galaxies in the local universe by \citet[][]{McGaugh2015} is also shown.
Note that galaxies without robust smooth fits are omitted. Also, the $V_{{\rm rot,25}}$ in this plot are derived from rotation curves that assume a mix of kinematic and photometric inclinations.  Thus the plotted values do not include outer $V_{{\rm rot}}$ measurements from the problematic fits (described in Section~\ref{Overview.sec}) that yield  unreliable \inck\ values and more than 100 $\mathrm{km~s^{-1}}$ difference between the outer $V_{{\rm rot}}$ and the Tully-Fisher value; instead, the plot shows the $V_{{\rm rot,25}}$ fitted to the rotation curve at the adopted \incp.   

The correlation recovered in Figure~\ref{Tully_Fisher.fig}, albeit with substantial scatter, suggests that the inner kinematics probed in our PHANGS-ALMA CO maps provide reasonably good constraints on the asymptotic behavior of rotation curves.  The scatter in our reconstructions of $V_{{\rm rot,25}}$ at fixed stellar mass is $\sim 0.07$ dex. 

The measurements in Figure~\ref{Tully_Fisher.fig} largely fall within the scatter around the stellar Tully-Fisher relation measured by \citet[][]{McGaugh2015}.  We refer the reader to their discussion of the uncertainties responsible for this scatter, as they are expected to contribute similarly here.  
An additional potential source of scatter is related to the extent that CO kinematics from the inner disk can be extrapolated to large radii. The color-coding of the data points in Figure~\ref{Tully_Fisher.fig} shows the fraction of the optical extent of the disk covered by CO emission. We find no significant relation between the scatter about the Tully-Fisher relation and the extent of the mapped area for the galaxies in our sample.  This emphasizes that the kinematics contained in CO maps with even limited radial coverage can provide constraints on the kinematics beyond the mapped edge using our smooth two-parameter model.  

\section{Summary and Conclusions}
\label{Sect6.sec}
In this work, we have introduced a novel method to constrain disk orientations and rotation curves for 67 galaxies with sensitive, high-resolution CO(2-1) observations from the PHANGS-ALMA survey of local star-forming galaxies.  This has allowed us to  systematically constrain the detailed shapes of rotation curves at $150$\,pc resolution across a wide range of galaxy properties.  
Our main results are the following: 

\begin{itemize}

\item Our methodology to constrain disk position angles and inclination angles utilizes all sampled lines-of-sight in a given velocity field, making it well-suited for high spatial resolution CO maps that can have low area filling factor.  

\item Our approach affords limited sensitivity to non-axisymmetric dynamical structures, which are common features in the central regions of galactic disks probed by CO emission.  In general, we find good agreement (within 3$\sigma$) between our kinematics-based galaxy orientations and those inferred from outer-disk photometry, implying that the impact of stellar bars and spirals on our derived orientations is very limited, except in the most prominently barred cases. For face-on disks, where the photometric position angle can be ill-constrained, our kinematics position angles yield a visibly better description of the CO kinematics. 

\item We derive galaxy rotation curves across the PHANGS-ALMA sample by fitting a basic model of pure circular motion to each observed velocity field.  These typically extend out to $\sim 0.7\,R_{\rm 25}$ and exhibit qualities that correlate with other galaxy properties.  We find that inner velocity gradient increases with both stellar mass and bulge prominence, for example.  

\item We introduce smooth analytic models of our rotation curves and use these to demonstrate that CO rotational velocities hold promise as tracers of both  inner dynamical mass and global dynamical mass, once several systematics (mass-to-light ratio variations, stellar geometry/distribution) can be accounted for. 

\item The flexibility of our technique for deriving galaxy orientations and rotation curves makes it well-suited for a variety of future applications.  This includes higher-order velocity field decompositions that can be used to constrain non-circular gas motions and inflows.  Our method can also be applied to velocity fields constructed from optical emission and/or absorption line tracers, as well as HI emission. 
\end{itemize}

The measurements presented in this paper form the basis for a wealth of future studies, including tests of the maximum disk hypothesis (and/or constraints on galaxy dark matter fractions) with detailed dynamical modelling and comparisons between global mass estimates via unresolved kinematics vs.\ highly resolved rotation curves.  

\section{Acknowledgements}
This work was carried out as part of the PHANGS collaboration.
The authors acknowledge fruitful discussions with  Arjen van der Wel and Erwin de Blok. PL, ES, CF, and DL acknowledge support from the European Research Council (ERC) under the European Union’s Horizon 2020 research and innovation programme (grant agreement No. 694343).  
ER acknowledges the support of the Natural Sciences and Engineering Research Council of Canada (NSERC), funding reference number RGPIN-2017-03987.
JMDK and MC gratefully acknowledge funding from the Deutsche Forschungsgemeinschaft (DFG) through an Emmy Noether Research Group (grant number KR4801/1-1) and the DFG Sachbeihilfe (grant number KR4801/2-1). JMDK gratefully acknowledges funding from the European Research Council (ERC) under the European Union's Horizon 2020 research and innovation programme via the ERC Starting Grant MUSTANG (grant agreement number 714907).
SCOG acknowledges support from the Deutsche Forschungsgemeinschaft via SFB 881 ``The Milky Way System'' (Project-ID 138713538; sub-projects B1, B2 and B8) and via  Germany's Excellence Strategy EXC 2181/1 - 390900948 (the Heidelberg STRUCTURES Excellence Cluster). CH, AH and JP acknowledge support from the Programme National ``Physique et Chimie du Milieu Interstellaire'' (PCMI) of CNRS/INSU with INC/INP co-funded by CEA and CNES, and from the Programme National Cosmology and Galaxies (PNCG) of CNRS/INSU with INP and IN2P3, co-funded by CEA and CNES. 
JP and FB acknowledge funding from the European Union's Horizon 2020 research and innovation programme (grant agreement No 726384). 

This paper makes use of the following ALMA data: ADS/JAO.ALMA\#2013.1.00650.S,; \#2013.1.00803.S; \#2013.1.01161.S; \#2015.1.00925; \#2015.1.00956; \#2017.1.00392.S; \#2017.1.00886.L; and \#2018.1.01651.S.  ALMA is a partnership of ESO (representing its member states), NSF (USA) and NINS (Japan), together with NRC (Canada), MOST and ASIAA (Taiwan), and KASI (Republic of Korea), in cooperation with the Republic of Chile. The Joint ALMA Observatory is operated by ESO, AUI/NRAO and NAOJ. The National Radio Astronomy Observatory is a facility of the National Science Foundation operated under cooperative agreement by Associated Universities, Inc.  
We acknowledge the usage of the Extragalactic Distance Database \citep{Tully2009}, the HyperLeda database \citep{LEDA}, and the NASA/IPAC Extragalactic Database. 

\appendix
\restartappendixnumbering

\section{Consistency between the kinematic orientations fitted in this work and independently-derived values}
\label{orientationcomp.sec}

In this section we provide a sense of the basic consistency between the orientations fitted with the approach developed here and alternative methods used in the literature.  To date, an abundance of kinematics surveys have yielded quality  measurements of position angle, inclination angle, and rotation curves using techniques similar to ours.  These offer the potential to examine the impact of differences in fitting strategy on the fitted kinematic parameters.  For now, given that the overlap between these surveys and PHANGS-ALMA targets is quite small (summarized below), the comparison we draw here mostly offers a confirmation of consistency between approaches.

In what follows we will consider only comparisons between kinematics orientations and defer a comparison of rotation curves to future work. This is as we expect measured rotation curves to be sensitive to the choice of tilted-ring fitter as well as the resolution of the kinematic tracer and the nature of its true three-dimensional organization (given variations in intrinsic velocity dispersion, vertical distribution and relation to structure in the disk plane; as highlighted recently by \citealt{Levy2018} and \citealt{Colombo2014}).

 Our PHANGS-ALMA sample has two galaxies in common with the GHASP survey modeled by \citealt{Epinat2008} (NGC~2775 and NGC~3596), two galaxies in common with SPARC (NGC~2903 and NGC~ 3521; \citealt{lelli2016}), 6 galaxies in common with the HI sample modeled by \citealt{Ponomareva2016} (NGC~ 1365, NGC~ 3351, NGC~ 3621, NGC~ 3627, NGC~ 4535 and NGC~ 4536) and five galaxies from THINGS with tilted-ring fitted orientations (but no tabulated uncertainties, see \citealt{deBlok2008}; NGC~2903, NGC~3521, NGC~3621, NGC~3627 and NGC~4826).  An additional fourteen galaxies covered by VIVA (NGC~4254, NGC~4293, NGC~4298, NGC~4321, NGC~4424, NGC~4457, NGC~4535, NGC~4536, NGC~4548, NGC~4569, NGC~4579, NGC~4654, NGC~4689 and NGC~4694; see \citealt{Chung2009}) have not yet had their HI velocity fields modeled.\footnote{For this subset, only photometric orientations are tabulated by \citealt{Chung2009}.}  
 
In the majority of these cases, the orientations determined by fitting to CO velocity fields are consistent with the published kinematic orientations within the published uncertainties, where tabulated.  Orientations for the galaxies in GHASP and SPARC (which tend to have larger uncertainties $\sigma_{\rm lit}$ than the uncertainties $\sigma$ fitted in this work) are less than 1$\sigma_{\rm lit}$ different from our orientations.  
The uncertainties on the orientations modeled by \citealt{Ponomareva2016} are more similar to our uncertainties. For this subset of galaxies, agreement is within 1.5$\sigma$ on average.

With two exceptions, our orientations are also largely consistent (within 1.5 $\sigma$) with those published by \citet{deBlok2008} using our uncertainties on both PA and inclination to measure agreement (given that the uncertainties associated with the values from \citeauthor{deBlok2008} are not published).  For two galaxies (NGC~ 3521 and NGC~ 4826) where our inclination differs by more than 2\,$\sigma$ from the THINGS value, we have found that our CO kinematical orientation agrees very well with the 3.6\,$\mu$m photometric orientation.  For one of these, NGC~ 4826, the PAs also differ substantially.  This stems from a well-known kinematic decoupling between the central 1 kpc traced by CO and the extended disk probed by HI \citep{casoli1993}.

\section{MCMC fit examples}
\label{Examples.sec}

\begin {figure*}[tb]
\centering
  \includegraphics[width=0.95\textwidth]{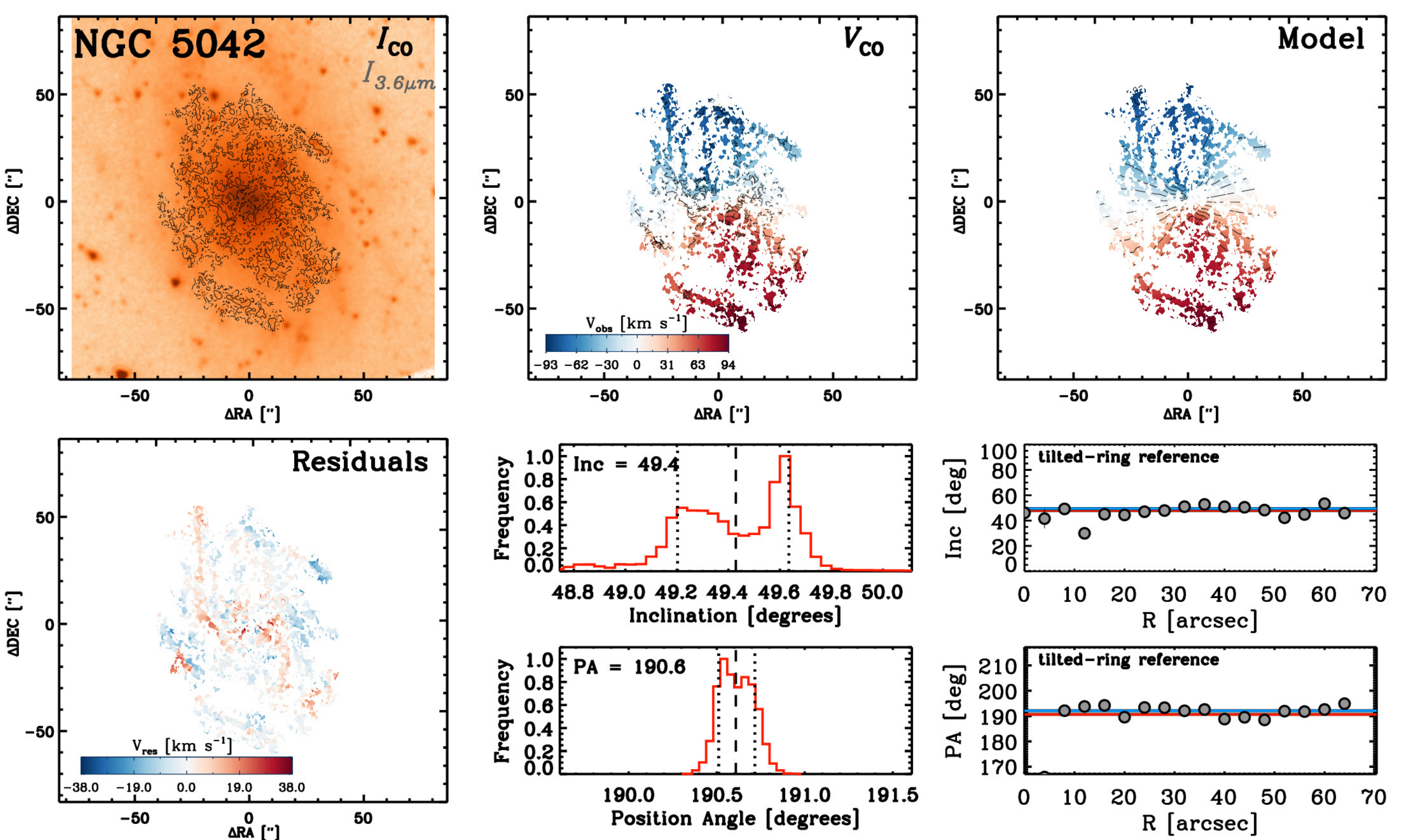}
      \vspace{10mm}
    \includegraphics[width=0.95\textwidth]{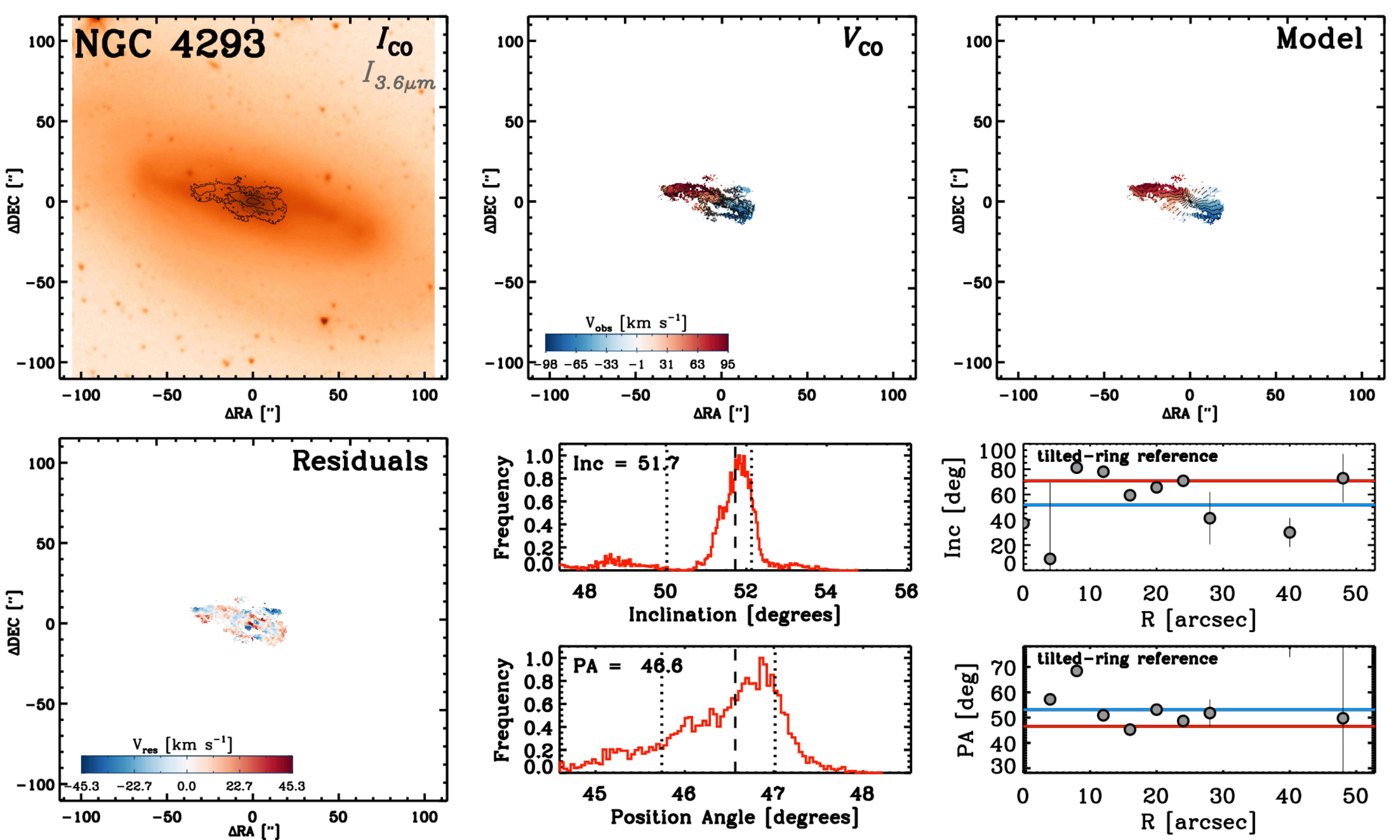}
\caption{Same as Figure\ \ref{Showcase.fig}, but showing two additional example galaxies to highlight the effect of variations in spatial sampling on our kinematic fit performance.}
\label{Showcase_sampling.fig}
\end {figure*}      

\begin {figure*}[tb]
\centering
  \includegraphics[width=0.95\textwidth]{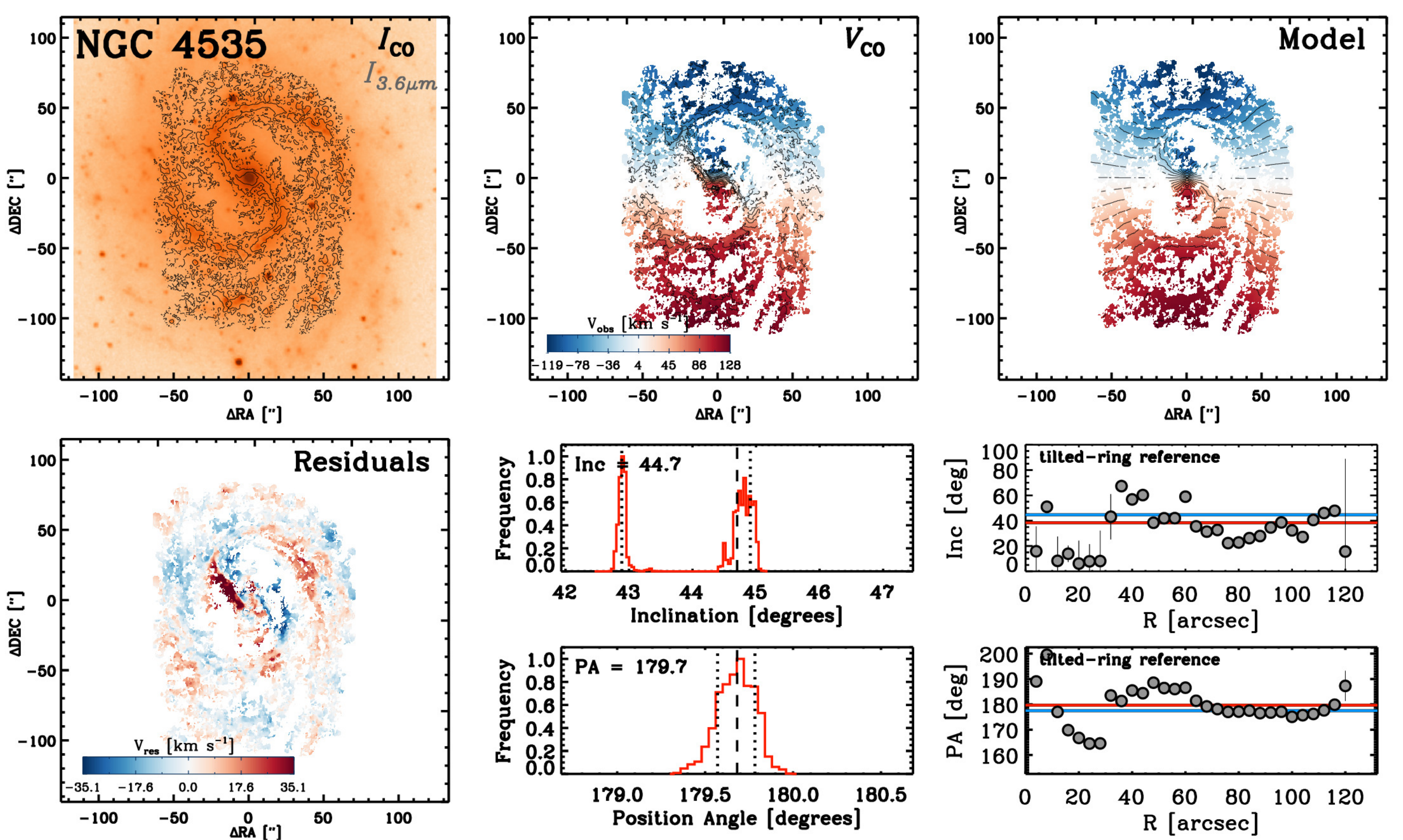}
    \vspace{10mm}
 \includegraphics[width=0.95\textwidth]{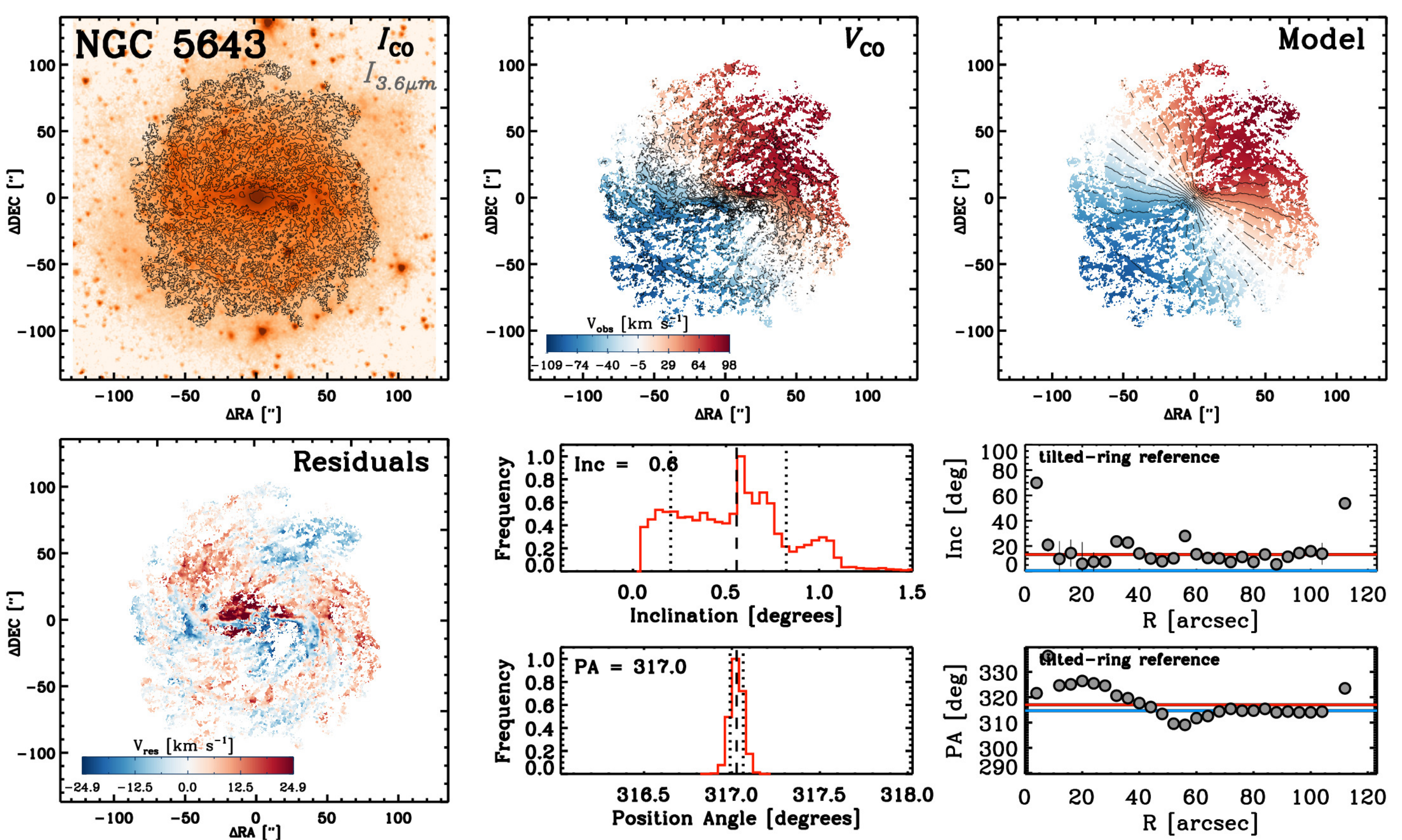}
\caption{Same as Figure\ \ref{Showcase.fig}, but showing two additional example galaxies to highlight the effect of stellar bars on our kinematic fit performance.}
\label{Showcase_bars.fig}
\end {figure*}      

\begin {figure*}[tb]
\centering
  \includegraphics[width=0.95\textwidth]{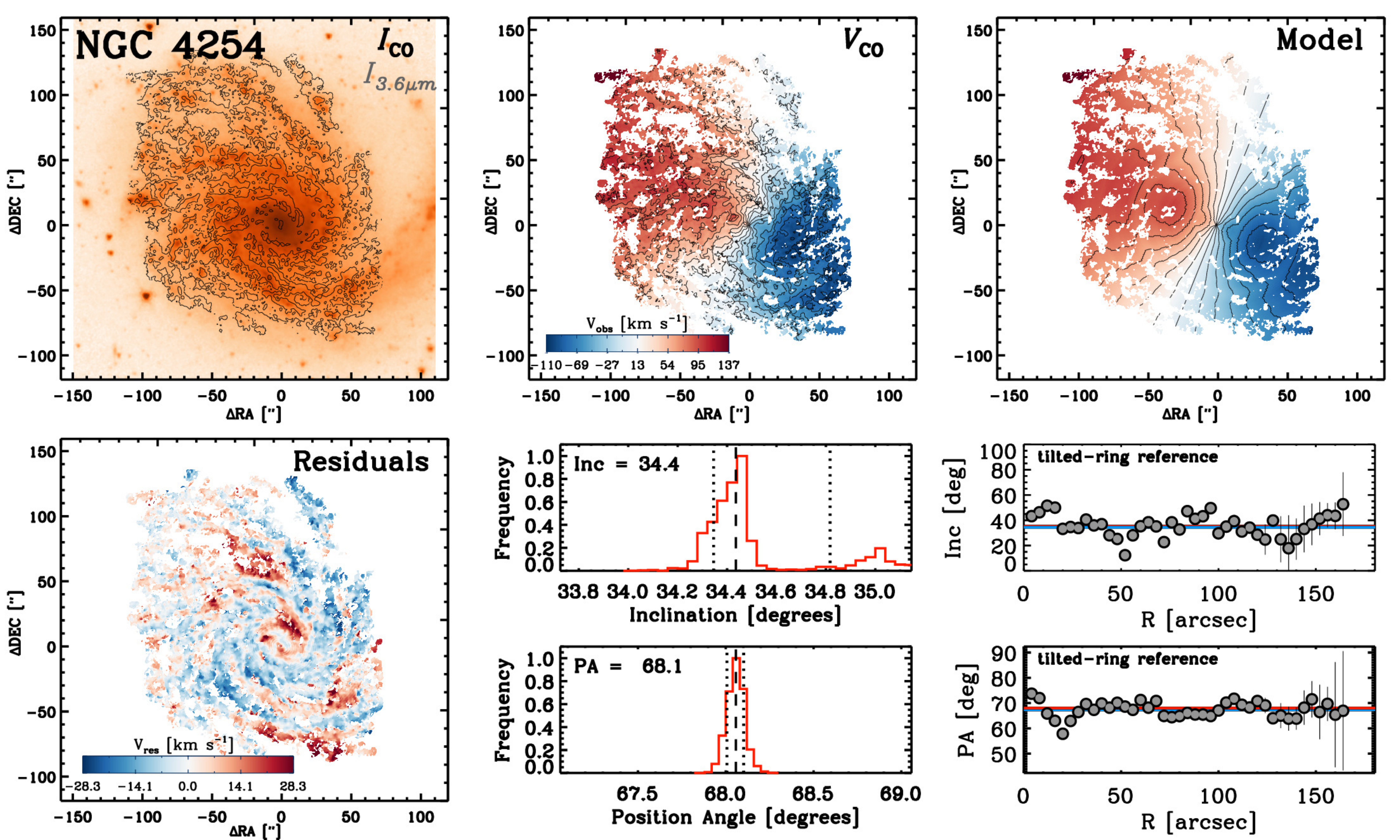}
    \vspace{10mm}
\caption{Same as Figure\ \ref{Showcase.fig}, but showing one additional example galaxy to highlight successful fit performance.}
\label{Showcase_4254.fig}
\end {figure*}     

\section{Rotation curves and smooth fits for the full sample}
\label{full_set.sec}

\begin {figure*}[tb]
\centering
  \includegraphics[width=0.80\textwidth]{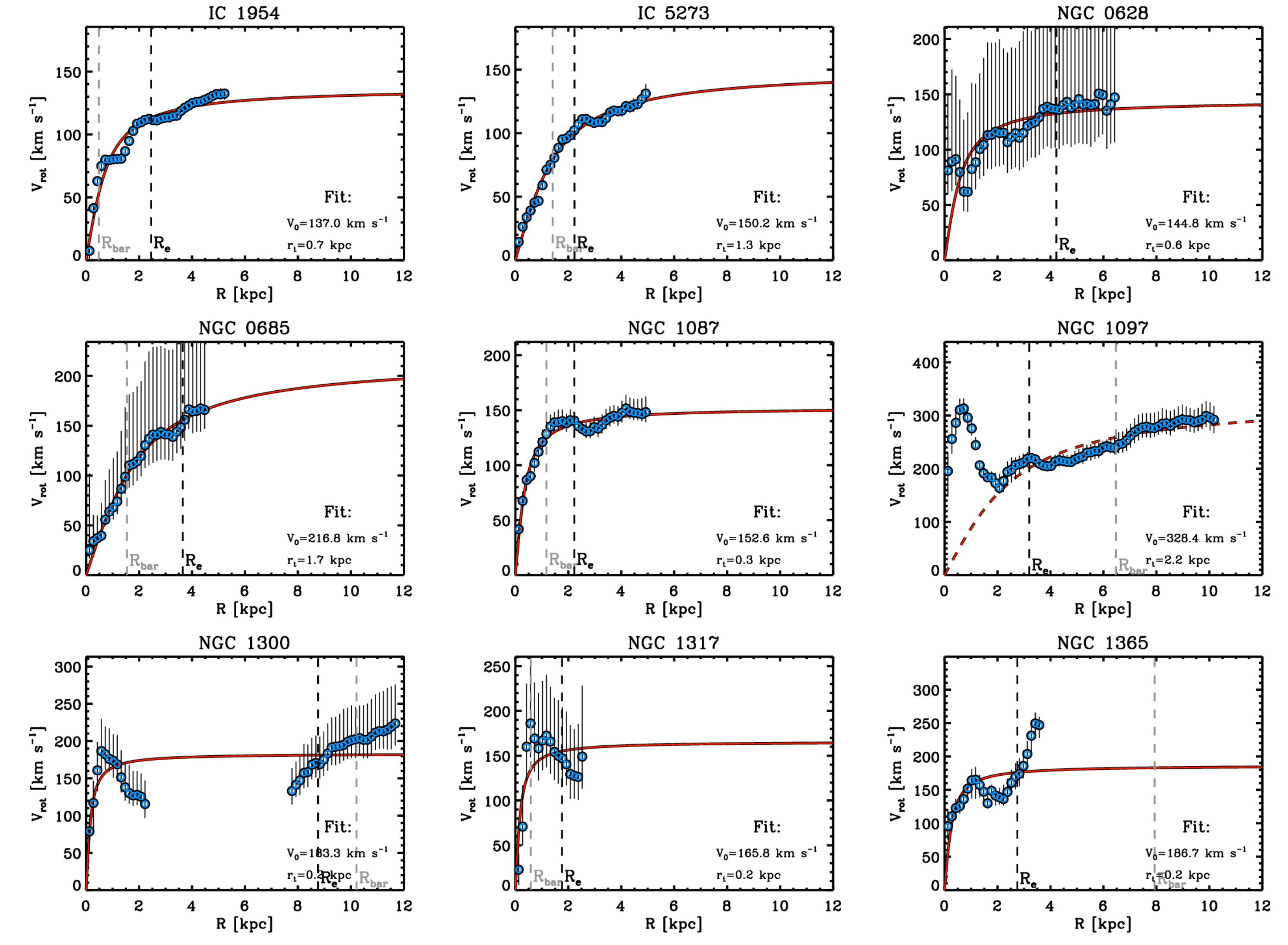}
  \includegraphics[width=0.80\textwidth]{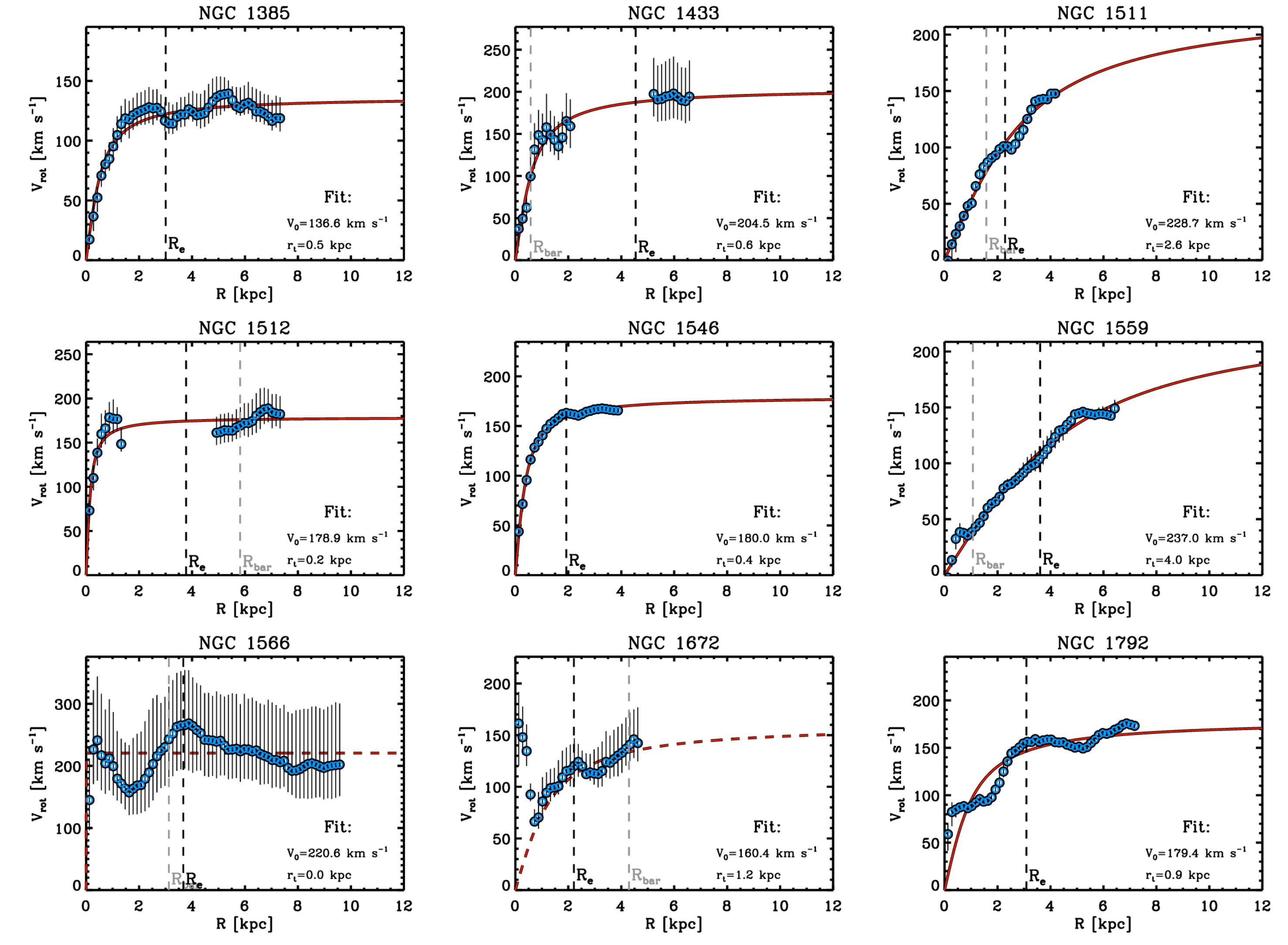}  
\caption{Same as Figure\ \ref{RC_examples.fig}, but showing the full galaxy sample}
\label{RC_all1.fig}
\end {figure*} 

\begin {figure*}[tb]
\centering
  \includegraphics[width=0.80\textwidth]{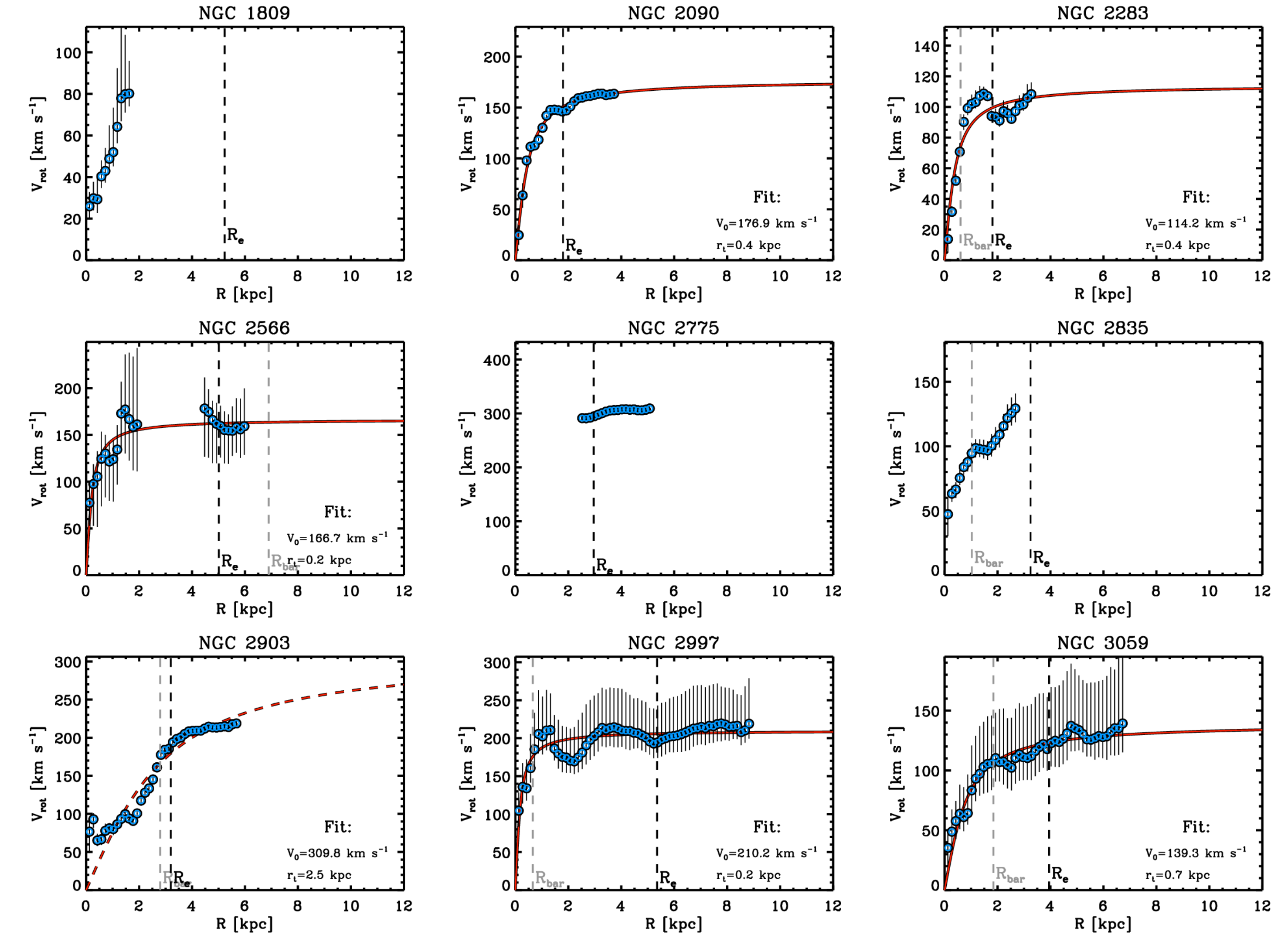}
  \includegraphics[width=0.80\textwidth]{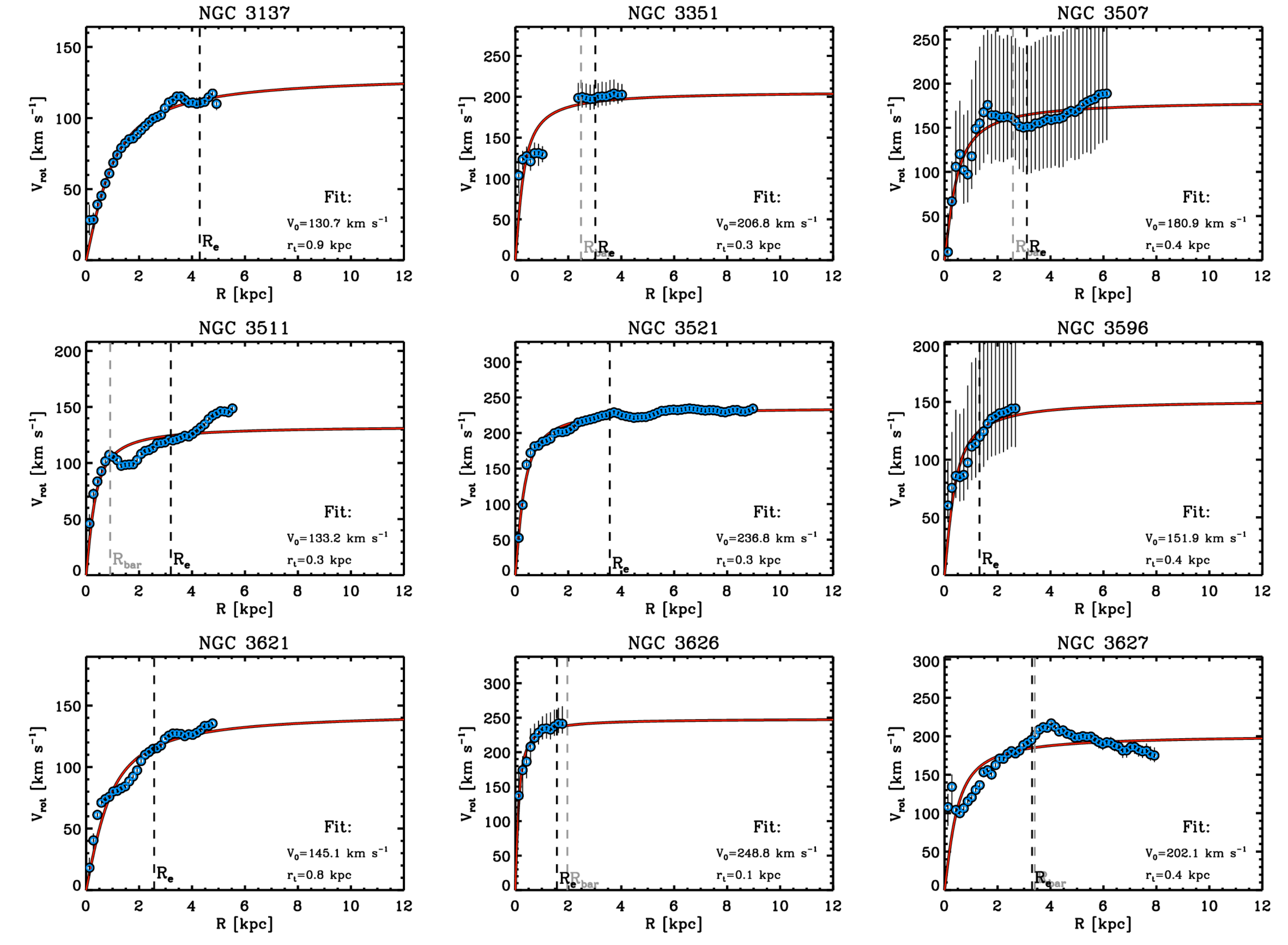}  
\caption{Same as Figure\ \ref{RC_examples.fig}, but showing the full galaxy sample}
\label{RC_all2.fig}
\end {figure*}

\begin {figure*}[tb]
\centering
  \includegraphics[width=0.80\textwidth]{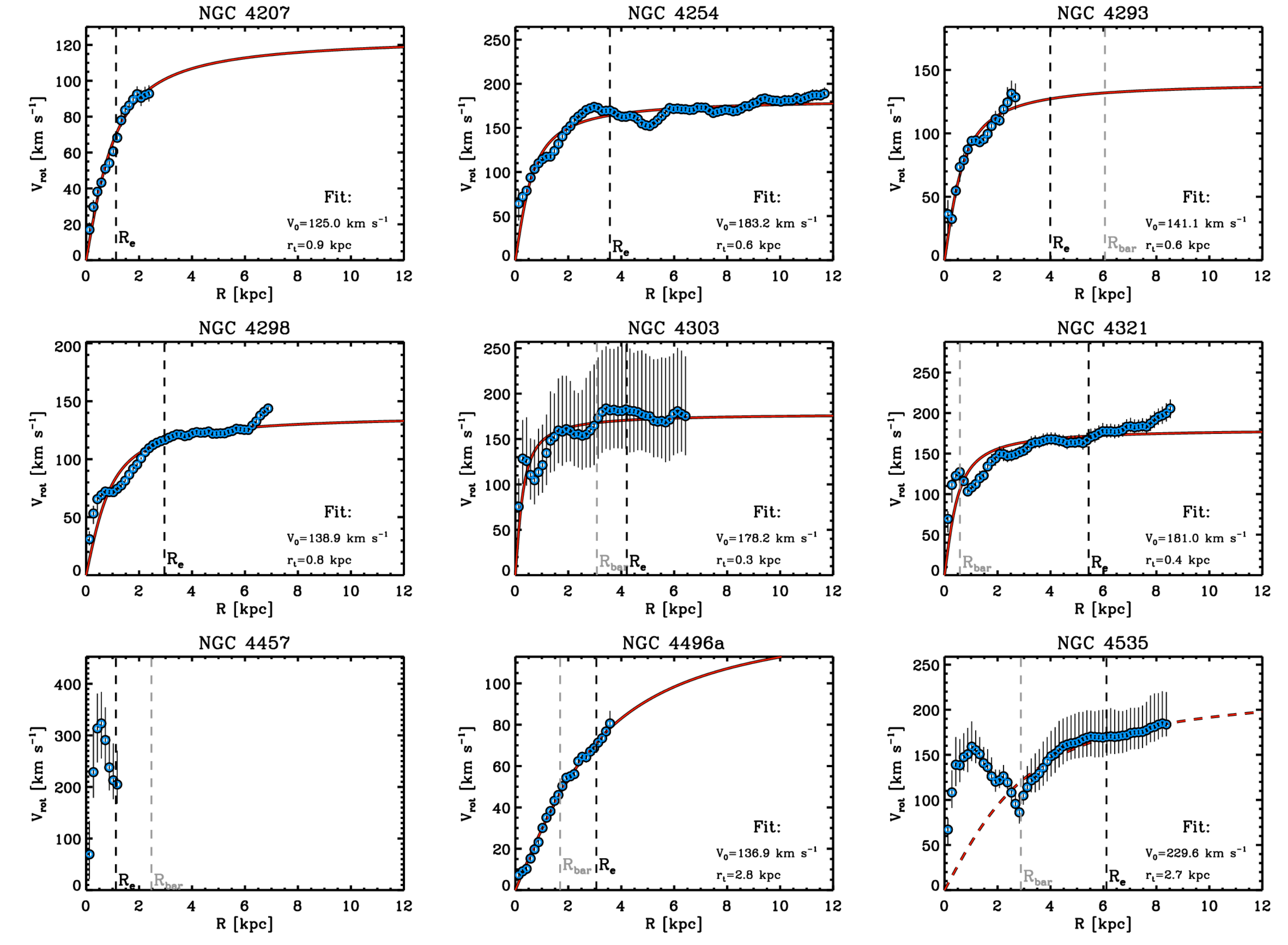}
  \includegraphics[width=0.80\textwidth]{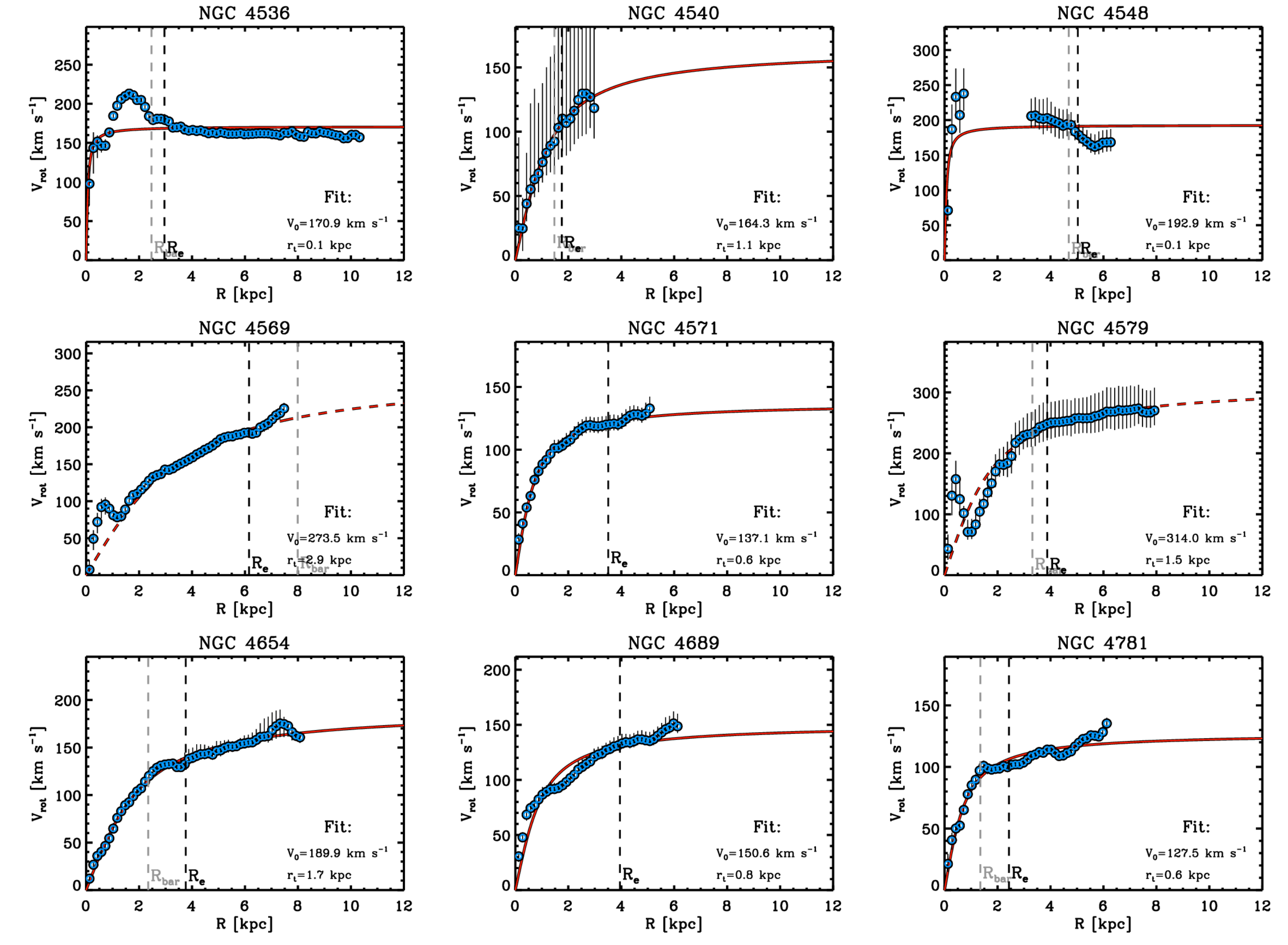}  
\caption{Same as Figure\ \ref{RC_examples.fig}, but showing the full galaxy sample}
\label{RC_all3.fig}
\end {figure*} 

\begin {figure*}[tb]
\centering
  \includegraphics[width=0.80\textwidth]{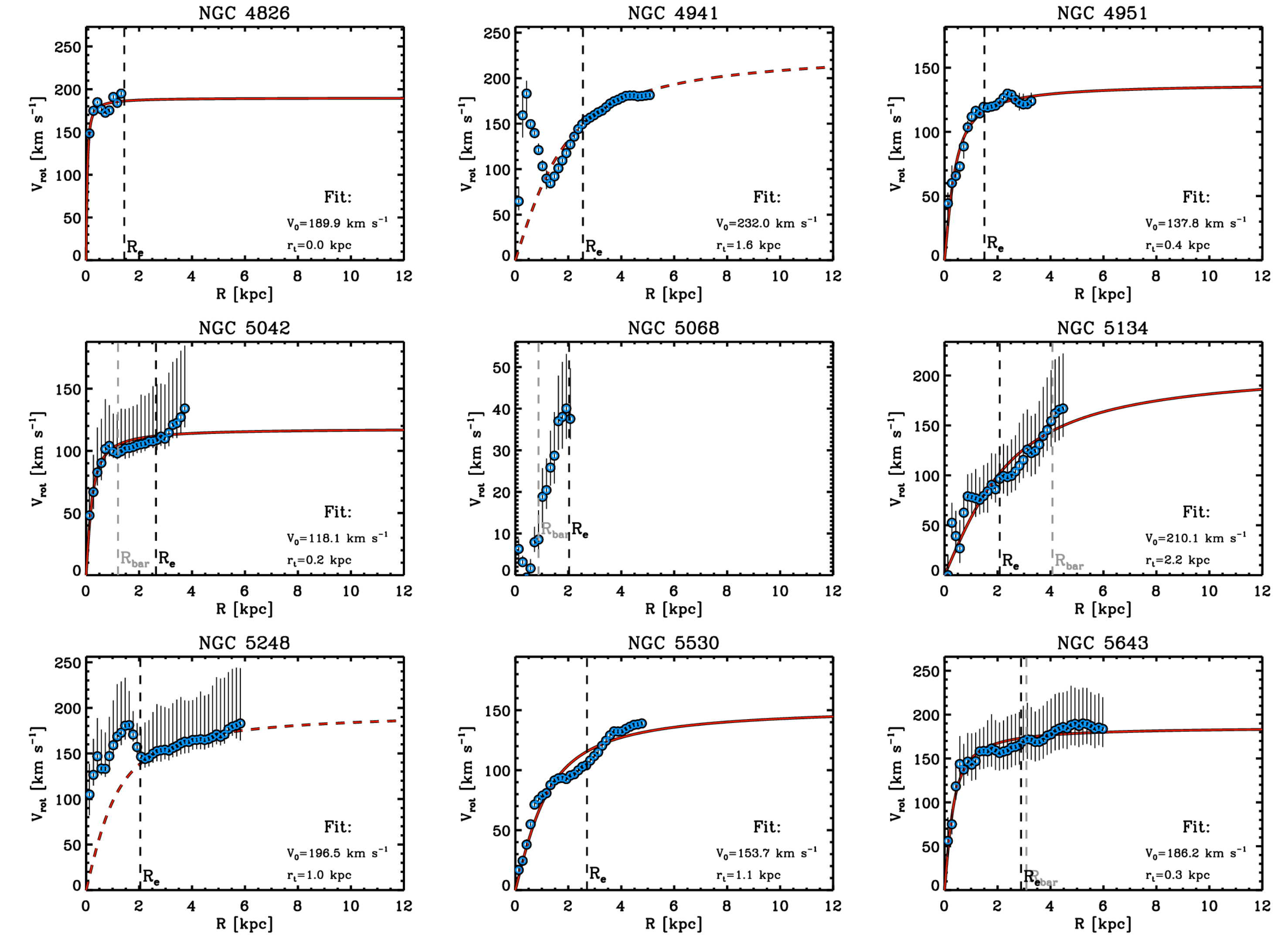}
  \includegraphics[width=0.80\textwidth]{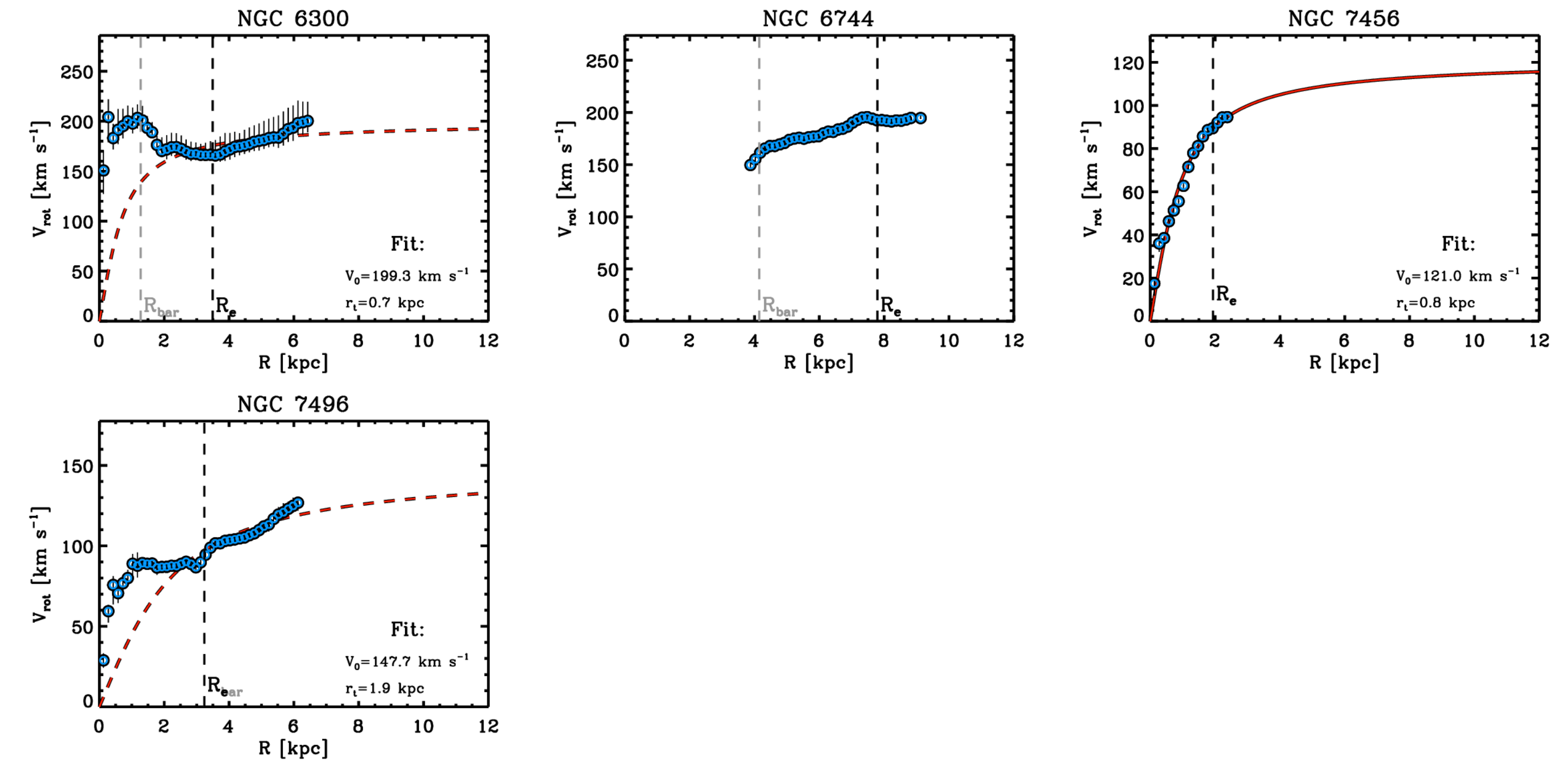}  
\caption{Same as Figure\ \ref{RC_examples.fig}, but showing the full galaxy sample}
\label{RC_all4.fig}
\end {figure*}

\begin{table}[tb]
\centering
\caption{Tabulated rotation curves for the full sample of 67 galaxies.}
\begin{tabular}{l*{5}{c}r}
\hline
\hline
R & Galaxy & $V_{{\rm rot}}^1$ & $V_{{\rm rot,err+}}^2$ & $V_{{\rm rot,err-}}^3$\\
 $[kpc]$ & & $[\mathrm{km~s^{-1}}$]  & $[\mathrm{km~s^{-1}}$] & $[\mathrm{km~s^{-1}}$] \\
 \hline
0.125  & IC~1954 &    7.4 &  5.0  &  8.8    \\
0.275  & IC~1954 &   41.3 &  2.5  &  7.6    \\
...                                        \\
5.075  & IC~1954  &  132.0 &   4.9  &  2.3   \\
5.225  & IC~1954  &  123.5 &   4.0  &  4.7   \\  
...                                         \\
0.125  & NGC~7496 &  28.9  &  4.2 &  4.6   \\  
0.275  & NGC~7496 &  59.4  &  4.1 &  7.1  \\  
...                                       \\
5.975  & NGC~7496 & 124.9  &  5.3 &   4.3  \\  
6.125  & NGC~7496 & 126.8  &  3.9 &  4.9  \\  
\hline
\hline
\end{tabular}
\tablecomments{$^1$: Rotation velocity $V_{{\rm rot}}$ per radial bin; $^2$: Upper velocity error; $^3$: Lower velocity error.  Velocity errors are based on the 16th and 84th percentiles of the 100 rotation curve realizations (see Section~\ref{final_rcs.sec} for details).}
\label{Vrot.tbl}
\end{table}

\section{Orientations based on 7m fits}
\label{7m.sec}

\startlongtable
\begin{deluxetable}{DDD}
\tablecaption{Orientation parameters based on fits to the 7m PHANGS-ALMA CO maps}
\tablehead{
\multicolumn2c{ID} & \multicolumn2c{$\phi$} & \multicolumn2c{$i$}\\ \multicolumn2c{} & \multicolumn2c{[deg]} & \multicolumn2c{[deg]}
}
\decimals
\startdata
${\rm IC~1954 }$  &   52.9  &    62.9     \\     
${\rm IC~5273 }$  &   49.0  &   231.8     \\     
${\rm NGC~0628}$  &  -  &    21.4     \\   
${\rm NGC~0685}$  &   35.0  &   101.1     \\     
${\rm NGC~1087}$  &   32.4  &    -1.3     \\     
${\rm NGC~1097}$  &  -  &   127.4     \\ 
${\rm NGC~1300}$  &  -  &   275.0     \\     
${\rm NGC~1317}$  &  -  &   210.6     \\     
${\rm NGC~1365}$  &  -  &   202.0     \\ 
${\rm NGC~1385}$  &   41.3  &   183.5     \\     
${\rm NGC~1433}$  &  -  &   198.8     \\ 
${\rm NGC~1511}$  &  -  &   293.7     \\     
${\rm NGC~1512}$  &  -  &   263.5     \\ 
${\rm NGC~1546}$  &   66.1  &   146.5     \\     
${\rm NGC~1559}$  &   52.7  &   245.9     \\     
${\rm NGC~1566}$  &   27.1  &   215.5     \\     
${\rm NGC~1672}$  &  -  &   128.2     \\ 
${\rm NGC~1792}$  &   58.2  &   318.4     \\     
${\rm NGC~1809}$  &   72.2  &   139.6     \\    
${\rm NGC~2090}$  &   65.4  &   192.4     \\     
${\rm NGC~2283}$  &   34.4  &    -4.0     \\     
${\rm NGC~2566}$  &  -  &   312.5     \\ 
${\rm NGC~2775}$  &   42.3  &   157.1     \\     
${\rm NGC~2835}$  &   46.0  &     0.5     \\     
${\rm NGC~2903}$  &   63.0  &   204.0     \\     
${\rm NGC~2997}$  &   32.8  &   107.5     \\     
${\rm NGC~3059}$  &  -  &   -13.0     \\     
${\rm NGC~3137}$  &   70.3  &     0.1     \\     
${\rm NGC~3351}$  &  -  &   192.8     \\ 
${\rm NGC~3507}$  &    7.8  &    57.0     \\     
${\rm NGC~3511}$  &   70.3  &   257.4     \\     
${\rm NGC~3521}$  &   67.4  &   343.3     \\     
${\rm NGC~3596}$  &  -  &    78.0     \\     
${\rm NGC~3621}$  &   63.7  &   343.7     \\     
${\rm NGC~3626}$  &  -  &   162.6     \\ 
${\rm NGC~3627}$  &   53.8  &   171.1     \\     
${\rm NGC~4207}$  &  -  &   119.2     \\     
${\rm NGC~4254}$  &   25.1  &    68.6     \\
${\rm NGC~4293}$  &  -  &    63.6     \\ 
${\rm NGC~4298}$  &   55.8  &   313.7     \\     
${\rm NGC~4303}$  &  -  &   314.4     \\     
${\rm NGC~4321}$  &   30.3  &   155.0     \\     
${\rm NGC~4457}$  &  -  &    76.8     \\ 
${\rm NGC~4496}$  &   30.2  &    50.6     \\     
${\rm NGC~4535}$  &   44.8  &   179.2     \\     
${\rm NGC~4536}$  &   65.9  &   304.1     \\     
${\rm NGC~4540}$  &   40.5  &    19.0     \\     
${\rm NGC~4548}$  &  -  &   136.5     \\ 
${\rm NGC~4569}$  &  -  &    17.6     \\ 
${\rm NGC~4571}$  &   27.7  &   217.2     \\     
${\rm NGC~4579}$  &   40.2  &    89.9     \\     
${\rm NGC~4654}$  &   57.1  &   123.3     \\     
${\rm NGC~4689}$  &   37.5  &   163.9     \\     
${\rm NGC~4781}$  &   55.2  &   289.7     \\     
${\rm NGC~4826}$  &   46.9  &   294.6     \\     
${\rm NGC~4941}$  &   46.6  &   203.1     \\     
${\rm NGC~4951}$  &   65.0  &    91.1     \\     
${\rm NGC~5042}$  &   48.5  &   191.8     \\     
${\rm NGC~5068}$  &  -  &   339.3     \\     
${\rm NGC~5134}$  &  -  &   319.2     \\    
${\rm NGC~5248}$  &   37.2  &   108.2     \\     
${\rm NGC~5530}$  &   57.0  &   306.3     \\     
${\rm NGC~5643}$  &  -  &   315.8     \\    
${\rm NGC~6300}$  &   51.1  &   107.5     \\     
${\rm NGC~6744}$  &   51.5  &    14.1     \\     
${\rm NGC~7456}$  &   51.7  &    15.6     \\     
${\rm NGC~7496}$  &   29.8  &   191.0     \\
\enddata
\label{7m.tbl}
\end{deluxetable}

\end{document}